\documentclass[pdftex,superscriptaddress,aps,onecolumn]{revtex4-1}
\usepackage{multirow,enumerate,epsfig,graphics,amssymb,amsmath,subeqnarray,mathrsfs,color,xcolor}
\usepackage{color,epsfig,graphics,amssymb,amsmath,subeqnarray,graphicx,amsthm,subfigure,mathrsfs}
\usepackage[colorlinks=true, citecolor=red, linkcolor=blue ]{hyperref}

\newcommand{\dd}{\mathrm{d}}
\newcommand{\ee}{\mathrm{e}}

\newcommand{\nb}{\mathbf{n}}

\newcommand{\sigmab}{\boldsymbol{\sigma}}

\newcommand{\ub}{\mathbf{u}}
\newcommand{\eb}{\mathbf{e}}

\newcommand{\xb}{\mathbf{x}}\newcommand{\rb}{\mathbf{r}}
\newcommand{\xib}{\boldsymbol\xi}
\newcommand{\Xb}{\mathbf{X}}

\let\grad\nabla
\let\grad\nabla

\newcommand{\pard}[2]{\frac{\partial #1}{\partial #2}}\newcommand{\totd}[2]{\frac{\mathrm{d}#1}{\mathrm{d}#2}}

\makeatletter
\def\sgn{\mathop{\operator@font sgn}}
\def\threevdots{\vbox{\baselineskip1\p@ \lineskiplimit\z@
  \kern6\p@\hbox{.}\hbox{.}\hbox{.}}}
\makeatother

\begin{document}
\title{Collective dissolution of  microbubbles}
\author{S\'ebastien Michelin}
\email{sebastien.michelin@ladhyx.polytechnique.fr}
\affiliation{LadHyX -- D\'epartement de M\'ecanique, Ecole Polytechnique -- CNRS, 91128 Palaiseau, France.}
\affiliation{Department of Applied Mathematics and Theoretical Physics, University of Cambridge, Cambridge CB3 0WA, United Kingdom.}
\author{Etienne Gu\'erin}
\affiliation{Department of Mechanical and Aerospace Engineering, University of California, San Diego, 9500 Gilman Drive, La Jolla, California 92093-0411, United States}
\author{Eric Lauga} 
\email{e.lauga@damtp.cam.ac.uk}
\affiliation{Department of Applied Mathematics and Theoretical Physics, University of Cambridge, Cambridge CB3 0WA, United Kingdom.}
\date{\today}

\begin{abstract}

A microscopic bubble of soluble gas always dissolves in finite time in an under-saturated fluid. This diffusive process is driven by the difference between the gas concentration near the bubble, whose value is governed by the internal pressure through Henry's law, and the concentration in the far field. The presence of neighbouring bubbles can significantly slow down this process by increasing the effective background concentration and reducing the diffusing flux of dissolved gas experienced by each bubble. We develop theoretical modelling of such diffusive shielding process in the case of small microbubbles whose internal pressure is dominated by Laplace pressure.  We first use an exact semi-analytical solution to capture the case of two bubbles and analyse in detail the shielding effect as a function of the distance between the bubbles and their size ratio. While we also  solve exactly for the Stokes flow around the bubble, we show that hydrodynamic effects are mostly negligible except in the case of almost-touching bubbles. 
In order to   tackle the case of multiple bubbles, we then derive  and validate two analytical approximate yet generic frameworks, first using the method of reflections and then by proposing a self-consistent continuum description. Using both modelling frameworks, we examine the dissolution of regular one-, two- and three-dimensional bubble lattices. Bubbles located at the edge of the lattices dissolve first, while innermost bubbles benefit from the diffusive shielding effect, leading to the inward propagation of a dissolution front within the lattice. We show that diffusive shielding  leads to severalfold increases in the dissolution time which  grows logarithmically with the number of bubbles in one dimensional lattices  and algebraically in two and three dimensions, scaling respectively as its square root and $2/3$-power.  We further illustrate the sensitivity of the dissolution patterns to initial fluctuations in bubble size or arrangement in the case of large and dense lattices, as well as non-intuitive oscillatory effects.

\end{abstract}
\maketitle

\section{Introduction}
Bubbles  are beautiful examples of the  interplay between  thermodynamics and physics   at the interface of two non-miscible phases and as such have long fascinated   physicists~\cite{degennes_book}.  Beyond the nucleation, growth, evolution and collapse of single bubbles, suspensions of many bubbles have attracted much attention for their interesting collective physical properties~\cite{leal80,manga1995collective,guazzelli2011physical}.  The flow of bubbles is    important in many industrial applications~\cite{brennen1995,pettigrew1994}, in geophysics \cite{llewellin2005bubble}, but also in the bio-medical world. For example, and thanks to our fundamental understanding of their acoustic forcing~\cite{plesset77,Leighton1994}, small bubbles can be used as contrast agents in ultrasound imaging~\cite{lindner04}. They may also have serious physiological consequences, such as embolism, a condition well-known to deep-sea divers subject to decompression sickness. More generally, microbubbles play important medical roles  in the blood stream~\cite{barak05} which further motivates   in-depth understanding of their individual and collective dynamics. Avoiding bubble nucleation and growth is also considered as a critical constraint for trees and other plants~\cite{tyree1991,cochard2006}.

From a fluid mechanics standpoint, two main points of view, or types of questions, have been considered in the dynamics of small bubbles. The classical approach, originating  from the work of Lord Rayleigh~\cite{rayleigh1917}, focuses   on the fluid mechanics {outside} the bubble, neglecting physico-chemical exchanges between the gas and liquid phases. The resulting  classical mathematical model for such  {inertial} bubble phenomena, namely the Rayleigh-Plesset equation~\cite{leal},  has been adapted to account for many different physical situations~\cite{plesset77}. Initially derived to capture  the axisymmetric collapse of an empty cavity as predicted by the Bernoulli equation~\cite{rayleigh1917,lamb1932}, this modelling framework has been extended to include the effects of surface tension and viscosity, and is the basis for classical studies on acoustic forcing, growth and collapse of cavitation (vapour) bubbles~\cite{nappiras80,brennen1995}. Studying these phenomena is crucial to understand, for example, the physics of bubble sonoluminescence~\cite{brenner02}. Further extensions   include non-spherical bubble oscillations and resonance under acoustic forcing~\cite{lamb1932,leal,lauterborn2010physics}.

A second class of problems, sometimes referred to as  {diffusive} bubble phenomena, focuses more specifically on (i) the heat and/or mass exchanges between the bubble and its liquid environment, (ii) the resulting bubble dynamics and (iii) the coupling between the bubble motion and  the diffusive dynamics in the liquid of the dissolved  quantity.  The nature and properties of the dissolved species, and the physical exchange on the surface of the bubble,  are the main distinguishing features between   vapour and gas bubbles~\cite{duda1971,plesset77}.   Specifically, for   vapour bubbles, interface processes are dominated by the vaporization/condensation of the liquid into the   bubble driven by the diffusion and transport of heat~\cite{prosperetti2017}, which is typically much faster than mass transport.   In contrast, for gas bubbles the liquid and gas phases are of two different chemical natures; the driving physical mechanisms are the dissolution of the gas into the liquid to maintain equilibrium at its surface as quantified by Henry's law~\cite{plesset77}, and the slow diffusion of the dissolved gas into the liquid phase.   Note that this second case  presents many formal similarities with the dissolution process of droplets~\cite{duncan2006}, their evaporation~\cite{carrier2016} or even dissolving solids~\cite{cable87}.

In the case of dissolving gas bubbles,  changes in the bubble radius are driven by mass transport and diffusion, and the general individual unsteady dynamics were described by Epstein \& Plesset~\cite{epstein1950}.   For most gases under ambient conditions, diffusion of the dissolved gas  is much faster than the evolution in the bubble radius due to the molar density contrast between the concentration of gas in the bubble and in the liquid~\cite{lohse2015}. As a result, transient and convective effects are essentially negligible except initially or when fluctuations in the background pressure are taken into account~\cite{duda1969,penaslopez2016,penas2017history}.   This  is in stark contrast with the heat-driven dynamics of vapour bubbles for which both time-scales are comparable and inertial effects can be significant~\cite{plesset1954}, or with gas bubbles with comparable concentrations in both phases~\cite{subramanian80}. For most dissolving gas bubbles, this separation of time-scales justifies the classical quasi-steady approximation~\cite{epstein1950} in  which   the diffusive dynamics of the dissolved gas takes place around a frozen bubble geometry.  This approach has been used in many extensions to this theory, including to multiple-component gas bubbles~\cite{ward75,weinberg1980}, and represents the classical framework to study the rapid dissolution of gas microbubbles in undersaturated environments~\cite{ljunggren97}.

For  micro- and nanobubbles~\cite{lohse2015}, inertia can be neglected and  the liquid flow  is viscous~\cite{happel,kimbook}. In the quasi-steady framework described above, the typical hydrodynamic pressure is also small in comparison with capillary pressures so that the bubble remains spherical. The resulting mathematical model, essentially identical to that of Epstein \& Plesset \cite{epstein1950}, was tested experimentally for the dissolution of microbubbles~\cite{duncan04} and microdroplets~\cite{duncan2006,su2013mass}. A critical ingredient in such dissolution dynamics is the description of the physico-chemical equilibrium at the interface (i.e.~Henry's law), which is most often simply approximated as a direct proportionality between the dissolved gas concentration in the liquid phase and its partial pressure in the bubble~\cite{plesset77}, thereby neglecting the role of surfactants~\cite{fyrillas1995} or complex molecular surface kinetics~\cite{yu2009}.

The dynamics and dissolution of small-sized bubbles   have attracted much attention because of their importance in industrial and biomedical applications and, recently,    as a result of the puzzling discovery of nanobubbles~\cite{lohse2015}. Indeed, the  Epstein \& Plesset framework predicts bubble dissolution times scaling as $R^2$ where $R$ is the bubble radius~\cite{epstein1950}, and thus free nanobubbles should in fact not be observable.  A series of recent studies resolved the mystery in the case of nanobubbles pinned on a surface in a supersaturated environment by showing that an equilibrium could be reached between the influx of gas due to the supersaturation and the outflux induced by the large diffusion rates occurring  on the edges of the bubble (coffee stain effect)~\cite{weijs2013surface,lohse2015b}. Surface pinning was further identified to play a key role in the coarsening process surface nanobubbles and droplets, in particular stabilizing them against Ostwald ripening~\cite{dollet2016,zhu2018}.

Most of the studies mentioned above consider the dissolution dynamics of a single isolated bubble. Collective effects have been investigated for cavitation problems~\cite{bremond06,bremond2006interaction}.  However, one expects collective effects to also play an important role in the dissolution of gas bubbles. Indeed, in a collection of bubbles, each bubble acts as a source releasing  gas into the liquid phase, thus reducing locally the undersaturation and slowing down the dissolution of neighbouring bubbles~\cite{weijs12}.  
Notably, a similar diffusive	 shielding effect   was   identified   for  bubbles   in contact with, or in the vicinity of, a solid surface~\cite{penaslopez2015}. Recent experiments and simulations on the dissolution of surface microdroplets also considered such shielding physics~\cite{laghezza2016,bao2018}.

In this work, we  address  the role of collective effects on the diffusion of dissolved gas  within the liquid phase. Specifically, we characterise  the   dissolution dynamics of gas microbubbles in under-saturated environments in the limit where the pressure inside the bubble is dominated by surface tension (i.e.~limit of small bubbles). We  quantify the impact of the arrangement of a group of $N$ bubbles  on their total dissolution time  as well as on the time-dependent dissolution pattern. We ignore confining surfaces to focus on the bulk dissolution problem, and follow the classical quasi-steady framework of Epstein \& Plesset,   well justified for most dissolved gases for which the bubble molecular gas concentration is   higher than the difference in dissolved gas concentration driving the   dissolution process~\cite{lohse2015}. 

After a short review of the fundamental physical assumptions behind the modelling framework in the case of a single bubble in \S~\ref{sec:single}, including a discussion of the relevant time scales, we focus in \S~\ref{sec:bispherical} on the two-bubble configuration as a test problem. An exact semi-analytical solution is obtained using bi-spherical coordinates, and we analyse the shielding effect as a function of the distance between, and size ratio of, the bubbles. We also analyse the effect of hydrodynamics  and show that it is mostly negligible except in the case of almost-touching bubbles. A generic analytical approximate framework using the method of reflections is then proposed and validated for the $N$-bubble problem  in \S~\ref{sec:asymptotic_models}. Using this framework, the dissolution of regular one-, two- and three-dimensional bubble lattices is addressed in \S~\ref{sec:results}. In particular, we obtain general results on the shielding effects and dissolution patterns   and a continuum model is proposed that emphasises the fundamental differences between one-, two- and three-dimensional lattices. Finally, \S~\ref{sec:conclusions} summarises our findings and offers some perspectives.

\section{Dissolution of  an isolated microbubble}
\label{sec:single}
We first focus on the reference problem of an isolated single-component gas bubble of radius $R(t)$ dissolving in an infinite incompressible fluid  of density $\rho$ and dynamic viscosity~$\eta$. The concentration of dissolved gas is noted $C(\xb,t)$ and its diffusivity in the fluid is $\kappa$. Far from the bubble, the fluid is at rest with pressure $p_\infty$ and a dissolved-gas concentration of $C_\infty$. At the surface of the bubble, thermodynamic equilibrium imposes  $ C_s=P_i/K_H$ (Henry's law)  where $C_s$ and $P_i$ are the uniform surface concentration and internal bubble pressure, respectively. 

The diffusion of gas out of the bubble is responsible for the dynamic evolution of $R(t)$. Mass conservation on the bubble is written as 
\begin{equation}\label{eq:massbalance}
\totd{M}{t}=\totd{}{t}\left(\frac{4 \pi R(t)^3P_i(t)}{3{\cal R}T}\right)=\kappa\int_{r=R(t)}\nb\cdot\grad C\,\dd S=4\pi \kappa R(t)^2\left.\pard{C}{r}\right|_{r=R(t)},
\end{equation}
with $M(t)$ the molar content of the single-component bubble and ${\cal R}$ the ideal gas constant. Note that the last  equality in Eq.~\eqref{eq:massbalance} exploits the spherical symmetry in the diffusive flux in the case of an isolated bubble. 

The inner bubble pressure, $P_i(t)$, is given by  the superposition of three contributions
\begin{equation}
P_i=P_\infty+P_h+P_\gamma,
\end{equation}
namely, the atmospheric pressure far from the bubble ($P_\infty$), the hydrodynamic pressure induced by the fluid motion ($P_h$), and the capillary pressure ($P_\gamma$). The latter is retained here, in contrast to most analytical derivations on bubble dissolution and growth~\cite{epstein1950,penas2017history}, effectively restricting their results to bubble radii greater than $1$--$10~\mu$m, and therefore excluding the final stages of the bubble's life.  In the following, we assume that hydrodynamic pressure is negligible; when the fluid motion results from the bubble collapse, this effectively assumes that the Capillary number, $\mbox{Ca}=\eta \dot{R}/\gamma$, is always small, i.e.~$\dot{R}\ll\,\gamma/\eta\approx 70$ m.s$^{-1}$ for water under normal conditions. In this limit, the bubble remains spherical at all times, and  its internal pressure is given by
\begin{equation}
P_i(t)=P_\infty+2\gamma/R(t).
\end{equation}

The diffusion of the dissolved gas around the bubble occurs on the typical time scale $\tau_\textrm{diff}\sim R_0^2/\kappa$, with $R_0$ the characteristic (initial) bubble radius. In contrast, by scaling
Eq.~\eqref{eq:massbalance}, we see that the   dissolution of the bubble occurs on the time scale  $\tau_\textrm{diss}\sim \tau_\textrm{diff}(K_H/\mathcal{R}T)$, which  is also the typical flow time-scale since  the motion of the fluid is  driven by the shrinking of the bubble. 
The ratio of these two time-scales is therefore also a relative measure of the importance of unsteady and convective effects in the dissolved gas dynamics compared to diffusion, and it is written as $\Lambda=\tau_\textrm{diff}/\tau_\textrm{diss}\sim\mathcal{R}T/K_H=C_s/\rho_g$, where $\rho_g$ and $C_s$ are the gas density and solubility (i.e.~surface concentration) in standard conditions~\cite{lohse2015}, i.e.~the ratio of the chemical species' interfacial concentration in the liquid and gas phases. It is essential to note here that this non-dimensional constant critically depends on the material properties of the gas species considered and therefore varies from one gas to another.
For most dissolved gases in ambient conditions  (including O$_2$, N$_2$, H$_2$), this ratio $\Lambda$ is small as the constant $K_H$ appearing in Henry's  law  is in the range $K_H\approx 8\times10^4$--$1.5\times10^5$~J.mol$^{-1}$ while $\mathcal{R} T\approx 2.4\times10^3 $~J.mol$^{-1}$ (e.g. $\Lambda_{\textrm{H}_2}\approx 2\,10^{-2}$, $\Lambda_{\textrm{O}_2}\approx 3\,10^{-2}$ and $\Lambda_{\textrm{N}_2}\approx 1.5\,10^{-2}$~\cite{lohse2015}). For other gases such as CO$_2$ or NH$_3$, $\Lambda$ is not small ($\Lambda_{\textrm{CO}_2}\approx 0.8$ and $\Lambda_{\textrm{NH}_3}\approx 250$~\citep{lohse2015}) and this separation of time-scales breaks down. 

In this paper, we focus exclusively on gases with $\Lambda\ll 1$ so that convective transport of the dissolved gas is negligible compared to diffusion. Furthermore, this limit ensures $\tau_\textrm{diff}\ll \tau_\textrm{diss}$  justifies the quasi-steady approximation considered throughout this paper: when considering the bubble dissolution process, the dissolved gas distribution around the bubble is at each instant equal to its concentration if the bubble radius was fixed. The validity of this quasi-steady framework is analyzed quantitatively in Appendix~\ref{app:quasisteady}.

Under this quasi-steady assumption, the concentration of dissolved gas around the bubble is  obtained by solving a steady diffusion problem for $C$ at each instant, $\kappa\nabla^2 C=0$, with time-dependent boundary conditions on the  surface of the bubble given by
\begin{equation}
C(r=R(t),t)=\frac{1}{K_H}\left(P_\infty+\frac{2\gamma}{R(t)}\right).
\end{equation}

Throughout the paper, we focus on this   limit of negligible hydrodynamic pressure ($\mbox{Ca}\ll 1$, spherical bubble) and quasi-steady diffusive concentration ($\Lambda\ll 1$), for both one or multiple bubbles. Notably, these two assumptions are related: from the time scales introduced above, we estimate $\mbox{Ca}\sim\Lambda(R^*/R_0)$, with $R^*=\eta\kappa/\gamma\approx\,10^{-10}$~m. The limiting assumption is therefore on $\Lambda\sim\mathcal{R}T/K_H=C_s/\rho_g$ which depends only on the ambient temperature and the nature of the gas considered (and not on the bubble size).

We use  $R_0$, $\tau_\textrm{diss}=4K_H R_0^2/(3\kappa \cal{R} T)$ and $2\gamma/K_HR_0$ as characteristic length, time and concentration scales, and from now on only consider non-dimensional quantities. Writing $a(t)=R(t)/R_0$ and $c=(C-C_\infty)/(2\gamma/K_HR_0)$, the spherically-isotropic concentration profile around the bubble is obtained explicitly as
\begin{equation}
c(r,t)=\frac{1+(1-\zeta)r_0 \,a(t)}{r},
\end{equation}
and the non-dimensional molar flux into the bubble, $q(t)=\displaystyle(K_H/4\pi\gamma)\int\nb\cdot\grad C\,\dd S$, is
\begin{equation}
q(t)=\left(1+\frac{3r_0 a}{2}\right)a\dot{a}=-2\Big[1+(1-\zeta)r_0 a\Big],\label{eq:single_bubble_dyn}
\end{equation}
where
\begin{equation}
r_0=\frac{P_\infty R_0}{2\gamma},\qquad \zeta=\frac{K_HC_\infty}{P_\infty}.
\end{equation}
The non-dimensional initial radius $r_0$ is a relative measure of the influence of background vs.~capillary pressure: for  $r_0\ll 1$, the internal pressure of the bubble is dominated by surface tension, while $r_0\gg 1$ corresponds to situations where the internal pressure (and therefore gas concentration) is independent of the bubble radius. Note that for $r_0\gg 1$, the dissolution equations above are formally identical to that of dissolving droplets. The saturation parameter $\zeta$ characterises the saturation of the environment, with $\zeta>1$ (resp.~$\zeta<1$) corresponding to an over-saturated (resp.~under-saturated) liquid while $\zeta=0$ corresponds to a solute-free environment. Note that $q(t)$ is counted positively (resp. negatively) for growing (resp. shrinking/dissolving) bubbles. 
In the following, we focus exclusively on $\zeta\leq 1$ corresponding to a fluid under-saturated (or exactly saturated) in gas.

Integrating Eq.~\eqref{eq:single_bubble_dyn} with initial conditions $a(0)=1$ leads to
\begin{equation}
\frac{1+2\zeta}{(1-\zeta)^2r_0}\left[\frac{1}{r_0(1-\zeta)}\log\left(\frac{1+(1-\zeta)r_0\,a(t)}{1+r_0(1-\zeta)}\right)-a(t)+1\right]+\frac{3}{2(1-\zeta)}(a(t)^2-1)=-4t,
\end{equation}
and the total dissolution time $T_f$ such that $a(T_f)=0$ is obtained as
\begin{equation}
T_f=\frac{3}{8(1-\zeta)}+\frac{1+2\zeta}{4r_0(1-\zeta)^2}\left(\frac{\log(1+(1-\zeta)r_0)}{(1-\zeta)r_0}-1\right).
\end{equation}

This result is generic with respect to the background conditions (pressure and concentration). Several  classical limits can be identified, namely
\begin{enumerate}[(i)]
\item{\emph{Capillary-dominated regime, $r_0\ll 1$},
\begin{equation}
a(t)=\sqrt{1-\frac{t}{T_f}},\qquad T_f=\frac{1}{4}\cdot
\end{equation}
}
\item{\emph{Equilibrium background conditions, $\zeta=1$}: the background pressure and concentration are at thermodynamic equilibrium, leading to
\begin{equation}
\frac{a(t)^2(1+r_0 a(t))}{4}=T_f-t,\qquad T_f=\frac{1+r_0}{4}\cdot
\end{equation}
In particular, when $r_0\gg 1$ (negligible capillary effects), $a(t)=(1-t/T_f)^{1/3}$.
}
\item{\emph{Negligible capillary effects,} $r_0\gg 1$: in that case, where the bubble inner pressure (and concentration) is independent of the bubble size, the solution for the bubble dissolution pattern takes the same form as the capillary-dominated regime, namely  $a(t)=\sqrt{1-t/T_f}$ with a modified final time $T_f=3/[8(1-\zeta)]$ which depends on the saturation of the environment.}
\end{enumerate}

\begin{figure}
\begin{center}
\begin{tabular}{cc}
\includegraphics[height=5.4cm]{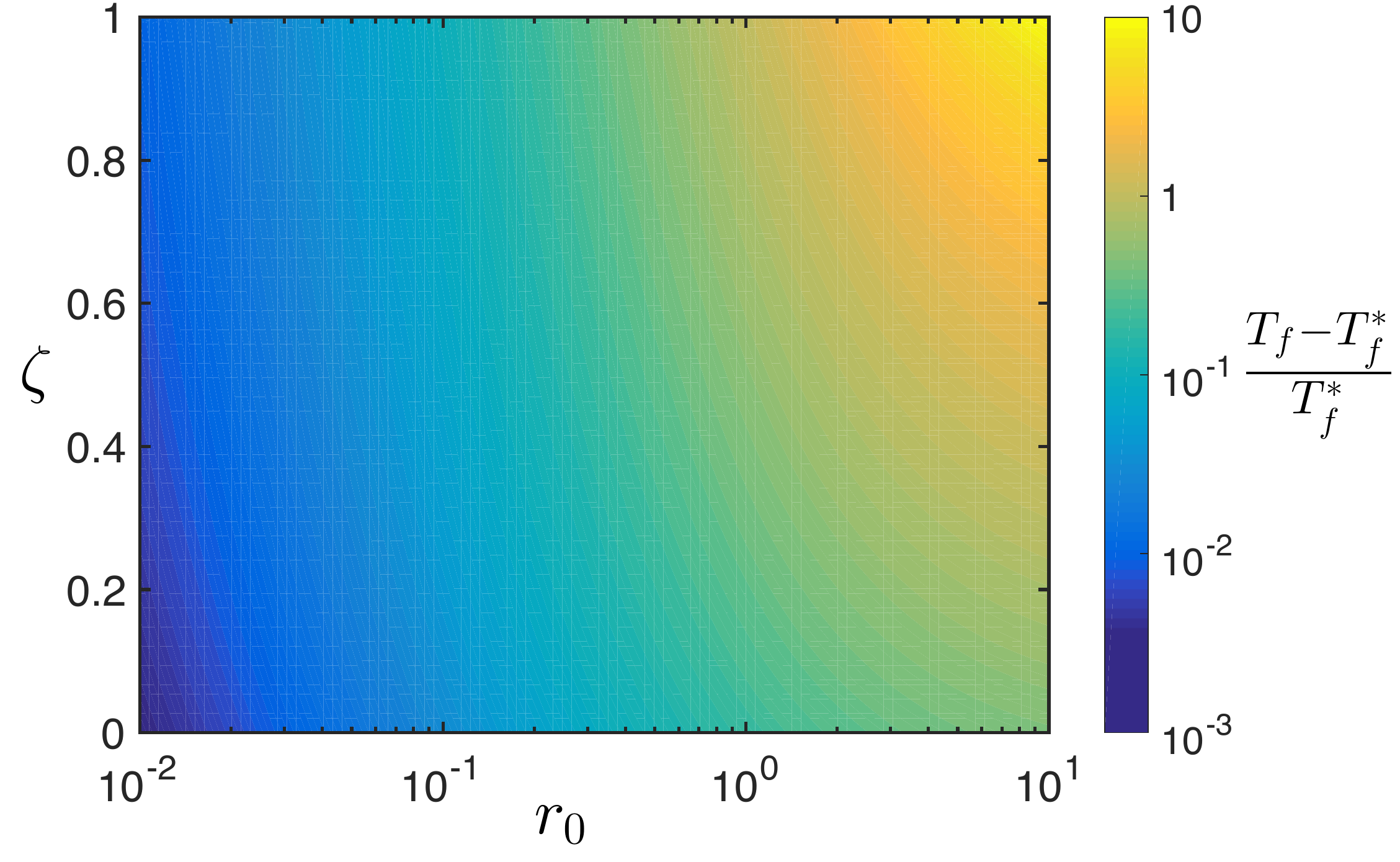}&
\includegraphics[height=5.4cm]{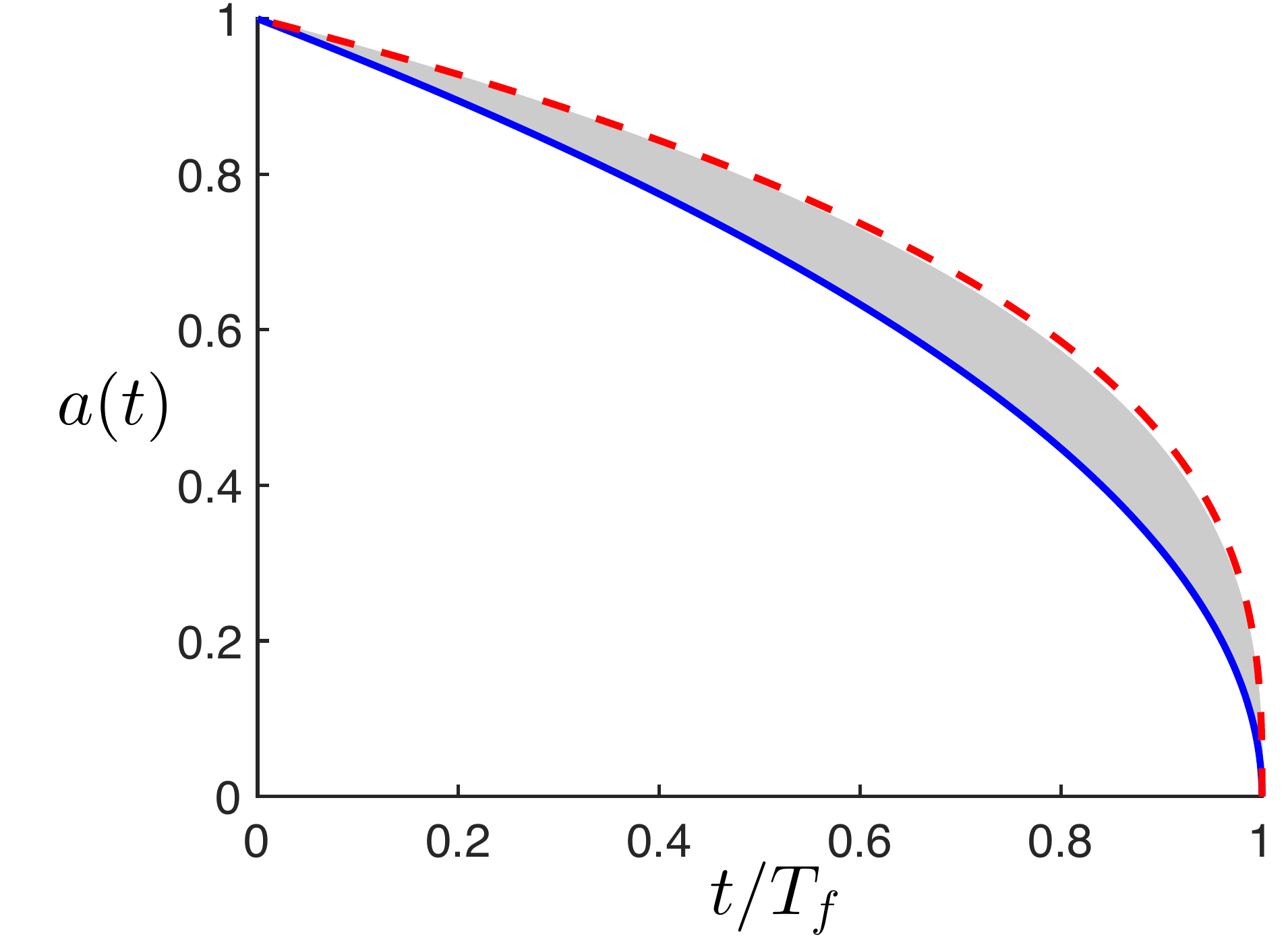}
\end{tabular}
\caption{Dissolution of a single bubble. Left: Relative change in the dissolution time, $(T_f-T_f^*)/T_f^*$, as a function of the relative bubble size ($r_0$) and chemical saturation of the environment ($\zeta$). The case $r_0\ll 1$ corresponds to the capillary-dominated regime (and serves as reference here with $T_f^*=1/4$) while for $r_0\gg 1$ the gas concentration in the bubble is independent of its size. Right: Temporal evolution of the radius of the bubble. The grey region corresponds to the envelope of all possible time-dependence for under-saturated regimes, $\zeta\leq 1$. Two limit cases are highlighted: (i) Capillarity-dominated regime $r_0\ll 1$ (solid blue line) and (ii) negligible capillarity $r_0\gg 1$ (dashed red line).
}\label{fig:tf}
\end{center}
\end{figure}

The complete bubble dynamics is illustrated on Fig.~\ref{fig:tf}, which shows that (i) atmospheric pressure (increasing $r_0$) increases the  bubble lifetime   as it increases the initial gas concentration in the bubble for fixed radius, and (ii) an undersaturated (resp.~over-saturated) background, $\zeta\leq 1$ (resp.~$\zeta>1$) tends to shorten (resp. extend) the bubble lifetime as it enhances (resp.~reduces) outward gas diffusion. 

In the rest of the paper, we focus on the capillary-dominated regime, i.e.~$r_0\ll 1$, and this single-bubble configuration, and its dissolution time $T_f^*=a_0^2/4=1/4$, will serve as a reference case against which the shielding effect of collective bubble dissolution is evaluated.

\section{Coupled dissolution of  two microbubbles}
\label{sec:bispherical}
 
\subsection{Exact solution in bispherical coordinates}
The dimensionless diffusion and hydrodynamic problems are formulated as follows. The concentration, velocity and pressure fields satisfy Laplace and Stokes equations in the fluid domain $\Omega_f$ outside the bubbles,
\begin{align}
\nabla^2 c=0,\qquad \nabla^2\ub=\grad p,\qquad \nabla\cdot\ub=0,\label{eq:laplace_stokes}
\end{align}
with decaying boundary conditions at infinity ($c,\ub,p\rightarrow 0$ for $r\rightarrow\infty$). At the surface of bubble $i$ ($i=1,\,2$), Henry's law, the impermeability condition and the absence of tangential stress are given by
\begin{align}
\left.c\right|_{r_i=a_i(t)}&=\frac{1}{a_i(t)},\label{eq:2b_bc1}\\
\nb\cdot\left.\ub\right|_{r_i=a_i(t)}&=\dot{a}_i(t)+\dot\Xb_i\cdot\nb,\label{eq:2b_bc2}\\
\left.\sigmab\cdot\nb\right|_{r_i=a_i(t)}&=0.\label{eq:2b_bc3}
\end{align}
where we note $\rb_i=\rb-\Xb_i$ and $r_i=|\rb_i|$ with $\Xb_i$ the position of the centre of mass of bubble $i$. Then,  dynamics of bubble $i$ results from the conservation of mass equation
\begin{equation}\label{eq:2b_mass}
a_i\dot{a}_i=q_i=\frac{1}{2\pi}\int_{r_i=a_i}\nb\cdot\grad c\,\dd S.
\end{equation}
Finally, the force-free conditions on each bubble provide a set of closure equations for their translation velocities $\dot{\Xb}_i=\hat{W}_i\eb_z$, where $\eb_z=(\Xb_1-\Xb_2)/|\Xb_1-\Xb_2|$ is the unit vector of the axis of symmetry in the problem. Note that while the bubbles are also torque-free, this does not provide additional information here due to the free-slip boundary condition and spherical symmetry of their boundary.

\begin{figure}
\begin{center}
\includegraphics[height=7.5cm]{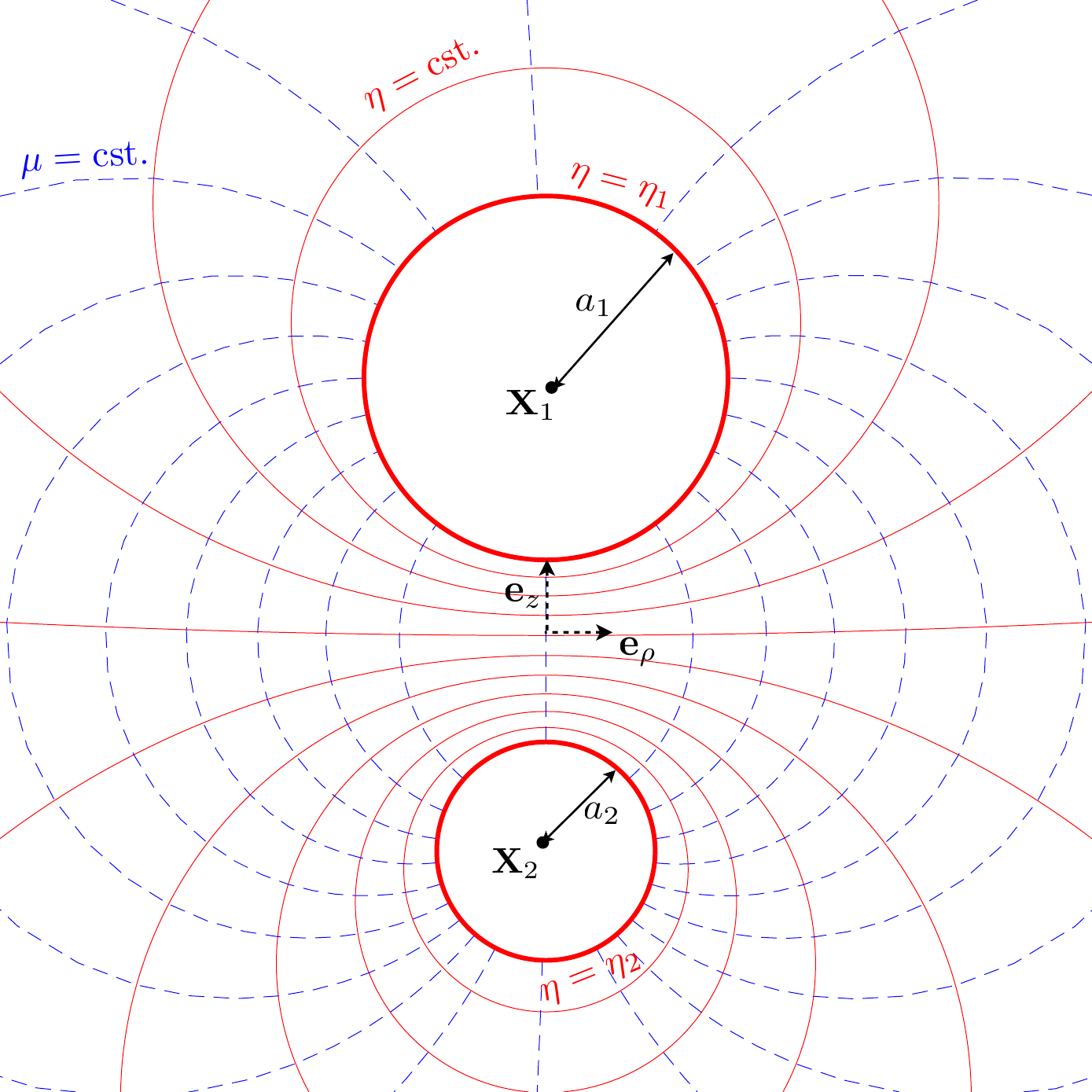}
\caption{Bispherical coordinate system for the dissolution of two bubbles of radius $a_1$ and $a_2$. Solid (red) lines are surfaces of constant $\eta$ and dashed (blue) lines correspond to constant $\mu$. The surfaces of the bubbles ($\eta=\eta_{1,2}$) are shown as thick red lines.}\label{fig:bispherical}
\end{center}
\end{figure}
A full analytical solution can be obtained for this two-bubble geometry using bi-spherical coordinates $(\eta,\mu,\phi)$, obtained from classical cylindrical polar coordinates ($\rho,\phi,z$) as 
\begin{equation}
\rho=\frac{k\sqrt{1-\mu^2}}{\cosh\eta-\mu},\qquad z=\frac{k\sinh\eta}{\cosh\eta-\mu}\cdot
\end{equation}
Surfaces of constant $\eta$ are spheres of radius $k/|\sinh\eta|$ centred in $k/\tanh\eta$ (Figure~\ref{fig:bispherical}). Here $k$ is a positive constant such that the surface of the two bubbles are given by $\eta=\eta_1>0$ and $\eta=\eta_2<0$, hence
\begin{equation}
\sinh\eta_1=\frac{k}{a_1},\quad \sinh\eta_2=-\frac{k}{a_2},\quad d=\sqrt{a_1^2+k^2}+\sqrt{a_2^2+k^2},\label{eq:2b_geom}
\end{equation}
with $d$ the distance between the centres of the spheres. Equation~\eqref{eq:2b_geom} defines $\eta_1$, $\eta_2$ and $k$ uniquely from the geometric arrangement of the two bubbles. In the following, the contact distance $d_c=d-a_1-a_2$ between the two bubbles is also used to characterise their proximity.

\subsubsection{Laplace problem}
The unique solution to the axisymmetric Laplace problem presented above for the dissolved gas concentration $c$ is~\citep{stimson1926,michelin2015}
\begin{equation}
c=\sqrt{\cosh\eta-\mu}\sum_{n=0}^\infty P_n(\mu)\left(\alpha_n\ee^{-(n+\frac{1}{2})\eta}+\beta_n\ee^{(n+\frac{1}{2})\eta}\right),
\end{equation}
with $P_n(\mu)$ the $n$-th Legendre polynomial, and $\alpha_n$ and $\beta_n$ uniquely determined to satisfy $c(\eta=\eta_i,\mu)=1/a_i$ for all $\mu$
\begin{equation}\label{eq:alphan_betan}
\alpha_n=\frac{a_1\ee^{(n+\frac{1}{2})\eta_+}-a_2\ee^{-(n+\frac{1}{2})\eta_-}}{a_1a_2\sqrt{2}\,\sinh(n+\frac{1}{2})\eta_-},\qquad \beta_n=\frac{a_2\ee^{-(n+\frac{1}{2})\eta_+}-a_1\ee^{-(n+\frac{1}{2})\eta_-}}{a_1a_2\sqrt{2}\,\sinh(n+\frac{1}{2})\eta_-},
\end{equation}
with $\eta_\pm=\eta_1\pm\eta_2$  (see details in Appendix~\ref{ap:laplace}). From this result, the total flux $q_i$ of dissolved gas into the two bubbles  can be computed as
\begin{equation}\label{eq:2bubble_fluxes}
q_1=a_1\dot{a}_1=-2k\sqrt{2}\sum_{n=0}^\infty \beta_n,\qquad q_2=a_2\dot{a}_2=-2k\sqrt{2}\sum_{n=0}^\infty \alpha_n.
\end{equation}

\subsubsection{Hydrodynamic problem}
The axisymmetric flow forced by the motion of two spherical particles or bubbles {of constant radii} is a classical problem~\cite{stimson1926,lamb1932,happel}. Its general solution, $\ub_\textrm{visc}$, can be  written in terms of a streamfunction $\psi(\eta,\mu)$ 
\begin{align}\label{eq:gensol_volcons}
\ub_\textrm{visc}&=-\frac{(\cosh\eta-\mu)^2}{k^2}\pard{\psi}{\mu}\eb_\eta+\frac{(\cosh\eta-\mu)^2}{k^2\sqrt{1-\mu^2}}\pard{\psi}{\eta}\eb_\mu,\\
\psi&=(\cosh\eta-\mu)^{-3/2}\chi(\eta,\mu),\quad \textrm{with\quad} \chi=\sum_{n=1}^\infty V_n(\mu)U_n(\eta),\label{eq:chi}
\end{align}
and
\begin{align}
V_n(\mu)&=P_{n-1}(\mu)-P_{n+1}(\mu)=\frac{2n+1}{n(n+1)}(1-\mu^2)P_n'(\mu),\\
U_n(\eta)&=A_n\cosh\left(n-\frac{1}{2}\right)\eta+B_n\sinh\left(n-\frac{1}{2}\right)\eta+C_n\cosh\left(n+\frac{3}{2}\right)\eta+D_n\sinh\left(n+\frac{3}{2}\right)\eta.\label{eq:undef}
\end{align}

In order to account for the change in radius (i.e.~the non-zero mass flux out of any closed surface that contains at least one of the bubbles), a potential flow solution, $\ub_\textrm{pot}=\grad\varphi$, must be added to the generic viscous solution above so that $\ub=\ub_\textrm{visc}+\grad\varphi$, with
\begin{equation}
\varphi=-\frac{(\cosh\eta-\mu)^{1/2}}{k\sqrt{2}}\left(Q_1\,\ee^{\eta/2}+Q_2\,\ee^{-\eta/2}\right),
\end{equation}
and $Q_i=a_i^2\dot{a}_i=a_iq_i$. 

The viscous solution is then uniquely determined by enforcing the impermeability and stress-free conditions, Eqs.~\eqref{eq:2b_bc2}--\eqref{eq:2b_bc3}, at the surface of each bubble. As a consequence we obtain (see details in Appendix~\ref{ap:hydro})
\begin{align}
U_n(\eta_i)=&\frac{3\sqrt{2}}{4(2n+1)}\left(- 2 Q_i\sinh\frac{\eta_i}{2}\sinh\frac{|\eta_i|}{2}-Q_1+Q_2\right)\left[\frac{\ee^{-(n+\frac{3}{2})|\eta|}}{2n+3}-\frac{\ee^{-(n-\frac{1}{2})|\eta|}}{2n-1}\right]\nonumber\\
&-\frac{\delta_{n1}\sqrt{2}}{3}\left(Q_1\ee^{\eta_i/2}-Q_2\ee^{-\eta_i/2}\right) + \frac{Q_i\sqrt{2}\sinh\eta_i\sinh|\eta_i|}{2(2n+1)}\ee^{-(n+\frac{1}{2})|\eta_i|}
\nonumber\\
&-\frac{k^2\hat{W}_i n(n+1)\sqrt{2}}{2(2n+1)}\left[\frac{\ee^{-(n-\frac{1}{2})|\eta_i|}}{2n-1}-\frac{\ee^{-(n+\frac{3}{2})|\eta_i|}}{2n+3}\right],\label{eq:Un}
\end{align}
\begin{align}
U_n''(\eta_i)=&-\frac{3Q_i\sqrt{2}\sinh\eta_i}{8}\left[-\frac{2\sinh^2\eta_i\,\ee^{-(n+\frac{1}{2})|\eta_i|}}{1+\cosh\eta_i}+(2n+3)\ee^{-(n-\frac{1}{2})|\eta_i|}-(2n-1)\ee^{-(n+\frac{3}{2})|\eta_i|}\right]\nonumber\\
&+\frac{3\sqrt{2}}{4}\ee^{-(n+\frac{1}{2})|\eta_i|}(Q_1-Q_2)-\frac{\delta_{n1}}{\sqrt{2}}(Q_1\ee^{\eta_i/2}-Q_2\ee^{-\eta_i/2})
\nonumber\\
&-\frac{k^2\sqrt{2}n(n+1)}{2}\hat{W}_i\sinh|\eta_i|\ee^{-(n+\frac{1}{2})|\eta_i|}-\left(n-\frac{1}{2}\right)\left(n+\frac{3}{2}\right)U_n(\eta_i).\label{eq:Unddot}
\end{align}
Applying Eqs.~\eqref{eq:Un}--\eqref{eq:Unddot} in $\eta=\eta_1$ and $\eta_2$, together with the definition of $U_n(\eta)$ in Eq.~\eqref{eq:undef} provides for each value of $n$ a $4\times 4$ linear system which can be solved uniquely for the four constants $A_n$, $B_n$, $C_n$ and $D_n$ in terms of the rate of change of the radius for each bubble  (determined by the diffusion problem) and their respective translation velocities ($\hat{W}_1,\hat{W}_2$).   The total axial force on each bubble is directly obtained from the viscous solution (the potential flow solution does not provide any contribution), so that  the total hydrodynamic force on each bubble is obtained  formally as \citep{stimson1926}
\begin{equation}\label{eq:2bubble_motion}
\left(\begin{array}{c}F_1\\F_2\end{array}\right)=\mathbf{R}\cdot\left(\begin{array}{c}\hat{W}_1\\\hat{W}_2\end{array}\right)+\mathbf{\tilde{R}}\cdot\left(\begin{array}{c}q_1\\q_2\end{array}\right)=\frac{2\pi\sqrt{2}}{k}\sum_{n=1}^\infty(2n+1)\left(\begin{array}{c}A_n+B_n+C_n+D_n\\A_n-B_n+C_n-D_n\end{array}\right)\cdot
\end{equation}
Enforcing that each bubble is  force-free ($F_1=F_2=0$)  determines implicitly their translation velocities $\hat{W}_1$ and $\hat{W}_2$ in terms of the rate of change of their radii.

\subsubsection{Numerical solution}
The initial value problem for two bubbles of initial radii ($a_1^0=1$ and $a_2^0\leq 1$) is  solved numerically. At each time step, Eq.~\eqref{eq:2bubble_fluxes} provides the mass flux into each bubble from their geometric arrangement and size. Then Eq.~\eqref{eq:2bubble_motion} is solved for $\hat{W}_1$ and $\hat{W}_2$ and the position of the bubbles is updated. For both the hydrodynamic and diffusion problems, a sufficiently  large number of Legendre modes is chosen to ensure the convergence of the results. A fourth-order Runge-Kutta scheme with adaptive time-step is used to solve this initial value problem and carefully resolve the final collapse of each bubble (for $a_i\ll 1$, $a_i\sim\sqrt{T_{f,i}-t}$ with $T_{f,i}$ the lifetime of bubble $i$).

\subsection{Collective dissolution of two identical bubbles}

\begin{figure}
\begin{center}
\begin{tabular}{cc}
\subfigure[~Dissolution time\label{fig:Tmax_2bubble}]{\includegraphics[height=6cm]{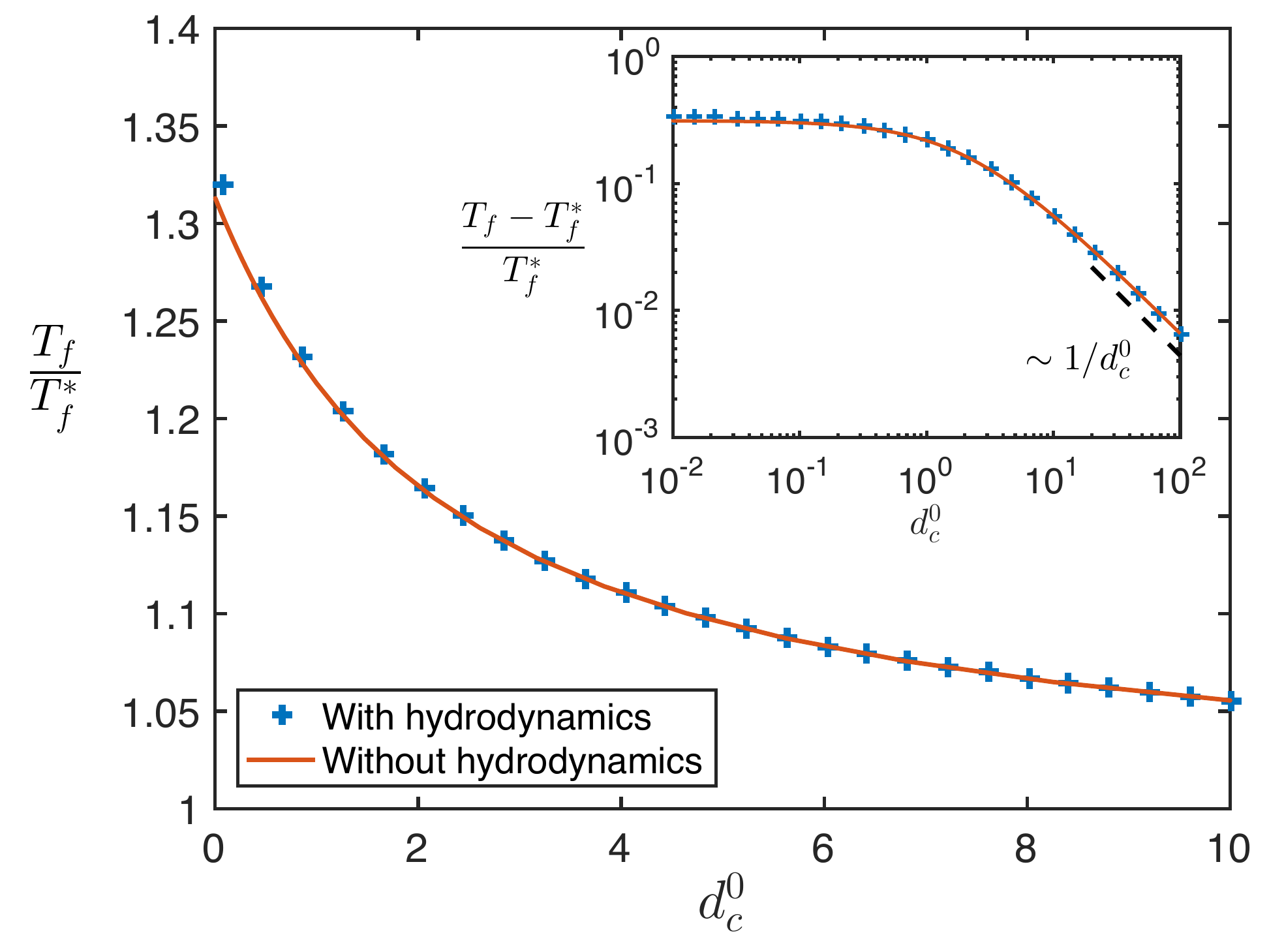}} &
\subfigure[~Instantaneous mass flux\label{fig:2bubble_flux}]{\includegraphics[height=6cm]{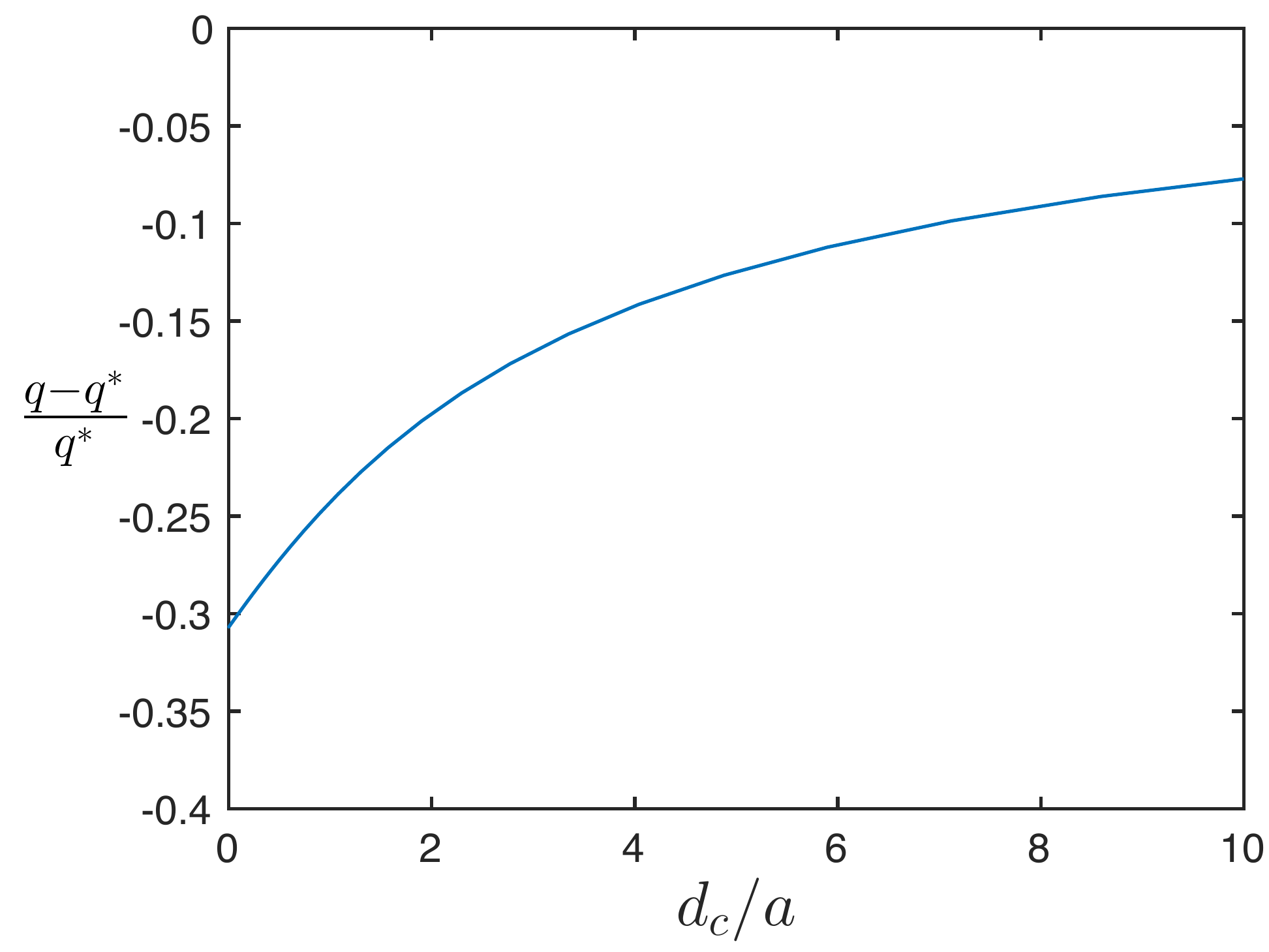}}
\end{tabular}
\caption{Shielding effect on the lifetime of two identical bubbles of initial radius $a_i^0=1$. (a): Relative dissolution time, $T_f/T^*_f$, of the bubbles (the value $T_f^*=1/4$ corresponds to the reference case of an isolated bubble) as a function of their initial relative distance. The inset displays the algebraic convergence rate in the far-field limit. In both cases, we show the  results   including hydrodynamics (i.e.~induced bubble motion, blue symbols) and without hydrodynamics (i.e.~bubbles whose centres are fixed, solid red line). (b): Instantaneous relative mass flux of two identical bubbles as a function of their instantaneous dimensionless relative distance, $d_c/a$; $q^*=-2$ is the reference of a single isolated bubble. 
}
\end{center}
\end{figure}

When the two bubbles are identical  $a_1^0=a_2^0=1$, the dissolution time is increased by the proximity of a second bubble (see ratio of lifetimes plotted in Fig.~\ref{fig:Tmax_2bubble}). Each bubble acts as a source of dissolved gas for its neighbour, effectively raising the background concentration seen by each individual bubble and slowing down its dissolution. 

This effect is significant and, as expected,  more pronounced for bubbles in close proximity. In the case of   bubbles in close contact, the lifetime of the bubbles is increased by more than $30\%$ (with an increase of the bubble lifetime $T_f$ of about $22\%$ for an initial contact distance $d_c^0\approx a_1^0$). For widely-separated bubbles, the relative increase in dissolution time decreases as $a_1^0/d_c^0$. This shielding effect, i.e.~the reduction in magnitude of the mass flux $|q_i|$ out of each bubble, does not  remain constant  throughout the dissolution of the bubble. Indeed, as the bubble   decreases in size, the instantaneous ratio $a/d$ decreases and the shielding effect of the second bubble becomes negligible (see Fig.~\ref{fig:2bubble_flux}).

\begin{figure}
\begin{center}
\begin{tabular}{c}
\includegraphics[height=6.4cm]{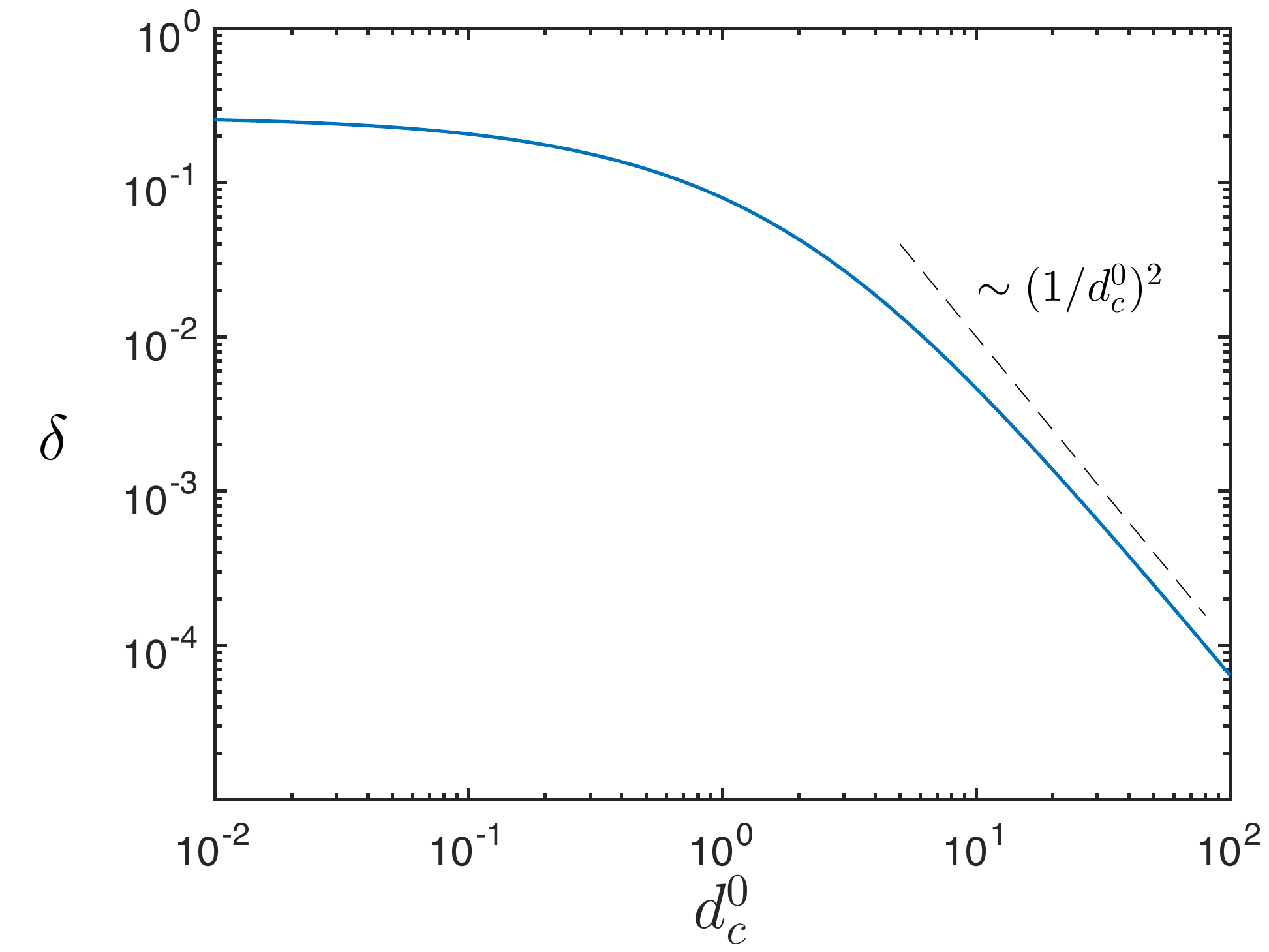}
\end{tabular}
\caption{Relative displacement of two identical bubbles due to their collective dissolution measured by  the    change in their center-to-center distance, $\delta$, over the full duration of the dissolution process, as a function of their initial relative distance, $d_c^0$. 
}\label{fig:2bubble_distance}
\end{center}
\end{figure}

Hydrodynamics tends to increase the lifetime of the bubbles by bringing them closer and thus increasing their instantaneous diffusive shielding. 
 This hydrodynamic effect is however quite insignificant unless the bubbles are initially closely packed (see Fig.~\ref{fig:Tmax_2bubble}). Neglecting the flow-induced motion of the bubbles (i.e.~keeping their centres fixed) leads to underestimating the lifetimes by only $1.5\%$ when the contact distance is $d_c^0=10^{-2}$, and by $0.2\%$ for $d_c^0=1$.  Two regimes can be identified for   the hydrodynamically-induced change in bubble distance $\delta$ (Fig.~\ref{fig:2bubble_distance}). In the lubrication limit, i.e.~for $d_c^0\ll 1$, $\delta$ is finite and equals $1/4$. In the far-field limit, $\delta\sim 1/(d^0_c)^2$, a signature of the source-type flow field generated by the shrinking bubble when the bubbles are far away from each other. These results therefore show that for two identical bubbles, hydrodynamics only plays a minor role, except in the lubrication limit ($d_c\ll 1$).

\subsection{Asymmetric dissolution of two bubbles of different radii}
\begin{figure}
\begin{center}
\begin{tabular}{cc}
\subfigure[~Dissolution time of bubble $1$]{\includegraphics[height=6cm]{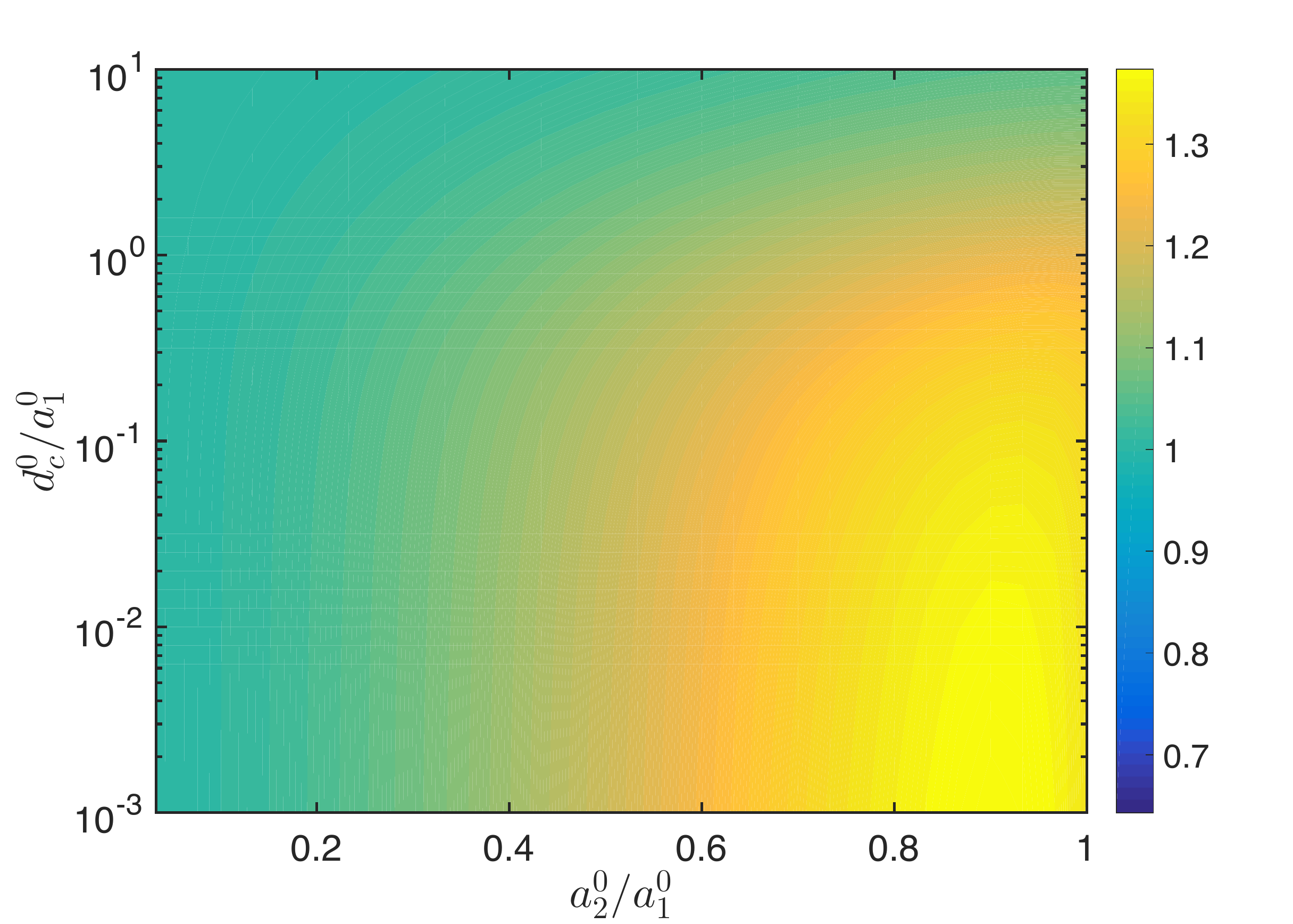}} & 
\subfigure[~Dissolution time of bubble $2$]{\includegraphics[height=6cm]{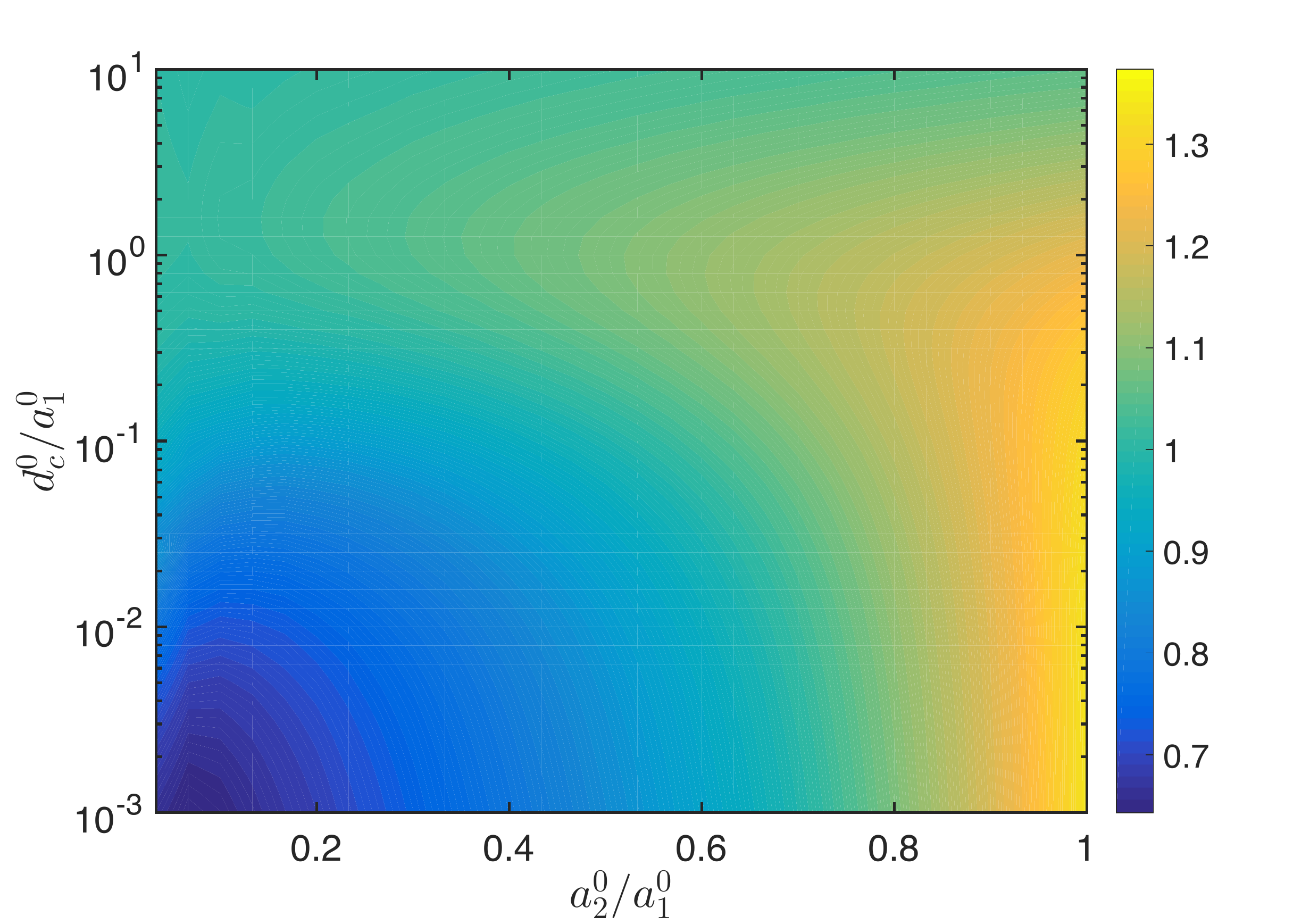}} \\
\subfigure[~Relative displacement of bubble $1$]{\includegraphics[height=6cm]{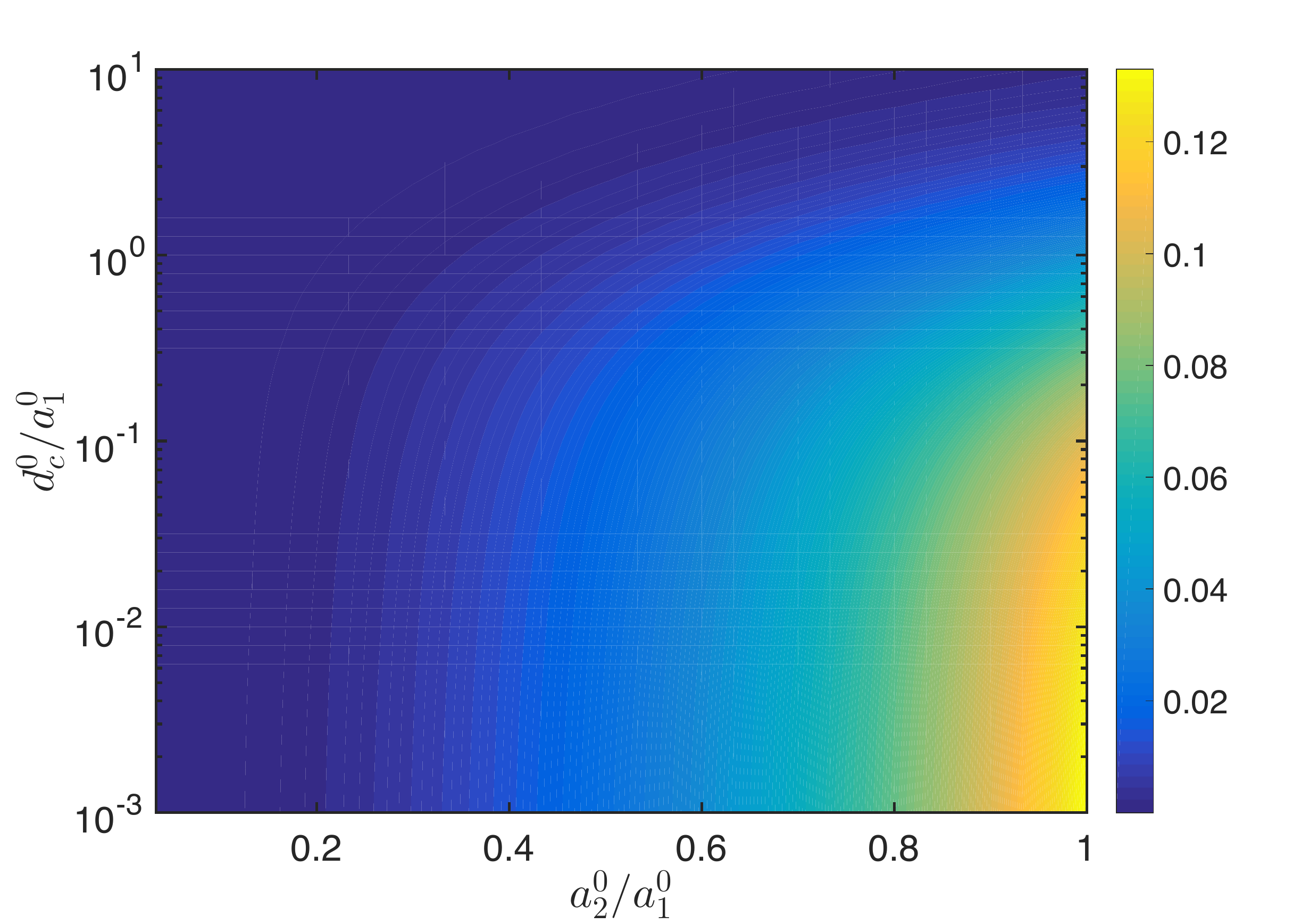}} &
\subfigure[~Relative displacement of bubble $2$]{\includegraphics[height=6cm]{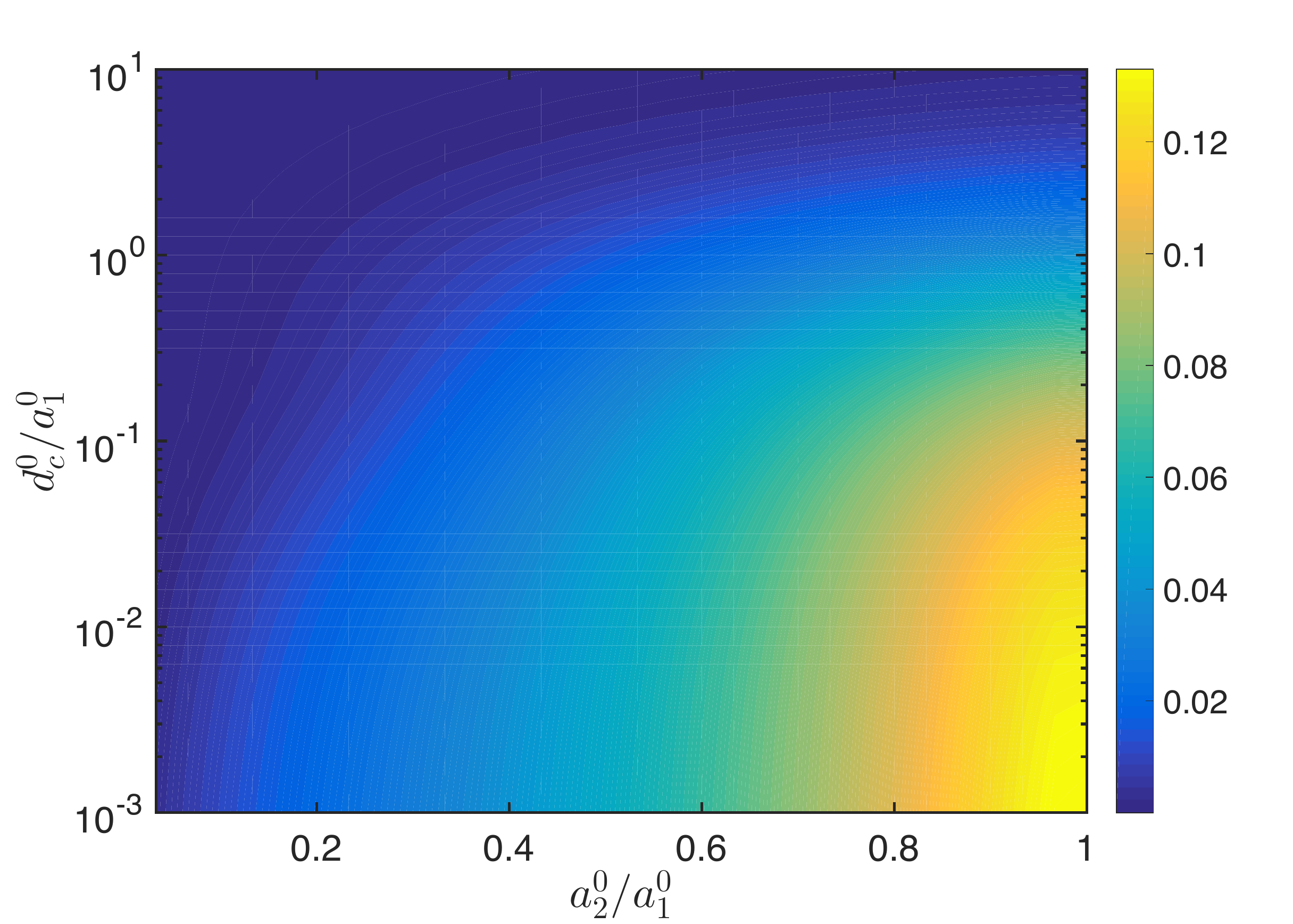}} 
\end{tabular}
\caption{(a,b) Normalised bubble lifetime, $T_{f,i}/T_{f,i}^*$, and (c,d) final displacement of bubble $1$ ($a^0_1=1$, left) and bubble $2$ ($a^0_2<1$, right) as a function of the initial size ratio, $a^0_2/a^0_1$, and the dimensionless  contact distance, $d_c^0/a_1^0$. Here we use  $T_{f,i}^*=(a_i^0)^2/4$ to denote the dissolution time of an isolated bubble of same initial radius $a_i^0$.}\label{fig:Tmax_2bubble_gen}
\end{center}
\end{figure}

The general case of two bubbles of arbitrary initial radii $a_2^0\leq a_1^0=1$ reveals the asymmetry of the shielding effect on the dissolution (see Fig.~\ref{fig:Tmax_2bubble_gen}). The lifetime of the larger bubble always increases (Fig.~\ref{fig:Tmax_2bubble_gen}a), but the dissolution of the smaller bubble can be either slowed down if either $a_2^0$ is sufficiently large  or the bubbles are initially far apart or accelerated when the neighbouring bubble is much larger and the contact distance is small (Fig.~\ref{fig:Tmax_2bubble_gen}b).
\begin{figure}
\begin{center}
\begin{tabular}{cc}
\subfigure[~Mass flux into bubble $1$]{\includegraphics[height=6cm]{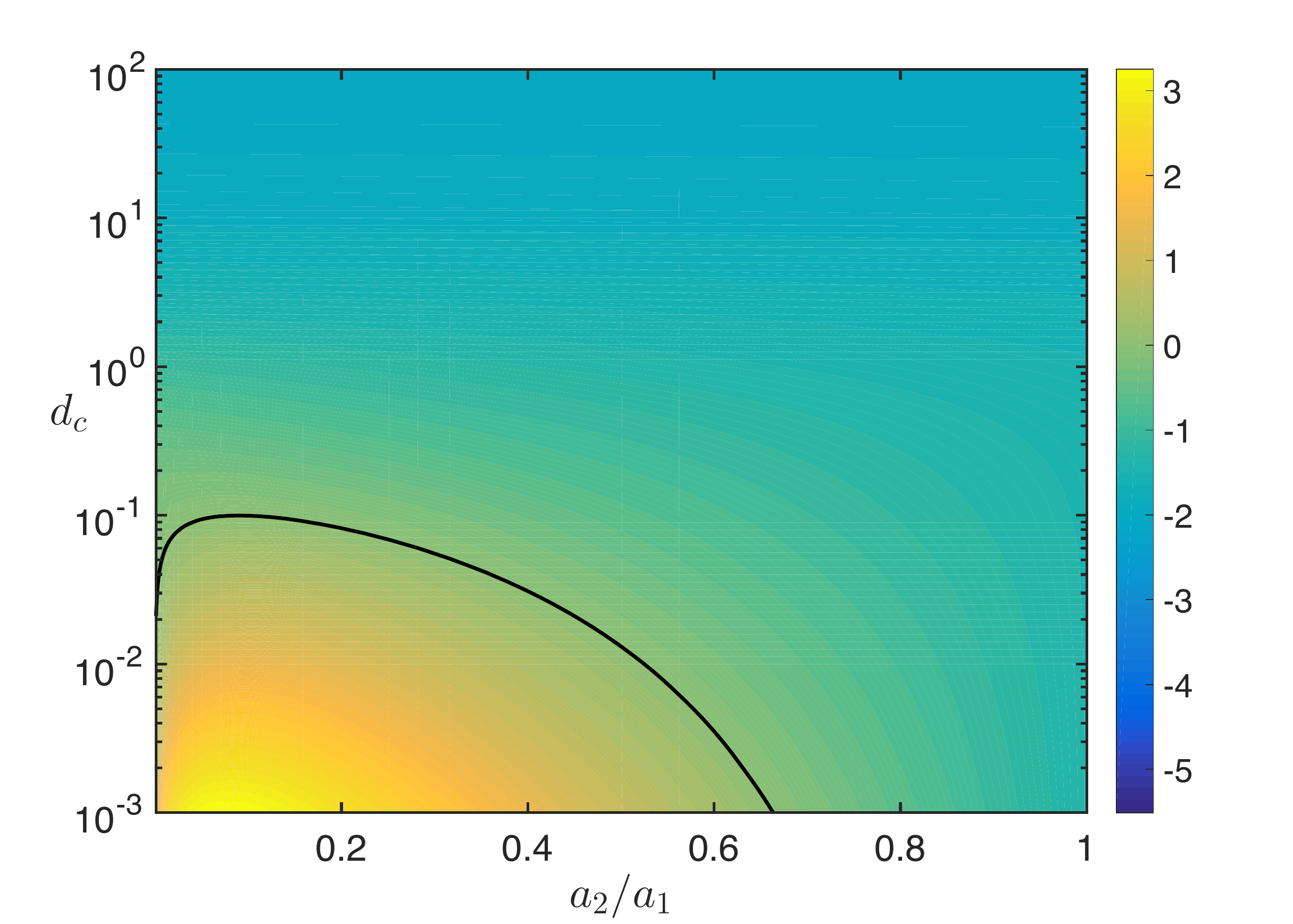}} &
\subfigure[~Mass flux into bubble $2$]{\includegraphics[height=6cm]{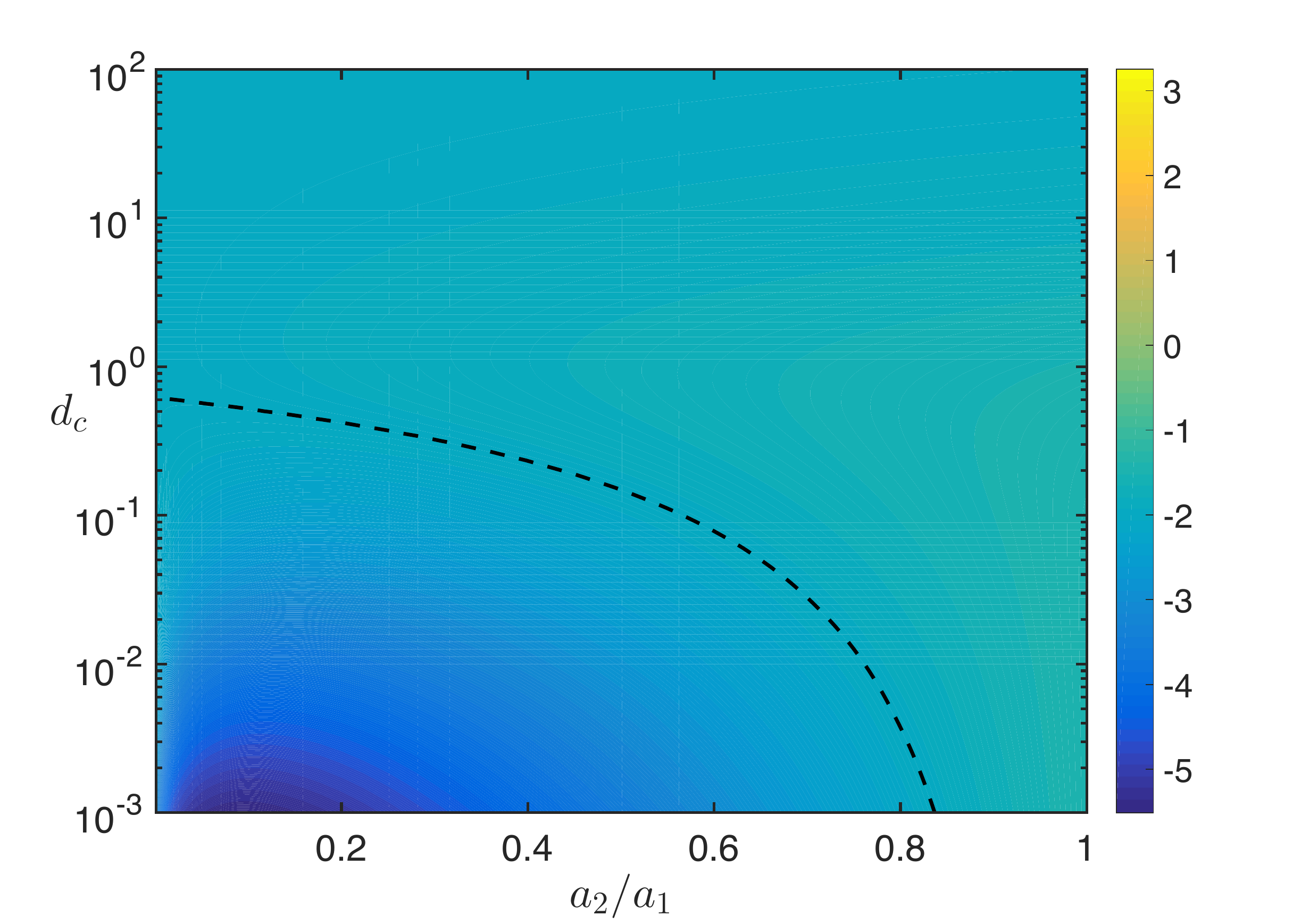}} 
\end{tabular}
\caption{Iso-values of the diffusive surface flux $q_i$ into (a) bubble $1$ (of instantaneous  radius $a_1=1$) and (b) bubble $2$ (of instantaneous radius $a_2<1$). The dashed line corresponds to the value $q_i=-2$  for an isolated bubble, while the solid line corresponds to $q_i=0$ (no  dissolution).}\label{fig:2bubble_flux_gen}
\end{center}
\end{figure}

This effect, which can be seen as the superposition and competition of classical Ostwald ripening \cite{voorhees1985theory,dollet2016} in a two-bubble system  with  the global dissolution of the bubbles, is confirmed by considering the instantaneous modification of the diffusive mass flux out of each bubble (Fig.~\ref{fig:2bubble_flux_gen}). Small bubbles located close to larger ones  show an increase in their dissolution rate (i.e.~a larger value of $|q_i|$), while the dissolution  of larger bubbles is always slowed down. This increased dissolution of small bubbles stems from the large capillary pressure inside them that  translates into a large dissolved gas concentration contrast between their surface and their environment. In that case, the larger bubble can actually experience negative dissolution rates: the smaller bubble acts as a   source of dissolved gas that is absorbed by the larger bubble.

This effect is however only transient  due to the global dissolution process. The increase in  size of  bubble $1$ is associated to an accelerated dissolution of bubble $2$, which is already smaller than its neighbour and therefore quickly disappears. The lifetime of the larger bubble is then only marginally impacted  even though its size may initially increase (see Fig.~\ref{fig:Tmax_2bubble_gen}). Such effect could however become significant when exerted cumulatively by multiple neighbouring bubbles.

The role of hydrodynamics and its impact on the relative arrangement of the bubbles is significant only when both bubbles have comparable sizes and are located in close contact (Fig.~\ref{fig:Tmax_2bubble_gen}, bottom). When one of the bubbles is very small, its lifetime is short, and therefore so is the period over which hydrodynamics can modify the position of the bubbles (for a single bubble, hydrodynamics plays no role).

\section{Asymptotic models of  collective dissolution}
\label{sec:asymptotic_models}
\subsection{Method of reflections}
\label{sec:MOR}
When the number of bubbles is greater than $N=2$, solving analytically the Laplace and Stokes equations is no longer possible. However, the method of reflections can be used for both mathematical problems in order to derive an asymptotic expansion of the solution.  

Firstly, the  dissolved gas concentration satisfies the   Laplace equation, $\nabla^2 c=0$, with boundary conditions on bubble $j$ of instantaneous radius $a_j(t)$
\begin{align}
\left.c\right|_{r_j=a_j}&=c_j^s=\frac{1}{a_j},\\
q_j=a_j\dot{a}_j&=\frac{1}{2\pi}\int_{r_j=a_j}\nb_j\cdot\grad c\,\dd S,\label{eq:MoR_fluxbc}
\end{align}
which provides a direct and linear relationship between the diffusive mass flux, $q_j$, and the uniform surface concentration, $c_j^s=1/a_j$, at the surface of each bubble. 

Secondly, the translation velocity $\dot\Xb_j$ of  bubble $j$ whose centre is located instantaneously at $\Xb_j(t)$ follows from solving for the  Stokes flow forced by the shrinking motion of the bubbles under the  conditions
\begin{align}
\nb_j\cdot\left.\ub\right|_{r_j=a_j}&=\dot\Xb_j\cdot\nb_j+\dot{a}_j,\\
(\mathbf{I}-\nb_j\nb_j)\cdot\left.\sigmab\right|_{r_j=a_j}\cdot\nb_j&=0,\\
\int_{r_j=a_j}\sigmab\cdot\nb_j\,\dd S&=0.
\end{align}

The geometric arrangement of the bubbles  is characterised by $d_{jk}=|\Xb_k-\Xb_j|$ and $\eb_{jk}=(\Xb_k-\Xb_j)/d_{jk}$. The fundamental idea of the method of reflections (for both Laplace and Stokes problems) is to construct an iterative expansion of the solution $c=c^0+c^1+...$ (or $\ub=\ub^0+\ub^1+...$), where each iteration is the superposition of solutions to a local  problem (Laplace or Stokes) around  each bubble considered isolated with boundary contributions defined so as to satisfy the correct boundary condition on that particular bubble, taking into account the extra contribution of other bubbles introduced at the previous order of the expansion~\cite{kimbook}. This iterative approach, described in more details below and in Appendix~\ref{ap:reflections}, provides an asymptotic estimate of the full solution as a series of increasing order in $\varepsilon=a/d$ with $a$ and $d$ the typical bubble radius and inter-bubble distance.

\subsubsection{Laplace problem}
The goal of this section is to express the surface concentration of each bubble, $c_j^s$, as a function of a prescribed diffusive mass flux $q_j$. This mathematical approach  may seem counter-intuitive as in practice, $c_j^s$ is fixed by the size of the bubble and Henry's law ($c_j^s=1/a_j$). However, this implicit approach guarantees a faster convergence of the reflection process, similarly to the classical mobility formulation of the method of reflections for Stokes' flow problems~\cite{kimbook}. 

The solution of the Laplace problem for a single bubble is trivial and is obtained as $c^0_j=-q_j/(2r_j)$. A critical step in the method of reflections  is to determine the value of $c_j^i$, the correction to the concentration field, that satisfies Laplace equation outside of bubble $j$ with no net flux (since the flux boundary condition, Eq.~\eqref{eq:MoR_fluxbc}, is accounted for by the solution $c_j^0$), and cancels out any non-uniformity in the surface concentration introduced at the previous $i-1$ iteration by the reflections at the other bubbles, i.e.~non-uniform $\displaystyle\sum_{k\neq j}c_k^{i-1}$ at the surface of bubble $j$. Defining $\rb_j=\rb-\Xb_j$, using a Taylor series expansion at the surface of bubble $j$, we have
 \begin{equation}
c_k^{i-1}(r_j=a_j)=\left.c_k^{i-1}\right|_{r_j=0}+\left.\grad c_k^{i-1}\right|_{r_j=0}\cdot(a_j\nb_j)+\frac{a_j^2}{2}\left.\grad\grad c_k^{i-1}\right|_{r_j=0}:(\nb_j\nb_j)+...,
\end{equation}
so that the solution for the $i$-th iteration is obtained as
\begin{align}
c_j^i(\rb_j)=&-\sum_{k\neq j}\left[\left(\frac{a_j}{r_j}\right)^3\left.\grad c_k^{i-1}\right|_{r_j=0}\cdot\rb_j+\frac{1}{2}\left(\frac{a_j}{r_j}\right)^5\left.\grad\grad c_k^{i-1}\right|_{r_j=0}:(\rb_j\rb_j)\nonumber \right.\\
&\left.+\frac{1}{3}\left(\frac{a_j}{r_j}\right)^7\left.\grad\grad\grad c_k^{i-1}\right|_{r_j=0}\threevdots(\rb_j\rb_j\rb_j)+...\right],
\end{align}
and the correction of the $i$-th reflection  to the surface concentration of bubble $j$ is
\begin{equation}
c_j^{i,s}=\sum_{k\neq j}\left.c_k^{i-1}\right|_{r_j=0}.
\end{equation}
Using these results and $c_j^s=1/a_j$, after two reflections, the diffusive flux $q_j$ must satisfy the following linear system (see details in Appendix~\ref{ap:reflections})
\begin{equation}\label{eq:mor_flux}
-2=q_j+\sum_{k\neq j}q_k\left(\frac{a_j}{d_{jk}}\right)-\sum_{\substack{k\neq j\\l\neq k}}q_l\left[\frac{a_ja_k^3}{d_{kl}^2d_{jk}^2}\eb_{kl}\cdot\eb_{kj}+
\frac{a_ja_k^5}{2d_{kl}^3d_{jk}^3}(3(\eb_{kl}\cdot\eb_{kj})^2-1)\right]+O\left(q\varepsilon^7\right).
\end{equation}
The advantage of expressing $c_j^s$ in terms of $q_j$ rather than the opposite appears now clearly. Keeping only the first two terms (i.e.~a single reflection) provides an estimate that is valid up to an error in  $O(\varepsilon^4)$. More specifically, each reflection can be seen as a multipole expansion of the $c_j^i$. Prescribing $q_j$ at the zeroth-iteration imposes that the slowest decaying singularity (i.e.~the source) is zero at all subsequent order. The dominant contribution in further reflections therefore arises from the gradient of concentration generated by a source dipole. Keeping only the first two terms on the right-hand side of Eq.~\eqref{eq:mor_flux} provides an estimate of $q$ up to an $O(\varepsilon^4)$ while keeping the first three or four terms provides an estimate at order $O(\varepsilon^6)$ or $O(\varepsilon^7)$, respectively. In the following,  estimates with such accuracies are referred to as $S_4$, $S_6$ and $S_7$, respectively.

\subsubsection{Hydrodynamic problem}
A similar approach is followed for the Stokes problem in order to determine the velocity of the different bubbles, denoted $\dot\Xb_j$, in terms of their mass flux, $q_j$. The isolated bubble problem is trivial since  a single bubble does not move by symmetry and it generates a radial velocity field, $\ub_j^0=(a_jq_j/r_j^3)\rb_j$.

From the flow field generated by bubble $k$ at iteration $i-1$, the result of the $i$-th reflection is now obtained using Faxen's law for a bubble~\cite{rallison1978}
\begin{equation}
\dot{\Xb}_j^i=\sum_{k\neq j}\left.\ub_k^{i-1}\right|_{r_j=0}.
\end{equation}
For $i=0$, the flow field $\ub_j^0$ is that of a simple source/sink, while for $i\geq 1$, the flow field $\ub_j^i$ generated by force- and torque-free bubble $j$ at that order is dominated by a symmetric force dipole, or stresslet $\mathbf{S}_j^i$, that can be computed directly in terms of the local gradient of the background flow~\cite{rallison1978}
\begin{equation}
\ub_j^i(\rb_j)=-\frac{3(\rb_j\cdot\mathbf{S}_j^i\cdot\rb_j)\rb_j}{8\pi r_j^5},\quad \mathbf{S}_j^i=\frac{4\pi a_j^3}{3}\sum_{k\neq j}\left(\left.\grad\ub_k^{i-1}\right|_{r_j=0}+\left.^T\grad\ub_k^{i-1}\right|_{r_j=0}\right).
\end{equation}
Using these results, the asymptotic expansion for $\dot{\Xb}_j$ in terms of $q_j$ is finally obtained as
\begin{equation}
\dot{\Xb}_j=\sum_{k\neq j}\frac{a_kq_k}{d_{jk}^2}\eb_{kj}-\sum_{\substack{k\neq j\\l\neq k}}\frac{a_k^3a_lq_l}{d_{kl}^3d_{jk}^2}\Big(1-3(\eb_{jk}\cdot\eb_{kl})^2\Big)\eb_{kj}+O\left(\varepsilon^7\right).
\end{equation}

\subsubsection{Validation: two-bubble problem}

\begin{figure}[h!]
\begin{center}
\begin{tabular}{cc}
\subfigure[~Diffusive flux ($a_2/a_1=1$)]{\includegraphics[height=5.3cm]{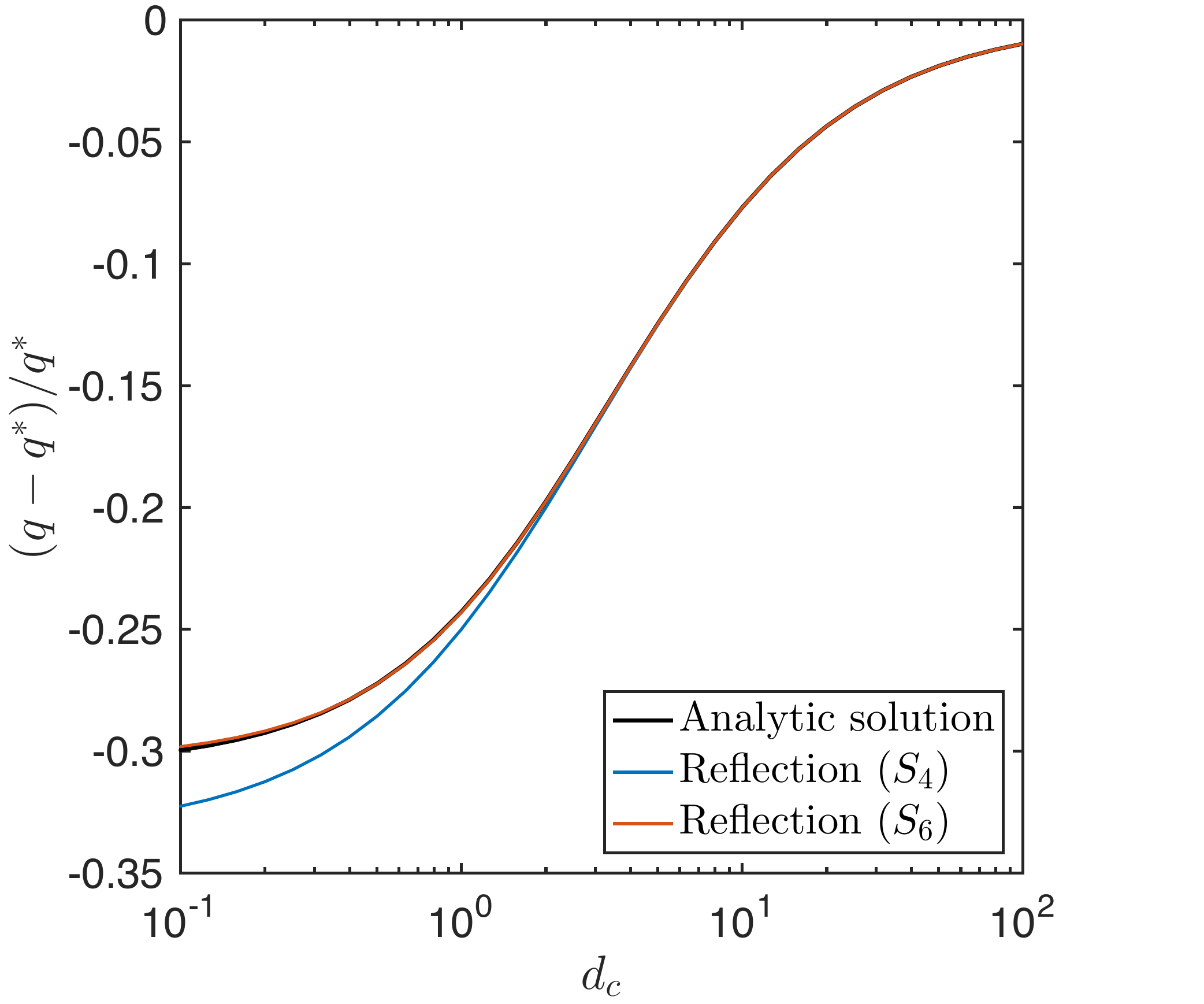}} &
\subfigure[~Diffusive flux ($a_2/a_1=1/4$)]{\includegraphics[height=5.3cm]{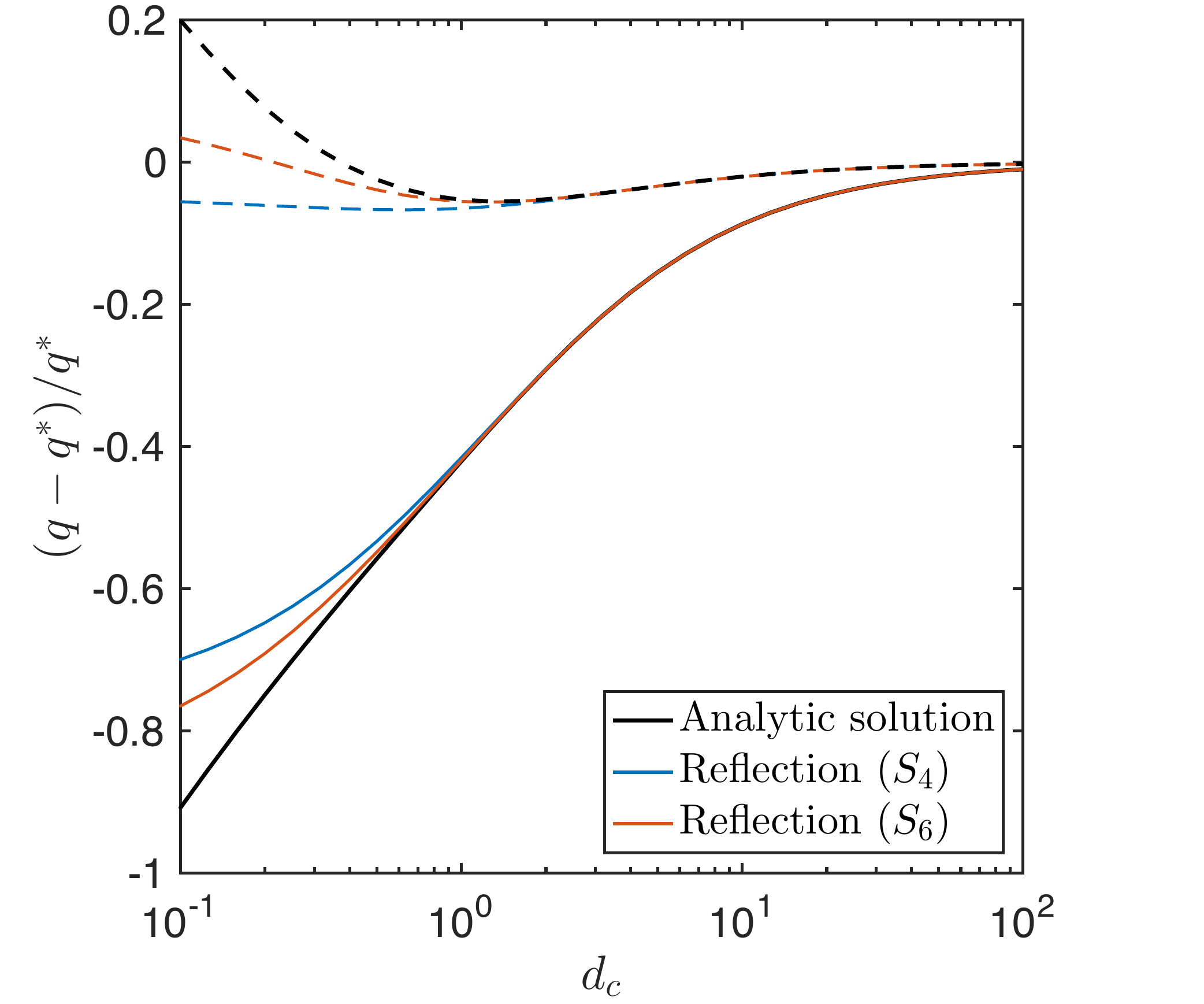}} \\
\subfigure[~Translation velocity ($a_2/a_1=1$)]{\includegraphics[height=5.3cm]{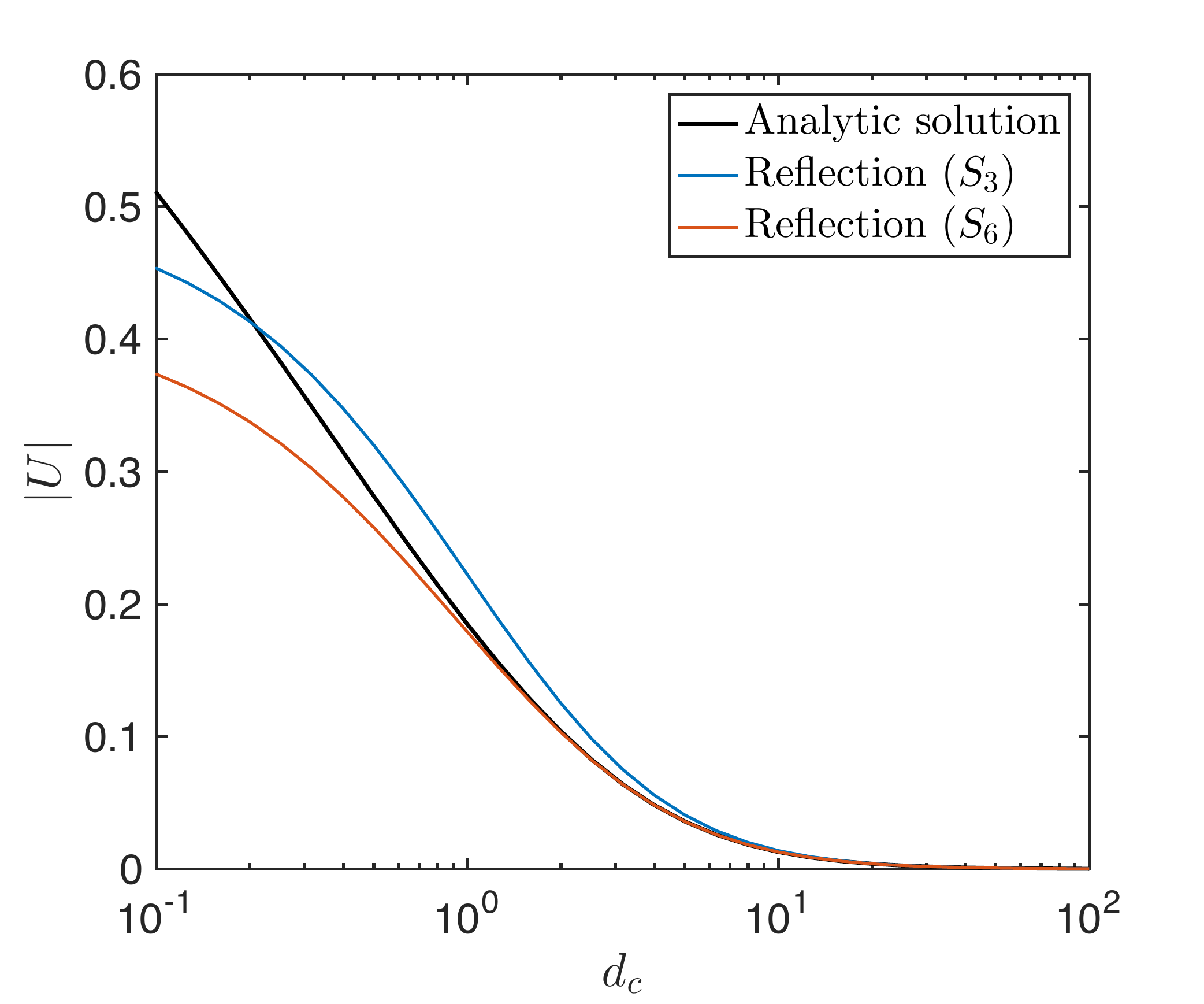}} &
\subfigure[~Translation velocity ($a_2/a_1=1/4$)]{\includegraphics[height=5.3cm]{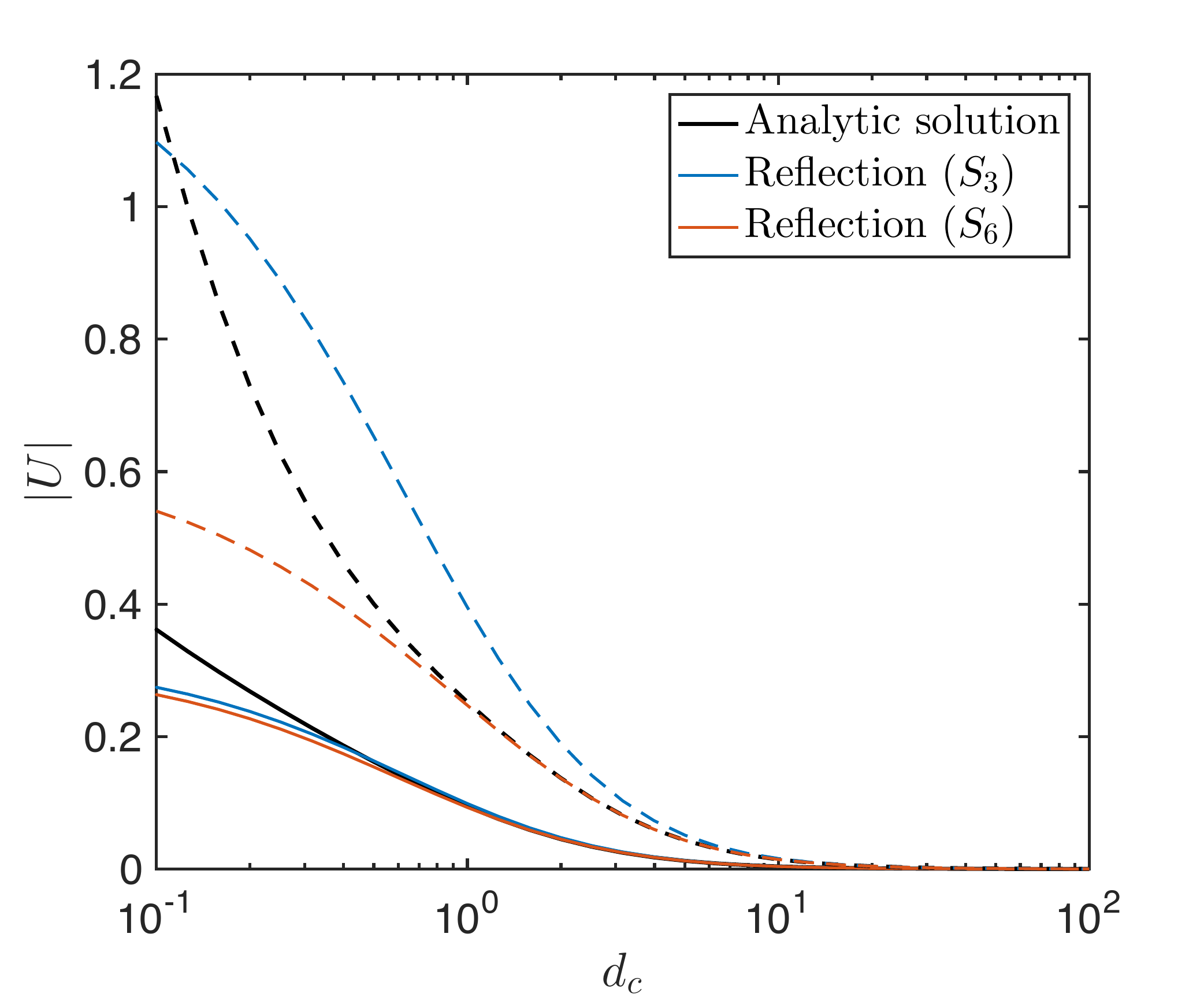}} \\

\subfigure[~Dissolution time ($a^0_2/a^0_1=1$)]{\includegraphics[height=5.3cm]{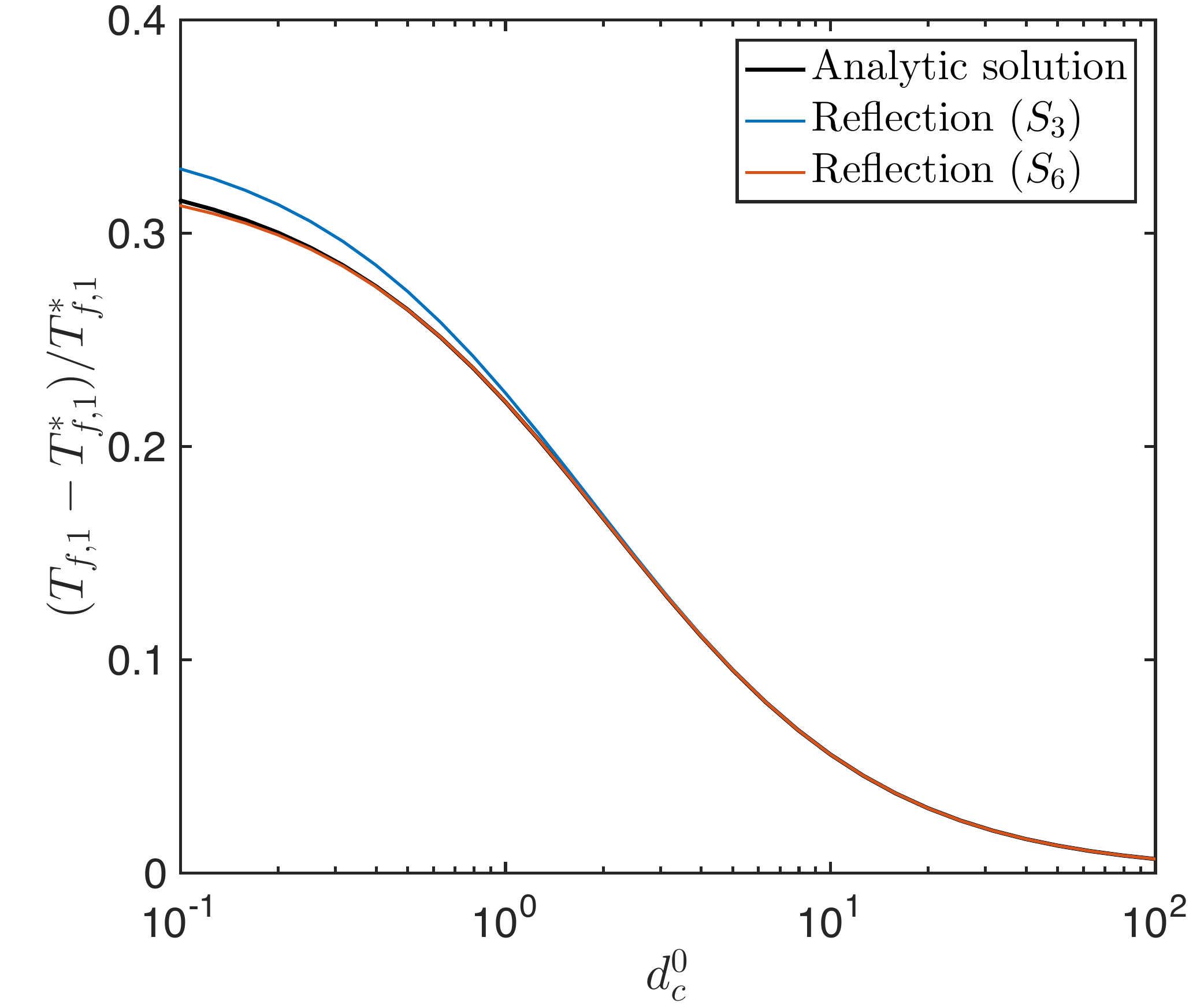}} 
&
\subfigure[~Dissolution time ($a^0_2/a^0_1=1/4$)]{\includegraphics[height=5.3cm]{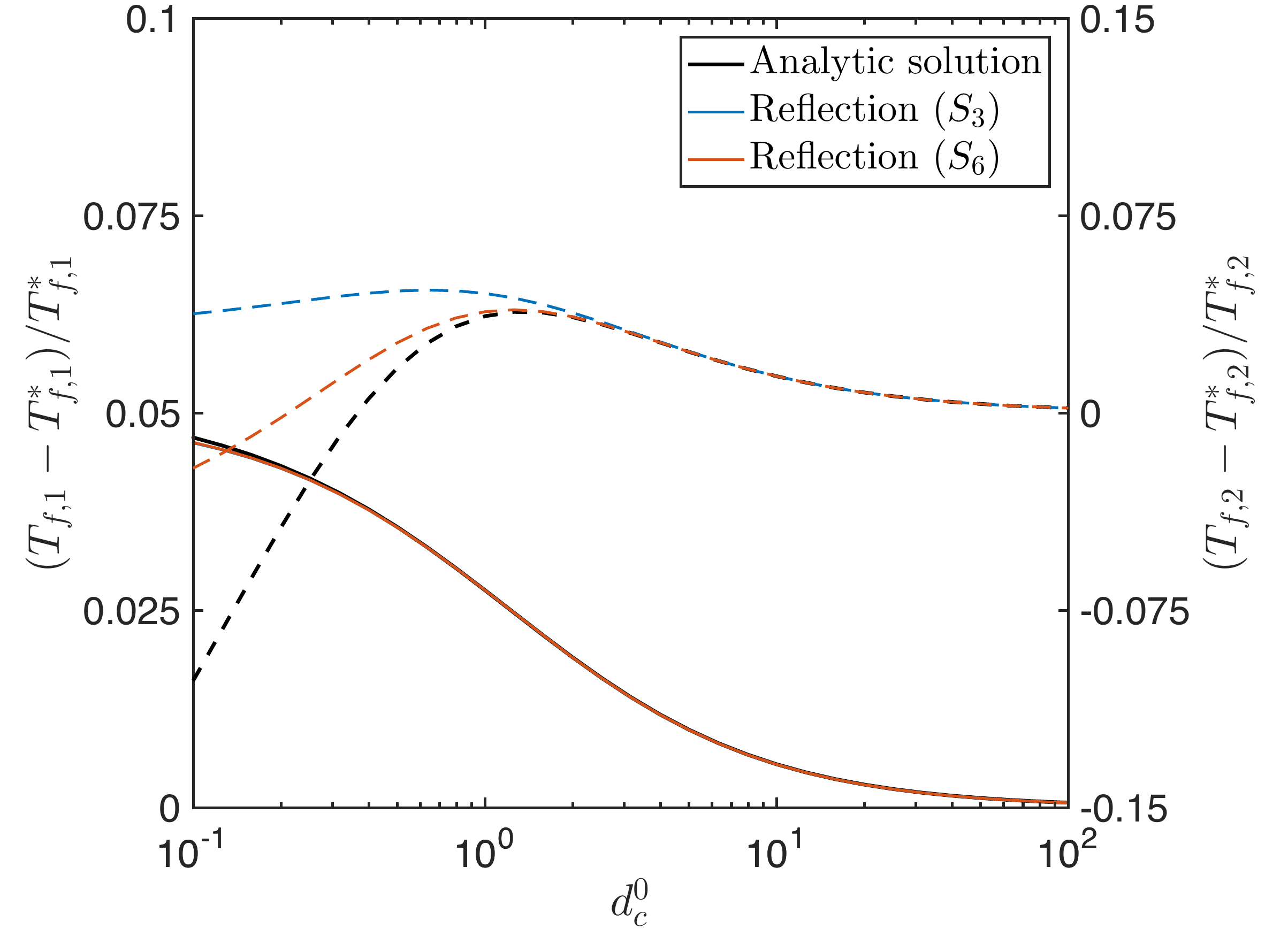}}
\end{tabular}
\caption{Validation of the method of reflections. (a,b) Relative diffusive flux at the surface of one of the bubbles, (c,d) resulting translation velocity magnitude and (e,f) dissolution time as a function of contact distance, $d_c$,  for two identical bubbles with unit radius (a,c,e) and  for two   bubbles with  radii $a_1=1$ (solid) and $a_2=1/4$ (dashed) (b,d,f). The results in (a)--(d) are instantaneous measurements (i.e.~$a_1$ and $a_2$ are the current radii of the bubbles), while (e)--(f) are measured over the lifetime of the bubbles (i.e.~$a_1^0$ and $a_2^0$ are the initial bubble radii).  Two successive approximations are shown in both cases and are compared to the exact solution (black). In both cases, solution denoted $S_n$ has an error in $O(\varepsilon^n)$.   $q^*=-2$ and $T_{f,i}^*=(a_i^0)^2/4$ are the corresponding reference diffusive flux and dissolution time for an isolated bubble. }\label{fig:valid_flux}
\end{center}
\end{figure}

The exact solution for two bubbles obtained  in \S~\ref{sec:bispherical} is used to validate the  approximation obtained using the method of reflections, its convergence and its accuracy, with results shown in Fig.~\ref{fig:valid_flux}. We see  that the method of reflections provides an extremely accurate estimate of the diffusive flux and resulting bubble velocity provided that the contact distance, $d_c$, is greater than $d_c>1/2$ (see also Appendix~\ref{ap:validation}). The agreement is better for two identical bubbles, for which the flux prediction using the $S_6$ approximation (first three terms in Eq.~\ref{eq:mor_flux})  captures  the correct flux with an error of less than $1\%$ in the case of almost-touching bubbles ($d_c\approx 10^{-2}$). The agreement on the total lifetime of the bubbles is also excellent.   These results therefore validate the present approach even in near-field conditions provided that the contact distance $d_c$ between the bubbles is at least of the order of their radius. 

We  note that the agreement on the instantaneous flux is much better than for the velocity of the bubbles. Nevertheless this does not seem to affect the validity of the prediction for the global dissolution process and is yet another indication of the limited role of hydrodynamic interactions on the overall dynamics. As a consequence, for the remainder of this paper, the motion of the bubbles  induced by their dissolution is neglected and we focus solely on the Laplace problem.

\subsection{Continuum model}
\label{sec:1dcontinuous}

Turning now to the case of many bubbles, and neglecting the role of hydrodynamics, the first reflection provides an estimate of the diffusive mass flux valid up to an $O(\varepsilon^4)$ error by superimposing the influence of each bubble as a simple source of intensity $q_j$.  For a large number of bubbles, when their typical radius, $a$, is small compared to the typical distance between bubbles,  $d$, and when  bubble size varies slowly across the bubble lattice, a simpler model may be obtained (i) by  considering the dynamics of a single bubble in a spatially-dependent background concentration $c_\textrm{back}(\xb)$, and (ii) by assuming that this background concentration is generated by a continuous distribution of bubbles. In that continuous limit, local bubble properties (radius, diffusive flux) are defined as $a(\xib)$ and $q(\xib)$,  where  $\xib$ is a spatial coordinate in the bubble cluster. An essential assumption of this model is the separation of length scales $a\ll d\ll L$ with $L\sim \textrm{max}(d^2/\Delta a,\bar{R}) $ with $\Delta a$  the typical difference in radius for two neighboring bubbles and $\bar{R}$ the size of the lattice. This restriction therefore   excludes  representing phenomena such as Ostwald ripening where the contrast in size between two neighboring bubbles must be large enough to be significant.

\subsubsection{Local continuum model for line distributions}

Assuming  that the distance $d$ between neighbouring bubbles is large compared to their radii (i.e.~$a/d\ll 1$), the dynamics of each bubble can be considered individually in the background concentration $c_\textrm{back}(\xb)$ from Eq.~\eqref{eq:conc_back}. The bubble dynamics in this case has already been solved in \S~\ref{sec:single} and one finds
\begin{equation}\label{eq:bubble_back}
q(\xb,t)=a(\xb,t)\dot{a}(\xb,t)=-2\bigg(1-c_\textrm{back}(\xb,t)a(\xb,t)\bigg).
\end{equation}
For a line distribution of bubbles with local density $\lambda(s)\approx 1/d$, the   ``background'' concentration field at position $\xb$ can be computed using the free-space Green's function of Laplace's equation~\citep{jackson1962}
\begin{equation}\label{eq:conc_back}
c_\textrm{back}(\xb,t)=-\frac{1}{2}\int\frac{\lambda(s',t)q(s',t)\dd s'}{|\xb-\xib(s')|}\cdot
\end{equation}

While  we consider in the following  a uniform density of bubbles with $\lambda=1/d$, the present model could be easily extended to account for density fluctuations. A similar approach can also be followed to treat two- and three-dimensional distributions in which case the bubble density scales  as $1/d^2$ and $1/d^3$, respectively (see \S~\ref{sec:2d_continuous} and \S~\ref{sec:3d}).

This continuum local model is valid under the assumption that (i) the bubbles are far apart from each other (i.e.~$\lambda a\ll 1$), (ii) there is a large number of bubbles (i.e.~$\lambda \bar{R}\gg 1$ with $\bar{R}$ the typical dimension of the cluster) and (iii) the bubble radius varies sufficiently slowly  that a continuum description is relevant ($d\ll d^2/\Delta a$). Under these assumptions, Eqs.~\eqref{eq:bubble_back} and \eqref{eq:conc_back}  provide an implicit determination of the rate of change in bubble radius as an integral equation for $q$. The integral kernel in Eq.~\eqref{eq:conc_back} is singular for $\xb=\xib$ and requires further treatment for line distributions. Isolating the self-contribution (logarithmic singularity) and  taking advantage of the locally discrete distribution of bubbles, the local background concentration on the line of bubbles is obtained as
\begin{align}
c_\textrm{back}(s)=&-\frac{\lambda}{2}\int_{s_\textrm{min}}^{s_\textrm{max}}\left[\frac{q(s')}{|\xb(s')-\xb(s)|}-\frac{q(s)}{|s'-s|}\right]\dd s'\nonumber\\
&-\lambda q(s)\left[\gamma_E+\log\left(\lambda\sqrt{(s_\textrm{max}-s)(s-s_\textrm{min})}\right)\right],\label{eq:cback}
\end{align}
where $s_\textrm{min}\leq s\leq s_\textrm{max}$ is the curvilinear coordinate along the line of bubbles and $\gamma_E$ denotes the Euler-Mascheroni constant,
\begin{equation}
\gamma_E=\textrm{lim}_{n\rightarrow\infty}\left[\sum_{k=1}^n\frac{1}{k}-\log n\right]\approx 0.57722.
\end{equation}

 \begin{figure}
 \begin{center}
 \begin{tabular}{cl}
 \multirow{-7}{*}{\vspace{-5cm}\includegraphics[height=10cm]{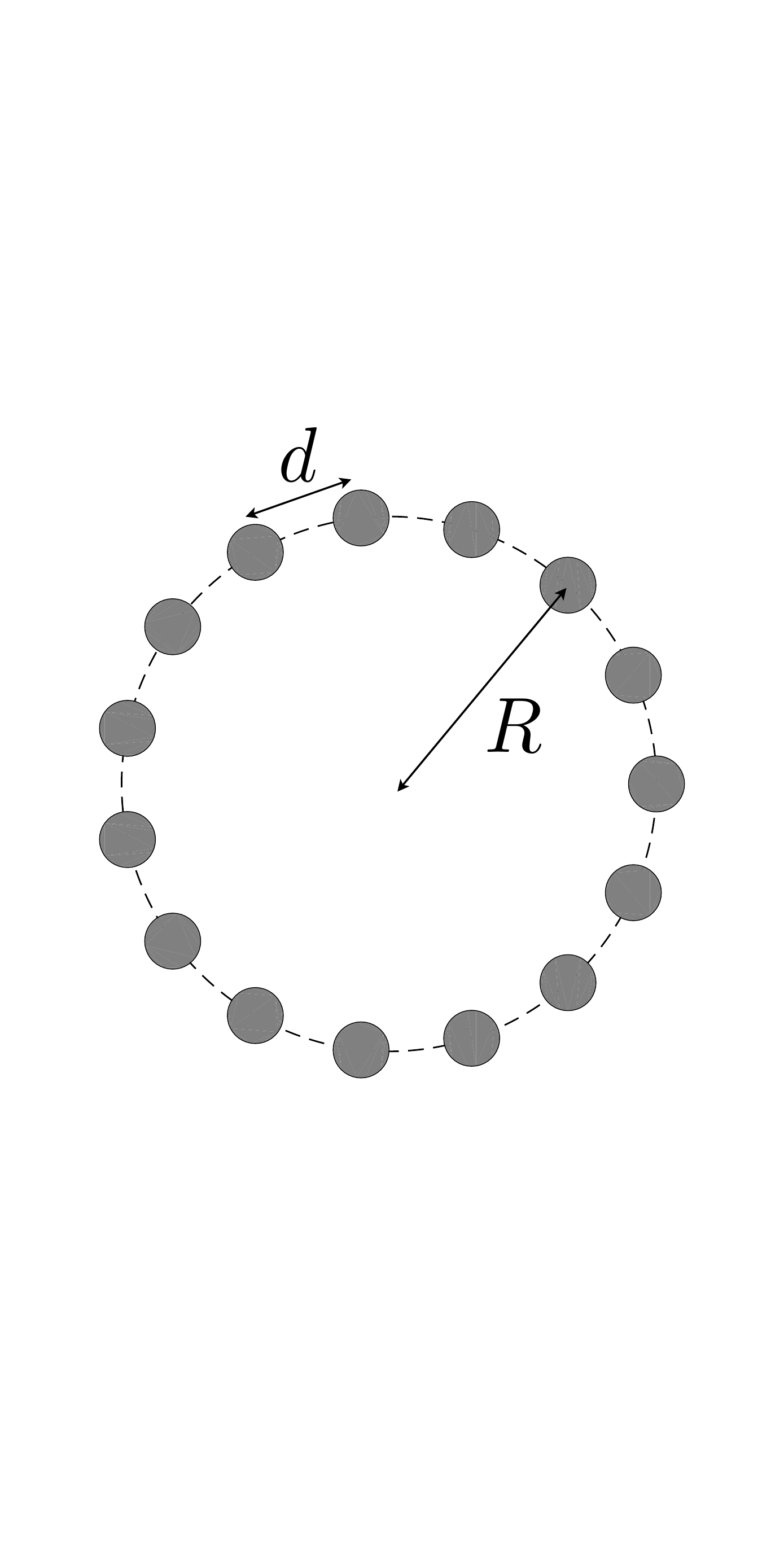} }& 
 \subfigure[~Bubble radius]{\includegraphics[height=6.5cm]{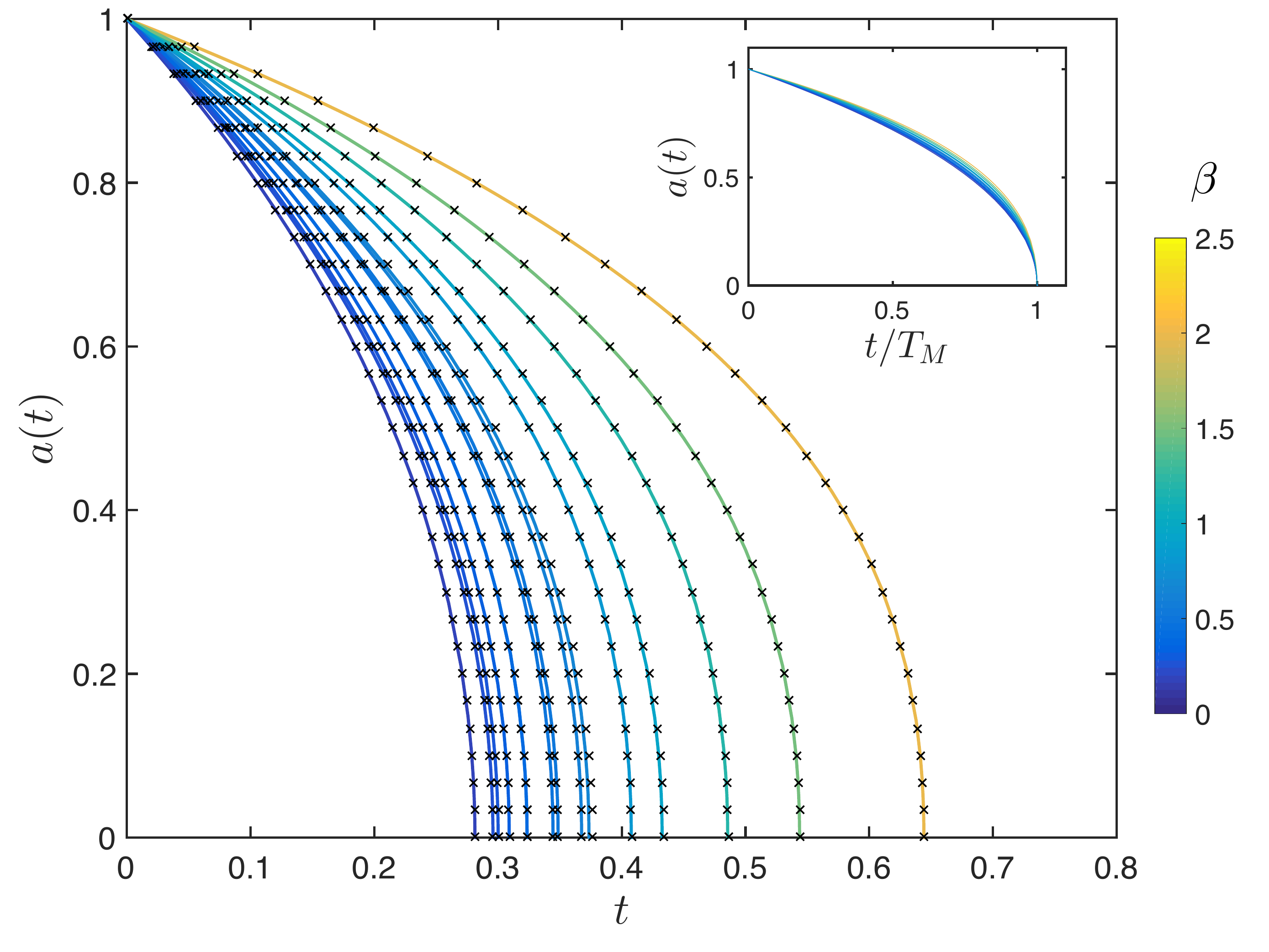}} \\ 
 &\subfigure[~Dissolution time]{\includegraphics[height=6.5cm]{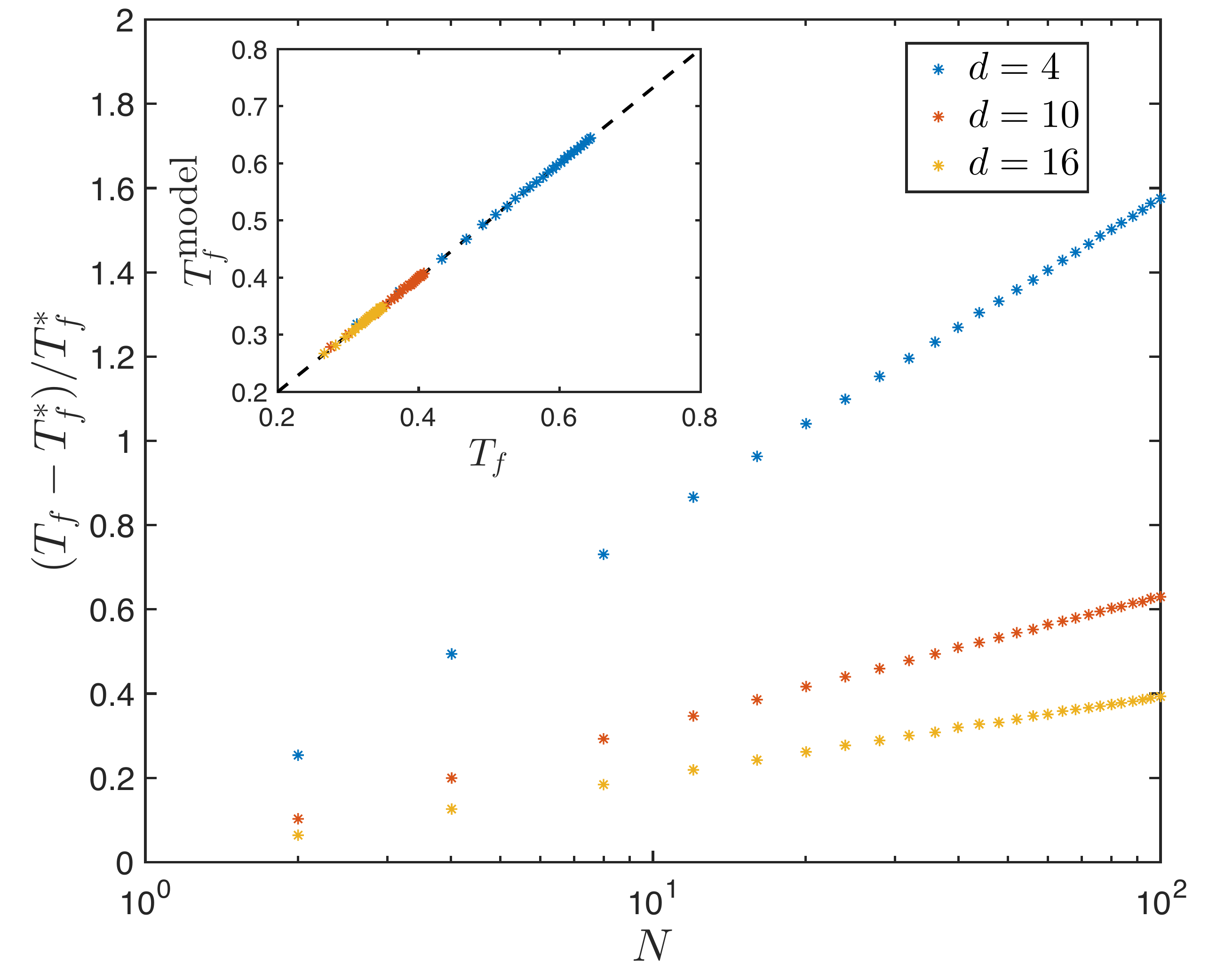}}
 \end{tabular}
 \caption{(a) Evolution of the bubble radius for a ring of identical bubbles (illustrated on the left) with initial unit radius. The bubble-bubble distance can take the values $d=[4,10,16]$  while we consider the cases of $N=[4,8,15,30,100]$ bubbles. Solid lines correspond to the complete model (using the method of reflections) and their colour is determined by the shielding parameter $\beta(N,d)$   defined in Eq.~\eqref{eq:ring_model}, while crosses correspond to the predictions of the local continuum model in Eq.~\eqref{eq:ring_Tmax} for the same value of $\beta$. (b) Relative dissolution time of the bubble ring as a function of the number of bubbles for three different values of the distance $d$ between the bubbles (inset: comparison to the continuum model predictions).}\label{fig:ring}
 \end{center}
 \end{figure}

\subsubsection{Validation: A circular ring of bubbles}
As an example, we consider $N$ identical bubbles uniformly distributed on a circular line of radius $R$. As all bubbles play equivalent roles, the radius $a$ and flux $q$ are functions of time only. In that case, Eq.~\eqref{eq:cback} simplifies and the dynamics of the radius  is governed by
\begin{equation}\label{eq:ring_model}
a\dot{a}=-\frac{2}{1+\beta a},\quad \beta=2\lambda\bigg(\gamma_E+\log\left(4\lambda R\right)\bigg).
\end{equation}
This equation is the same as Eq.~\eqref{eq:single_bubble_dyn} for the dynamics of a single bubble in a saturated environment with non-negligible background pressure ($\zeta=1$ and $r_0\neq 0$ in \S~\ref{sec:single}). 
The  evolution in time of the  radius of the bubbles, and the dissolution time of the assembly, are therefore  given by
\begin{equation}\label{eq:ring_Tmax}
\frac{a(t)^2}{4}+\frac{\beta a(t)^3}{6}=T^\textrm{model}_f-t,\qquad T_f^\textrm{model}=\frac{1}{4}+\frac{\beta}{6},
\end{equation}
and $\beta$ appears now explicitly as a quantitative measure of the collective shielding effect.

As shown in Fig.~\ref{fig:ring}, this local continuum model is in excellent agreement with the full solution even for small numbers of bubbles, with an error smaller than $0.1\%$ if $d\geq 5$ and $N\geq 8$ (and even $1\%$ in the case of only two bubbles). Quantitatively, for $\lambda=0.2$ and $\lambda R=10$ (i.e.~10 bubbles distributed on a circle at a distance of $5$ radii  from each other), $T_f/T_f^*\approx 1.7$, and collective effects provide a $70\%$ increase in the lifetime of the bubbles in the cluster.   More generally, the relative increase in dissolution time is observed to scale as $\log (N)/d$, a generic result for one-dimensional lattice as  confirmed in the next section.

\section{Collective dissolution of  microbubbles}
\label{sec:results}
\subsection{Dissolution of a line of microbubbles}

\begin{figure}
\begin{center}
\includegraphics[width=.95\textwidth]{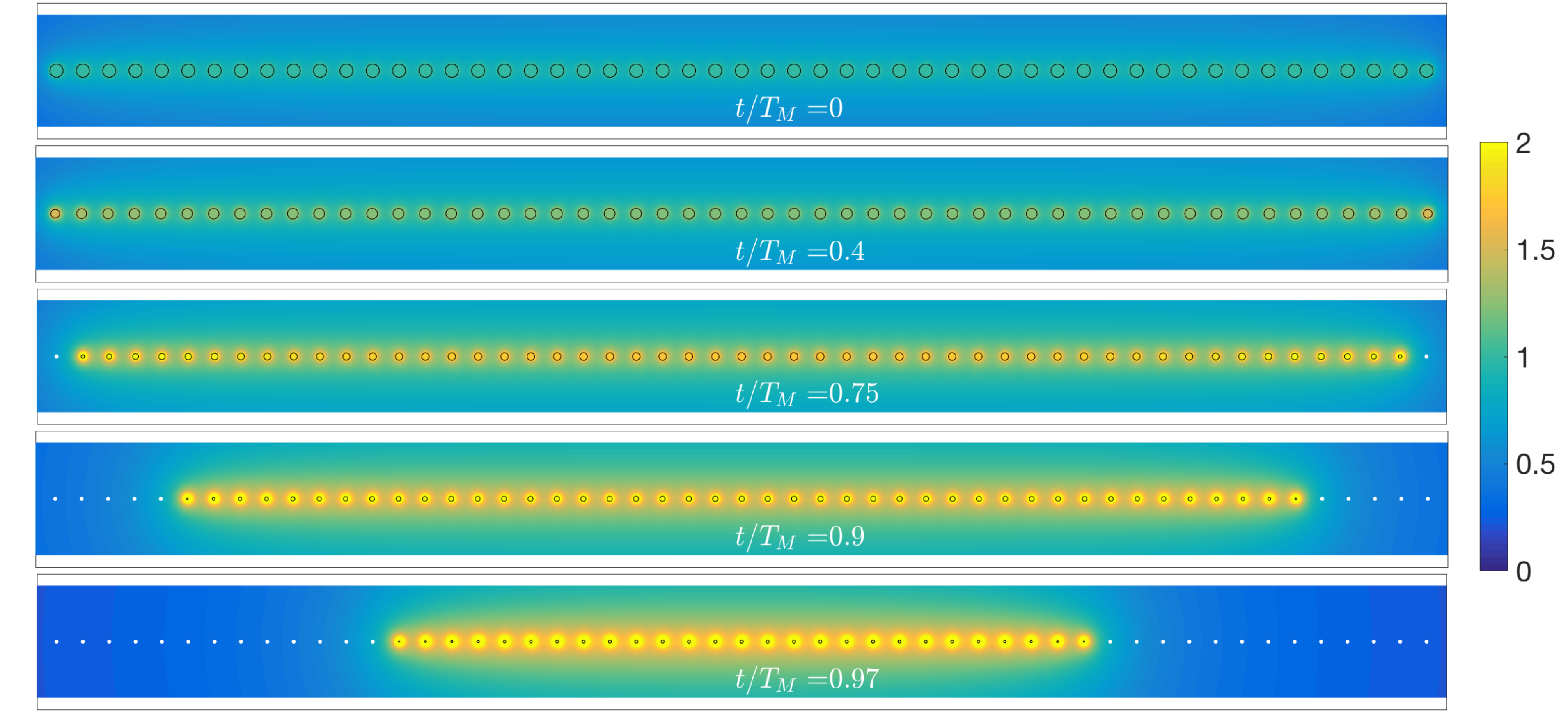}
\caption{Dissolution of a line of $N=53$ bubbles with initial unit radius and initial distance $d=4$. The surface of each bubble and the dissolved gas concentration in the fluid are shown at five different instants.  By convention, a constant concentration $c_j^s=1/a_j$ is shown inside the bubbles. The centres of bubbles that have   fully dissolved are indicated by a white dot.}\label{fig:line_movie}
\end{center}
\end{figure}

\begin{figure}
\begin{center}
\begin{tabular}{cc}
\includegraphics[height=6.5cm]{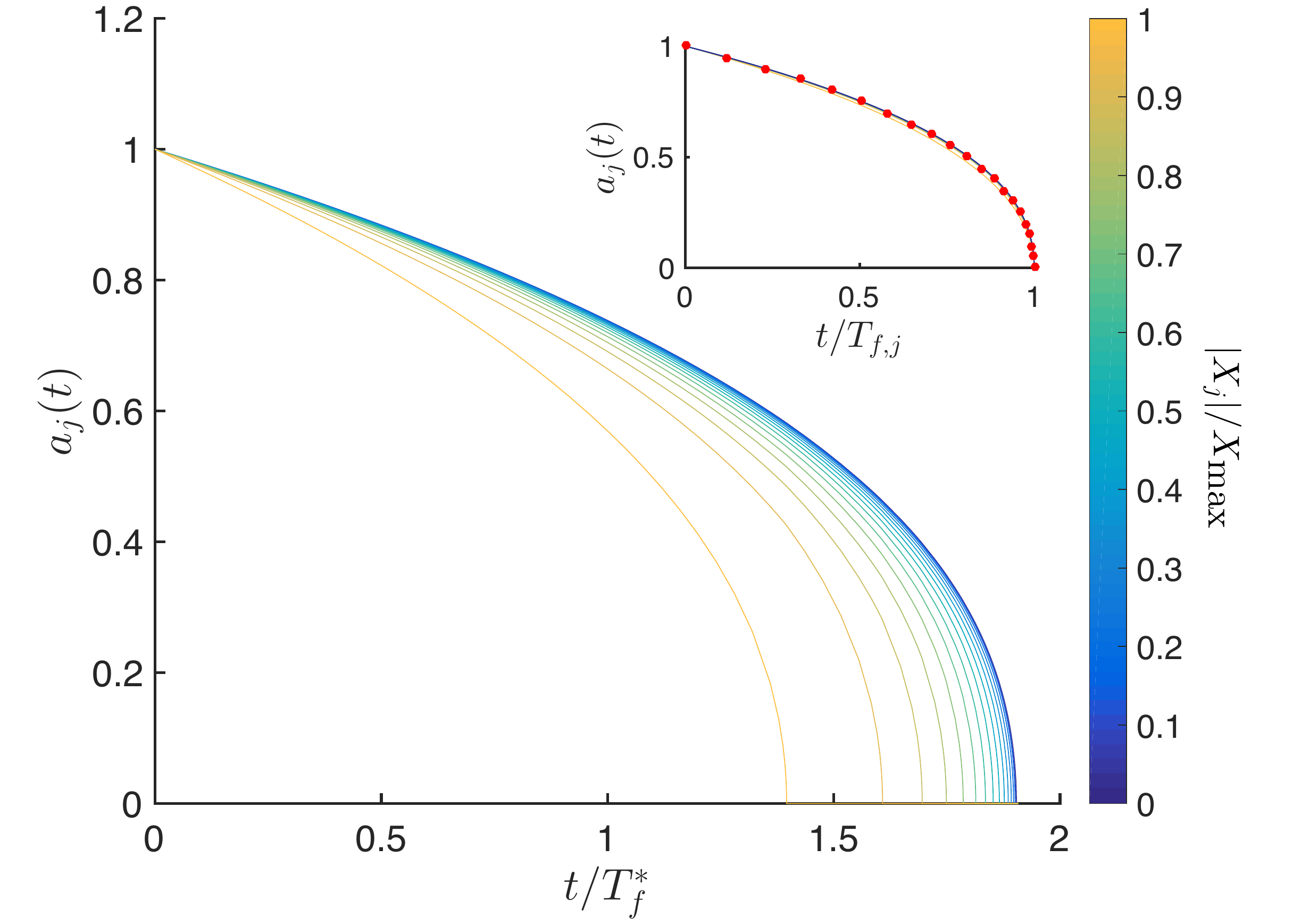}&
\includegraphics[height=6.5cm]{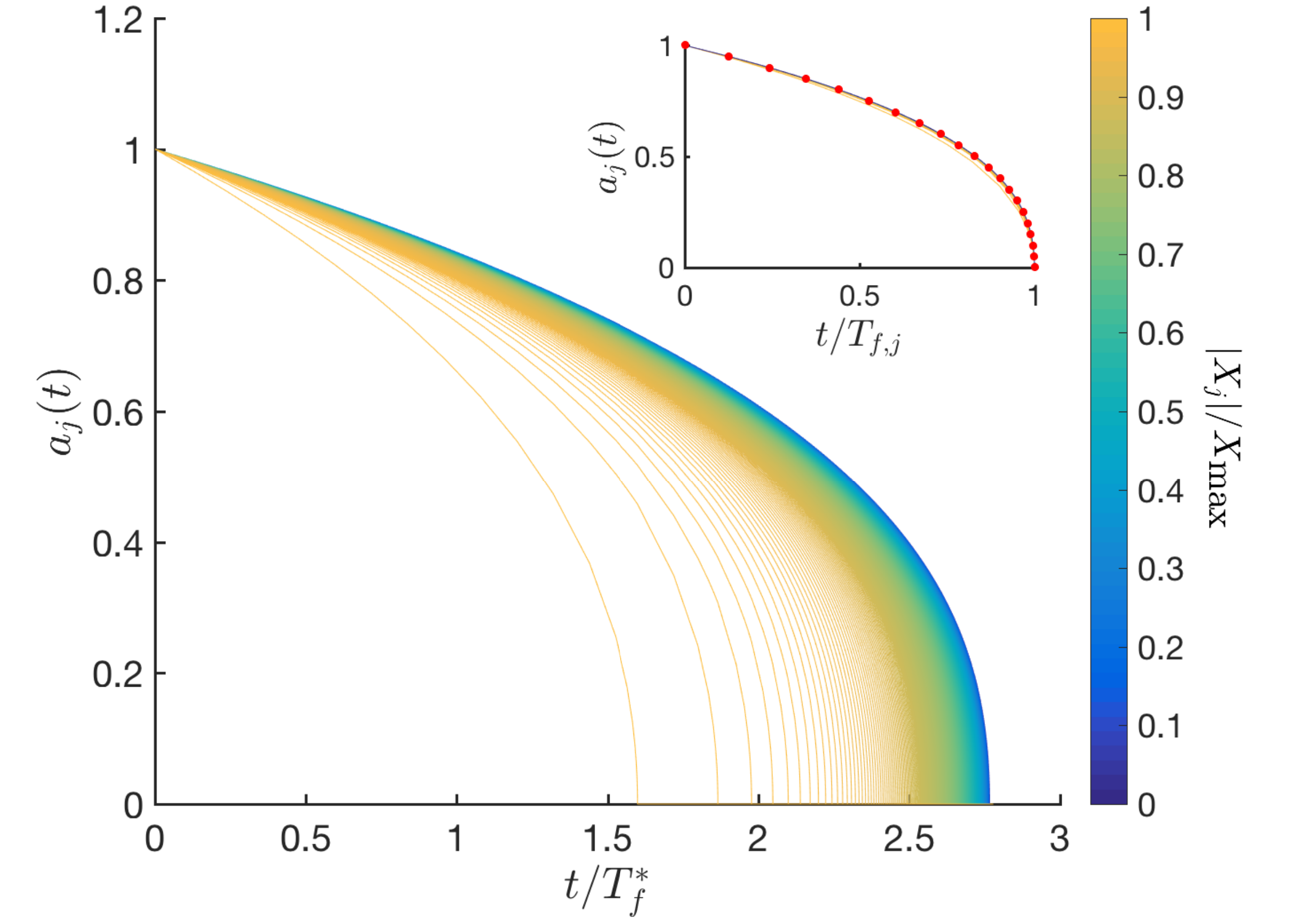}\\
\multicolumn{2}{c}{\includegraphics[height=7.5cm]{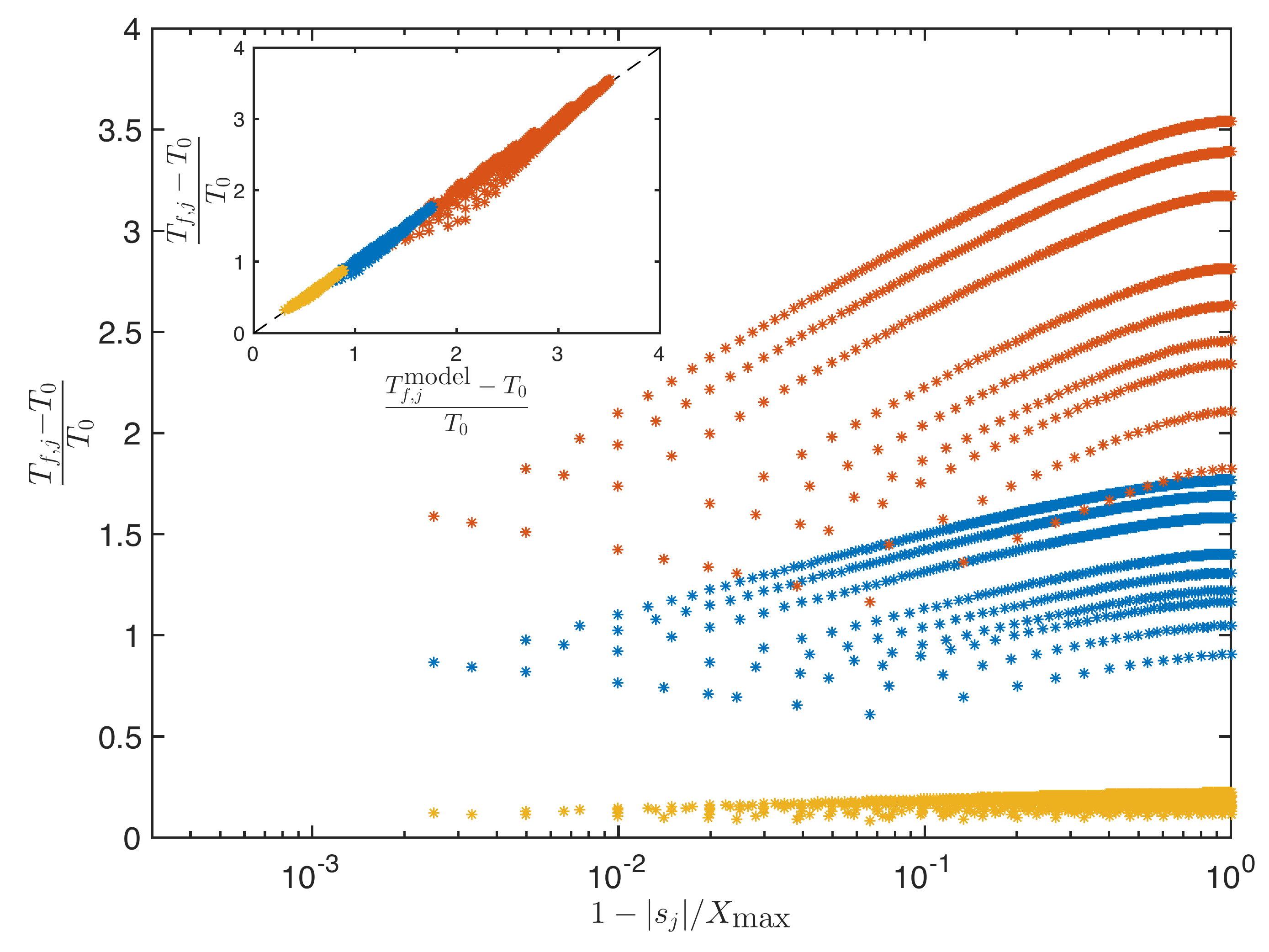}}
\end{tabular}
\caption{(Top) Time-evolution of bubble radius in a line of $N=31$ (left) and $N=803$ (right) equally-spaced bubbles with $d=5$ and initial unit radius. 
The different line colours show the normalised distance of the bubbles from the centre (i.e.~the value of $|s_j|/X_\textrm{max}$). $T_f^*=1/4$ is the dissolution time for the reference case of a single isolated bubble. The insert shows the rescaled dynamics in terms of dimensionless time, $t/T_{f,j}$, where $T_{f,j}$ is the final dissolution time of bubble $j$. The red dots indicate the dynamics obtained using the local continuum model in Eq.~\eqref{eq:line_model_time} with the shielding parameter $\beta$ adjusted to its value $\beta_0$ for the central bubble, i.e.~$\beta_0=1.33$ (left, $N=31$) and $\beta_0=2.63$ (right, $N=803$). No hydrodynamic effects are included. (Bottom) Relative increase in dissolution time as a function of the normalized distance of the bubble from the assembly edge, $1-|s_j|/X_\textrm{max}$, for $\lambda=0.1$ (yellow), $\lambda=0.2$ (blue) and $\lambda=0.4$ (red) with $31\leq N\leq 803$ bubbles (inset: comparison with the prediction of the local continuum model).}\label{fig:line_N403D5}
\end{center}
\end{figure}

We now use our asymptotic models to first address  a linear arrangement of $N$ bubbles equally-spaced by a distance $d_0=1/\lambda$. The end bubbles are therefore located at a distance $s=\pm X_\textrm{max}=\pm (N-1)d_0/2$ from the centre ($s=0$). The dissolution dynamics are illustrated  in  Fig.~\ref{fig:line_movie} in the case of $N=53$ bubbles with initial unit radius and initial distance $d=4$ at four different times showing both the sizes of the bubbles and the levels of dissolved gas concentration. A more precise quantification of the dissolution dynamics is offered in Fig.~\ref{fig:line_N403D5} where we plot the time-evolution of the bubble radii (top) and the spatial dependence of the relative increase in dissolution time (bottom). 

As expected the lifetimes of all the bubbles on the line are increased over  that of an isolated  bubble (up to a factor of three for the cases illustrated), and this effect is strongest for the bubbles located at the centre of the segment (dark blue) than for those at the end (yellow). Consequently, a dissolution front propagates from the extremal least-shielded bubbles toward the centre bubble that is most affected by its neighbours, with an exponentially-growing velocity. The lifetime $T_{f,j}$ of bubble $j$  grows logarithmically with its distance to the edge of the segment. Further, the local dissolution dynamics follows a self-similar pattern where the radius  $a_j(t)=f(t/T_{f,j})$ is seen to be identical for all the bubbles except for those located near the extremity of the segment.

This reduced dynamics is well captured using the local continuous model of \S~\ref{sec:1dcontinuous}. Direct simulations using the full model (\S~\ref{sec:MOR}), show that the nonlocal integral term in Eq.~\eqref{eq:cback} accounting for diffusive flux inhomogeneities is much smaller than   the local logarithmic term, with only two exceptions: (i)  bubbles located near the very end of the segment; (ii)  most bubbles in the final stages of their collapse. The latter can be understood by the fact that the effective ends of the segment are moving as the bubbles collapse, an effect that is not accounted for by  the local term. Neglecting these non-local effects, the continuum model simplifies into
\begin{subeqnarray}\label{eq:continuous_local}
q(s,t)&=&a(s,t)\dot{a}(s,t)=-\frac{2}{1+\beta(s)a(s,t)},\\
 \beta(s)&=&2\lambda\left(\gamma_E+\log\left[\lambda\sqrt{X_\textrm{max}^2-s^2)}\right]\right),
\end{subeqnarray}
and the dynamics of the different bubbles then reduces to that of isolated bubbles with a locally modified background forcing accounted for in the non-uniform shielding factor $\beta(s)$. The ODE in Eq.~\eqref{eq:continuous_local} can be integrated to obtain the local bubble dynamics and an estimation of the dissolution time $T_{f,j}^\textrm{model}$ of bubble $j$ as
\begin{subeqnarray}\label{eq:line_model_time}
\frac{a_j(t)^2}{4}&+&\frac{\beta_ja_j(t)^3}{6}=T_{f,j}^\textrm{model}-t,\\
T^\textrm{model}_{f,j}&=&\frac{1}{4}+\frac{\beta_j}{6}, \\
  \beta_j&=&2\lambda\left[\gamma_E+\log\left(\frac{N}{2}\right)+\frac{1}{2}\log\left(1-\frac{s_j^2}{X_\textrm{max}^2}\right)\right],
\end{subeqnarray} 
which is in excellent quantitative agreement with the observed dynamics and dissolution times (Fig.~\ref{fig:line_N403D5}).  This simple model provides a fast, yet accurate, estimate of the shielding effect introduced in a line distribution of bubbles, predicting in particular that   the final dissolution time, $T_f^\textrm{max}$,  is given by
\begin{equation}
T_f^\textrm{max}=\frac{1}{4}+\frac{\lambda}{3}\left[\gamma_E+\log\left(\frac{N}{2}\right)\right],
\end{equation}
which shows that the dissolution time grows logarithmically with the number of bubbles, and linearly with the bubble density (i.e.~$T_f^\textrm{max}\sim \log(N)/d$).

\subsection{Dissolution of two-dimensional bubble arrangements}
The results of the previous section emphasised the peculiarity of line distributions of bubbles. Due to the logarithmic behaviour of the dissolved gas concentration  near the line of bubbles, the dynamics of each bubble are governed by   its own properties and by its position within the lattice, such that other non-local effects are sub-dominant (e.g.~the size distribution of the bubbles within the lattice). This local dominance disappears for two- and three-dimensional distributions of bubbles. As an example, we now consider  the dissolution dynamics  in regular two-dimensional arrangements. 

\subsubsection{Hexagonal and circular lattices of mirobubbles}
\begin{figure}
\begin{center}
\begin{tabular}{cc}
\subfigure[~Circular lattice\label{fig:2d_circ_lattice}]{\includegraphics[height=6.5cm]{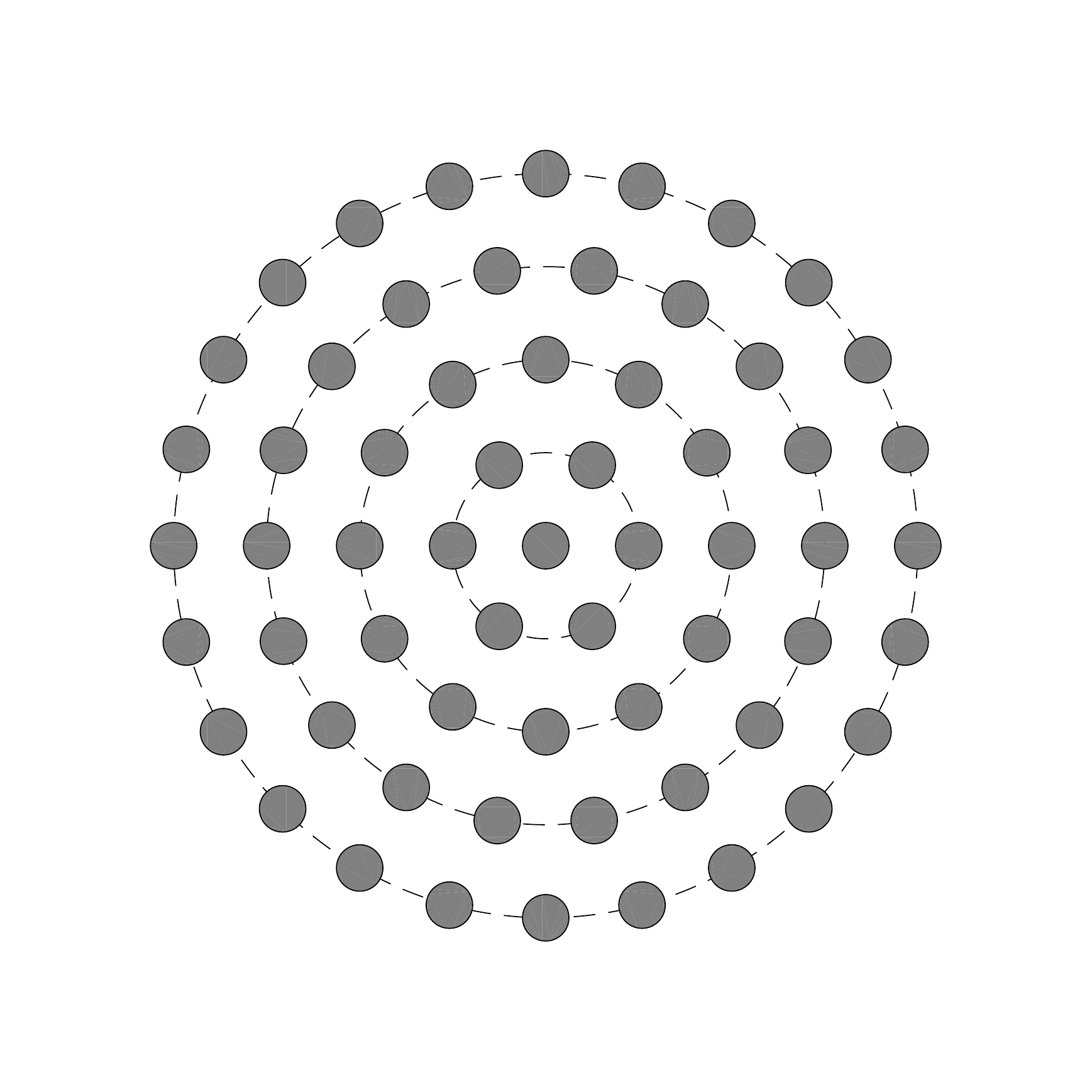}}&
\subfigure[~Hexagonal lattice\label{fig:2d_hex_lattice}]{\includegraphics[height=6.5cm]{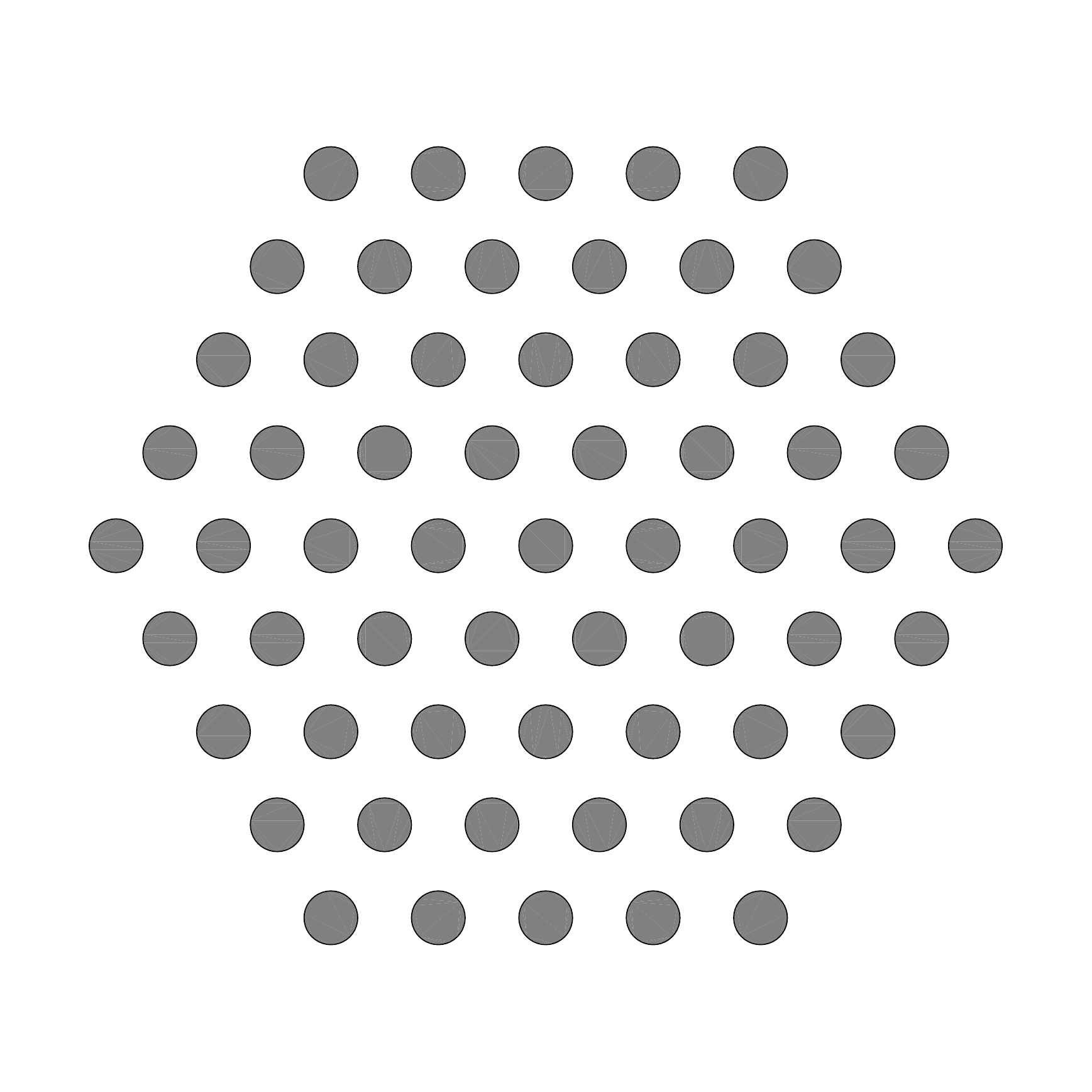}}
\end{tabular}
\caption{(a) Circular and (b)  Hexagonal lattices of bubbles with inter-bubble distance $d=4$ and $N_l=4$ layers. Initially, each bubble has unit radius. In the circular lattice, $6n$ bubbles are regularly spaced on a circle of radius $nd$ away from the central bubble. 
}\label{fig:2d_lattices_schema}
\end{center}
\end{figure}

We consider here  two different two-dimensional (2D) lattices characterised by a typical spacing $d$ between neighbouring bubbles (Fig.~\ref{fig:2d_lattices_schema}). The first  arrangement is circular, with a central bubble and $N_l$ concentric circular layers of radius $nd$ consisting of $6n$ equidistant bubbles ($1\leq n\leq N_l$), and is therefore characterised by the number of layers $N_l$ (or equivalently the radius of the lattice $\bar{R}=dN_l$), as shown in Fig.~\ref{fig:2d_circ_lattice}. The second geometry is a regular hexagonal lattice consisting of $N_l$ bubble layers around the central one; see Fig.~\ref{fig:2d_hex_lattice}. In analogy with the circular arrangement, the mean lattice radius can then be computed as the mean distance of the outer layer to the central bubble, i.e. $\bar{R}_\textrm{hex}=(3dN_l\log 3)/4\approx 0.82\,dN_l $. Although their mean density $\sigma$ (and mean radius $\bar{R}$) are slightly different, namely 
\begin{equation}
\sigma_\textrm{hex}=\frac{2}{d^2\sqrt{3}},\quad \sigma_\textrm{circ}=\frac{3N_l^2+1}{\pi N_l^2d^2},\label{eq:sigma}
\end{equation}
both lattices have the same total number of bubbles, $N=3N_l(N_l+1)+1$, and typical bubble distance, $d$.

\begin{figure}
\begin{center}
\begin{tabular}{c}
\subfigure[~Hexagonal lattice]{\includegraphics[width=.95\textwidth]{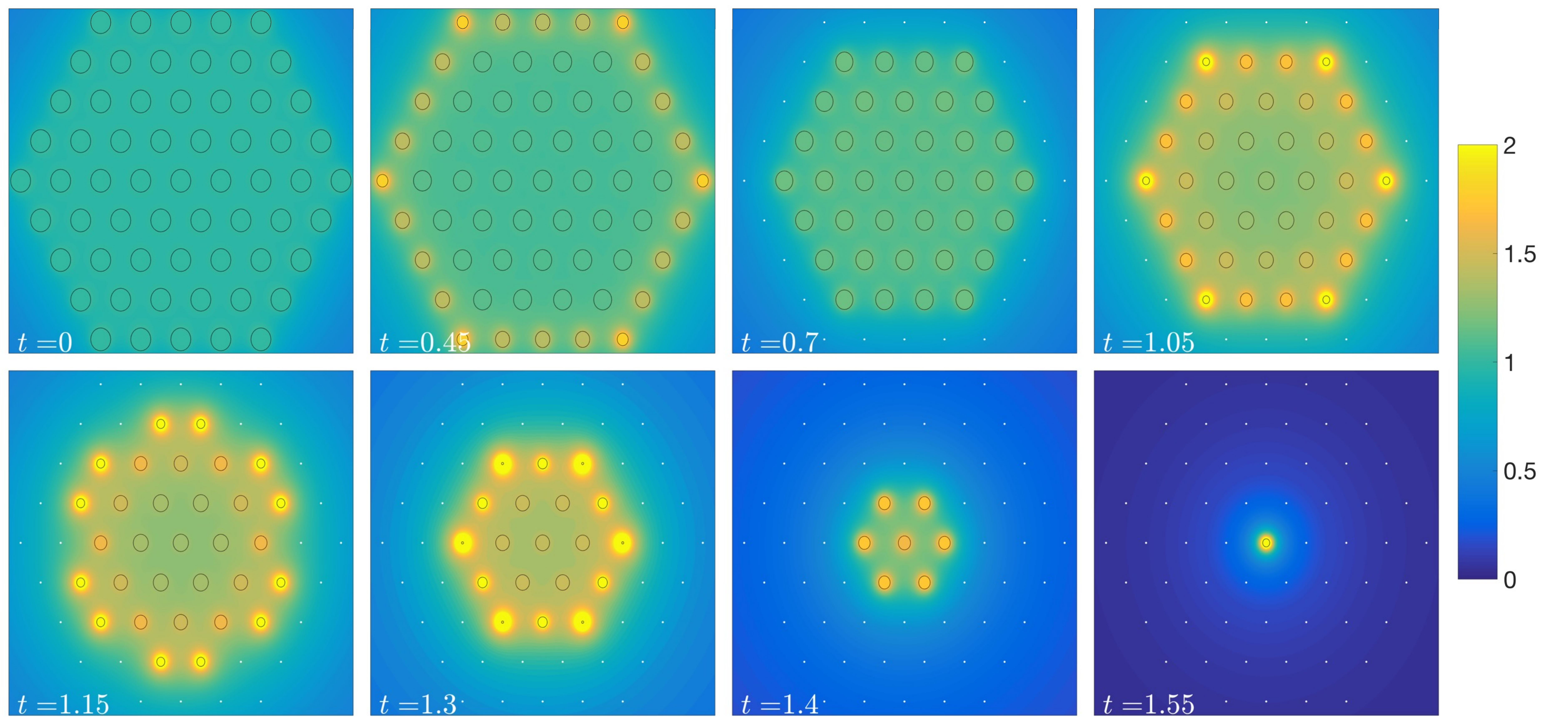}}\\
\subfigure[~Circular lattice]{\includegraphics[width=.95\textwidth]{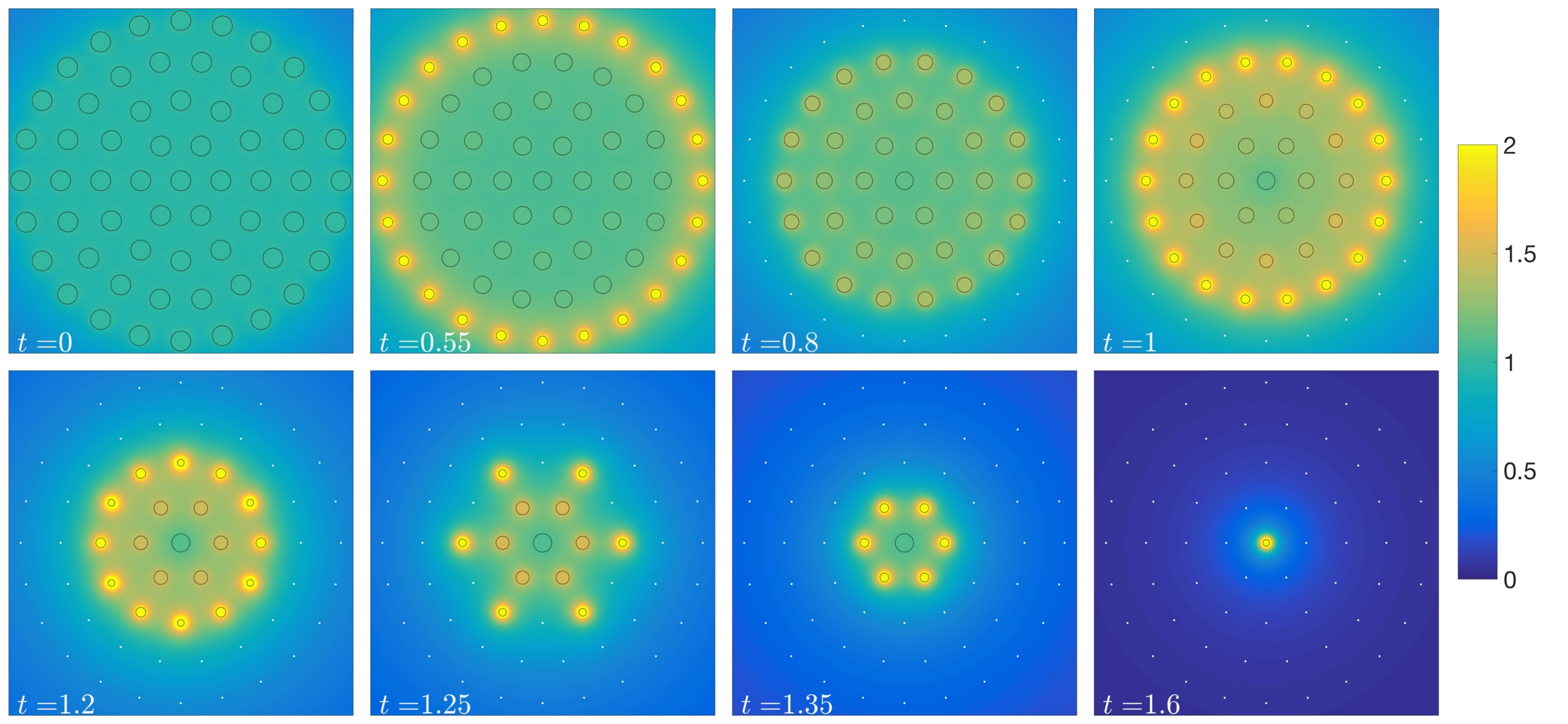}}
\end{tabular}
\caption{Dissolution of hexagonal and circular lattices with $N=61$ bubbles with initial radius $a_0=1$ and separation $d=4$ (the bubbles are distributed along $N_l=4$ layers). The surface of each bubble  and the levels of dissolved gas concentration   at different times are shown, using the convention that the constant surface concentration is assigned to the inside of each bubble. The centres of bubbles that have   fully dissolved  are indicated by a white dot. The propagation of the dissolution front is reported in Fig.~\ref{fig:2d_max_time}. Corresponding videos of the dissolution process are available as supplementary material~\cite{SI}.}\label{fig:timelapse_Nl4}
\end{center}
\end{figure}

\begin{figure}
\begin{center}
\begin{tabular}{cc}
\subfigure[~Lifetime distribution within the lattice]{\includegraphics[height=6.3cm]{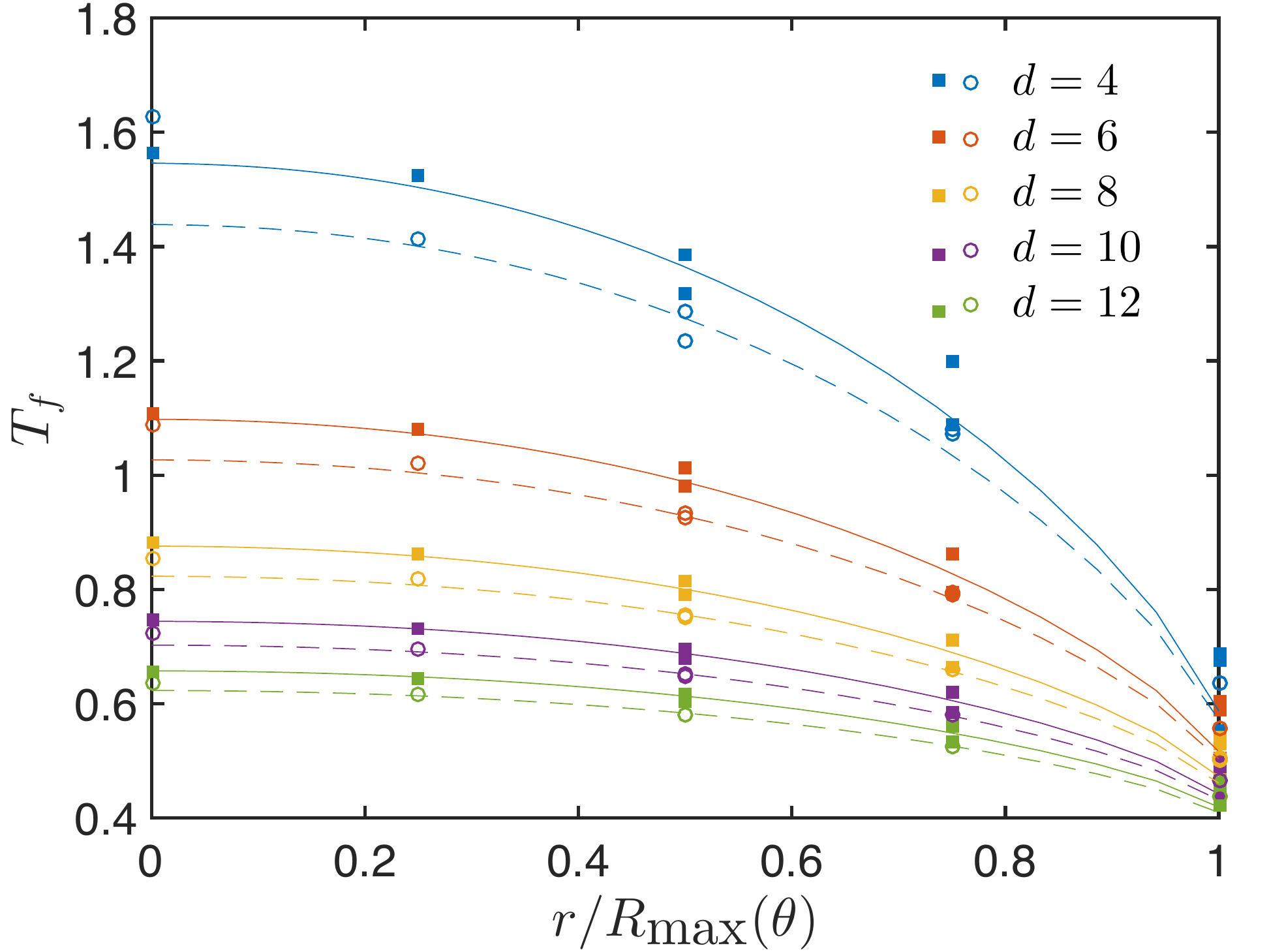}} & 
\subfigure[~Final dissolution time]{\includegraphics[height=6.3cm]{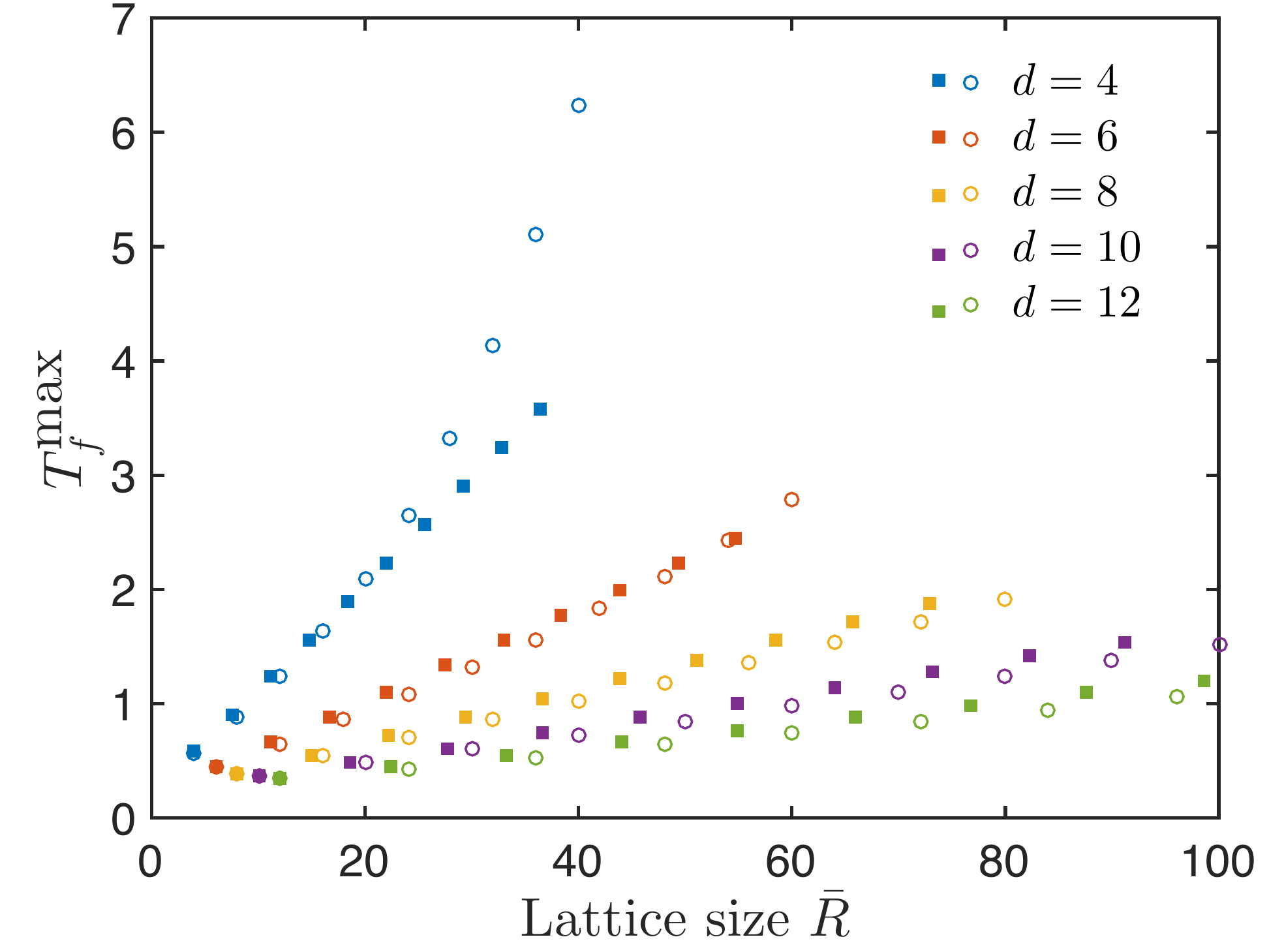}} \\
\multicolumn{2}{c}{\subfigure[~Final dissolution time (Model)]{\includegraphics[height=6.3cm]{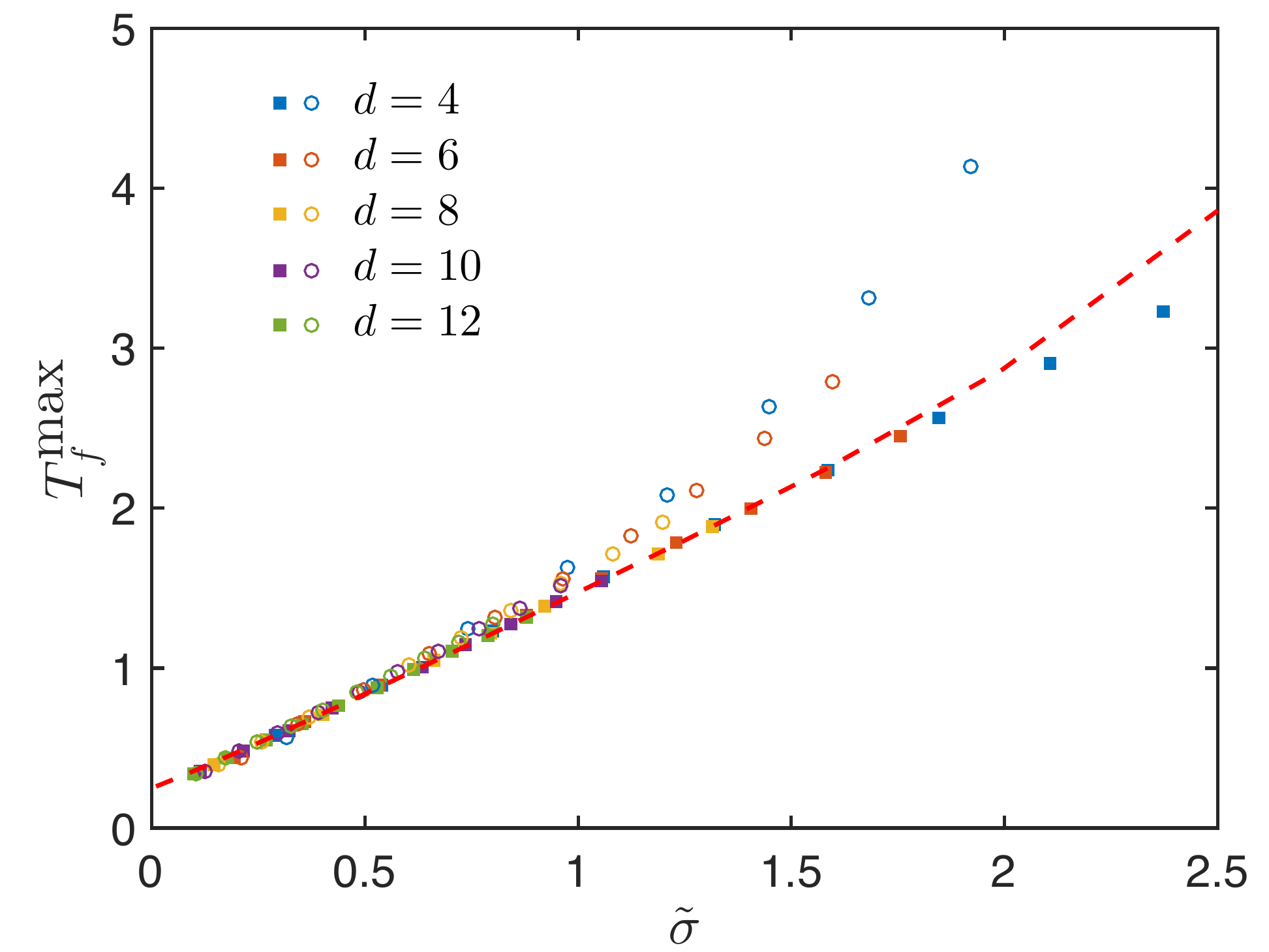}}}
\end{tabular}
\caption{(a) Distribution of individual bubble lifetime $T_f$ within the 2D lattice for $N_l=4$ layers, and (b) dependence  of the total dissolution time $T_f^\textrm{max}$ of the lattice on its mean radius $\bar{R}$ (for varying $d$ and $N_l=\bar{R}/d$). In (a) and (b), results are for the  hexagonal (solid square) and circular lattices (open circles)  and inter-bubble distances $4\leq d\leq 12$.  In (a), the results are compared to the predictions of the axisymmetric continuum model (solid line: hexagonal lattice; dashed line: circular lattice) computing the reduced density $\tilde\sigma=\sigma\bar{R}$ from Eq.~\eqref{eq:sigma} for each case (no fit). The relative radial position of the bubble is defined as $r/R_\textrm{max}(\theta)$ where $R_\textrm{max}(\theta)$ is the position of the outer edge of the lattice in terms of the polar angle $\theta$ (for the circular lattice, $R_\textrm{max}(\theta)=\bar{R}$). (c) The results of (b) are replotted in terms of the lattice reduced density, $\tilde{\sigma}=\sigma \bar{R}$,  and compared to the predictions from the continuum axisymmetric model  (dashed-red line). }\label{fig:2d_max_time}
\end{center}
\end{figure}

We obtain that the dissolution pattern is similar for  both lattices (see Fig.~\ref{fig:timelapse_Nl4} and corresponding video of the dissolution pattern~\cite{SI}), with  the outer most bubbles disappearing first and a dissolution front propagating inwards. In Fig.~\ref{fig:2d_max_time} we further show the final dissolution time of the bubbles as a function of their radial position which characterises the inward propagation of the dissolution front.  The total dissolution time of the lattice is an increasing function of the mean lattice radius, $\bar{R}$, and increases with the number of   layers. Beyond this qualitative similarity, the dynamics of both lattices can be quantitatively predicted by a single axisymmetric model (see next section) despite their local geometric differences, which emphasises that for such lattices the local arrangement   has a minor role in setting the global dynamics  at least for  moderate values of the educed density, $\tilde\sigma=\sigma \bar{R}$.  In contrast with a linear arrangement of bubbles, these results show that the increase in bubble lifetime induced by collective effects is linear in the reduced density $\tilde\sigma\sim \bar{R} a_0/d^2$ and therefore scales like $\sqrt{N}/d$.

\subsubsection{Axisymmetric continuum model}
\label{sec:2d_continuous}
Similarly to the one-dimensional model derived in \S~\ref{sec:1dcontinuous}, in the limit where the bubbles are far   from each other ($d\gg 1$) and the number of bubbles is large ($N\gg 1$), a two-dimensional model can be constructed by defining a local bubble radius $a(\xb,t)$, flux $q(\xb,t)$ and density $\sigma(\xb,t)$. For simplicity, we consider only the case of  a uniform density,  $\sigma$. In this two-dimensional case, the radius and flux distributions $a$ and $q$ satisfy
\begin{equation}
q(\xb,t)=-2(1-a(\xb,t)c_\textrm{back}(\xb,t)),\quad c_\textrm{back}(\xb,t)=-\frac{\sigma}{2}\int_S\frac{q(\xib,t)\dd S(\xib)}{|\xb-\xib|},\label{eq:2d_continuous}
\end{equation}
where the integral is now taken over the entire (planar) surface of the bubble assembly. The main difference with the one-dimensional (1D) situation is the integrability of the kernel  singularity in Eq.~\eqref{eq:2d_continuous}.  Scaling $\xb$ and $\xib$ with the typical size $\bar{R}$ of the bubble assembly, we see that the problem is governed by one parameter namely the reduced density $\tilde\sigma=\sigma \bar{R} a_0\sim \sqrt{N}(a_0/d)$. The validity of this model imposes $d,N\gg 1$ but no particular assumption on the value of $\tilde\sigma$.

Consider now an axisymmetric assembly of bubbles. In that case, all properties now solely depend on the radial coordinate, $0\leq r\leq \bar{R}$, and Eq.~\eqref{eq:2d_continuous} can be rewritten as
\begin{equation}
q(r,t)+4\tilde\sigma a(r,t)\int_0^1\frac{\rho}{\rho+r/\bar{R}}K\left(\frac{4\rho r/\bar{R}}{(r/\bar{R}+\rho)^2}\right)q(\rho\bar{R},t)\dd \rho=-2,\quad q(r,t)=a(r,t)\dot{a}(r,t),
\end{equation}
with $K(x)$ the complete elliptic integral of the first kind~\cite{abramowitz1964}. For given $a(r,t)$, the previous integral equation can be solved numerically for the diffusive flux $q(r,t)$ (and therefore $\dot{a}$) using classical quadrature methods,  and the system is then marched in time using a fourth-order accurate explicit Runge-Kutta time-stepping scheme. 
The predictions of this model are found in excellent agreement with the computations (Fig.~\ref{fig:2d_max_time}), both in terms of the final dissolution time $T_f^\textrm{max}$ and dissolution pattern (i.e.~distribution of $T_f(r)$ within the lattice).

 \subsubsection{Large-density lattices and local sensitivity}
The results obtained so far for low-to-moderate reduced density $\tilde\sigma$ showed that (i) the dissolution of regular two-dimensional lattices is characterized by the inward-propagation of a dissolution front (i.e.~the outer most bubble layers disappear fastest, shielding the inner bubbles from excess diffusion), (ii)  this process  depends only weakly on the local structure of the lattice and (iii)  these dynamics are very well represented by an axisymmetric continuum model.

\begin{figure}
\begin{center}
\includegraphics[width=.95\textwidth]{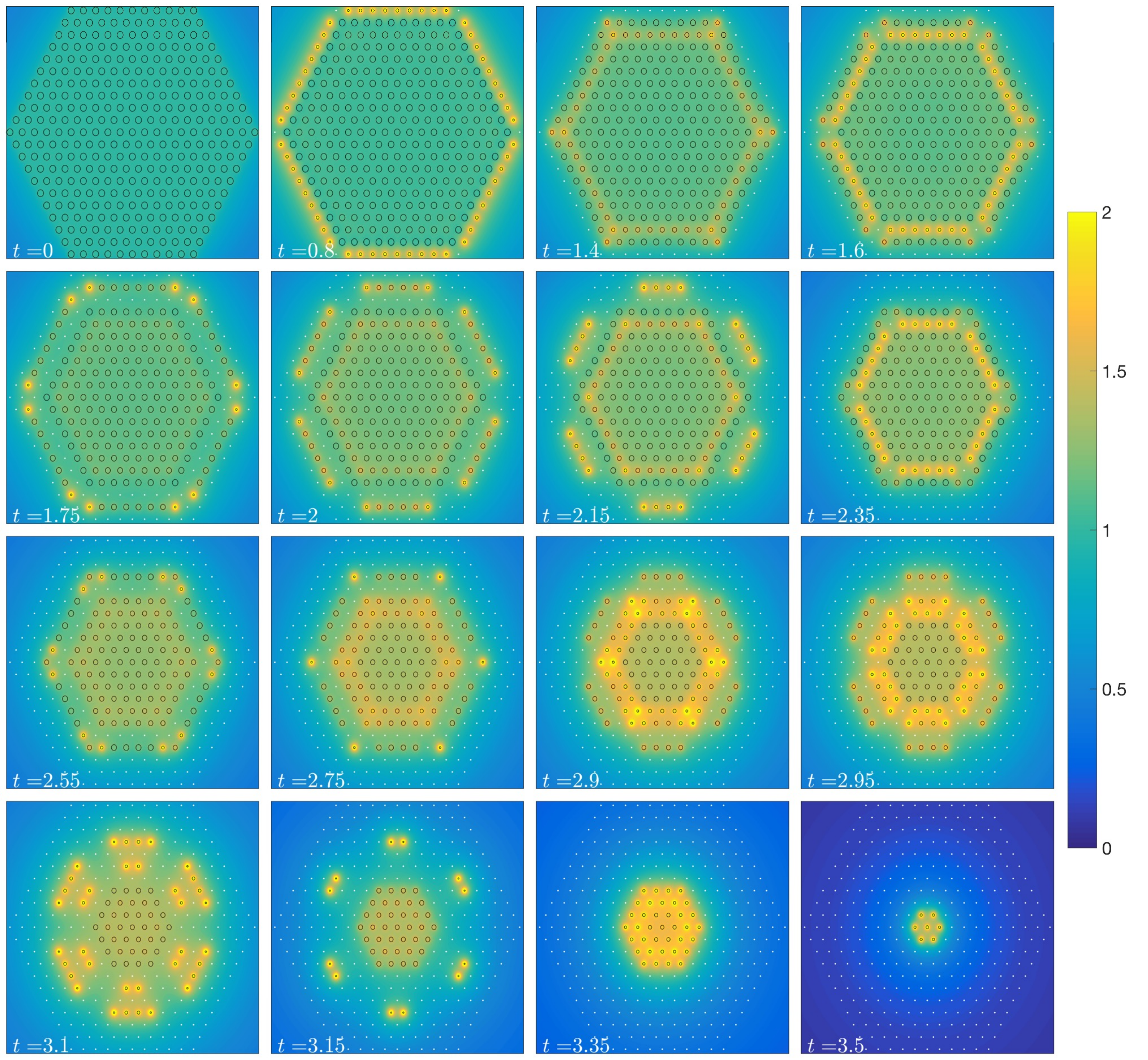}
\caption{Evolution of an hexagonal cluster of initially identical bubbles with unit radius ($a_0=1$) and $N_l=10$ layers (a total of $N=331$ bubbles). The distance between closest neighbors is $d=4$. The colour shows the dissolved gas concentration (taken as uniform within the bubbles and equal to their surface concentration) and the initial positions of dissolved bubbles are shown as  white dots.  Corresponding videos of the dissolution process are also available as supplementary material~\cite{SI}.}\label{fig:timelapse_Nl10}
\end{center}
\end{figure}

\begin{figure}
\begin{center}
\begin{tabular}{cc}
\subfigure[~$N_l=4$, $d=4$ ($\tilde\sigma=0.97$)]{\includegraphics[height=6.3cm]{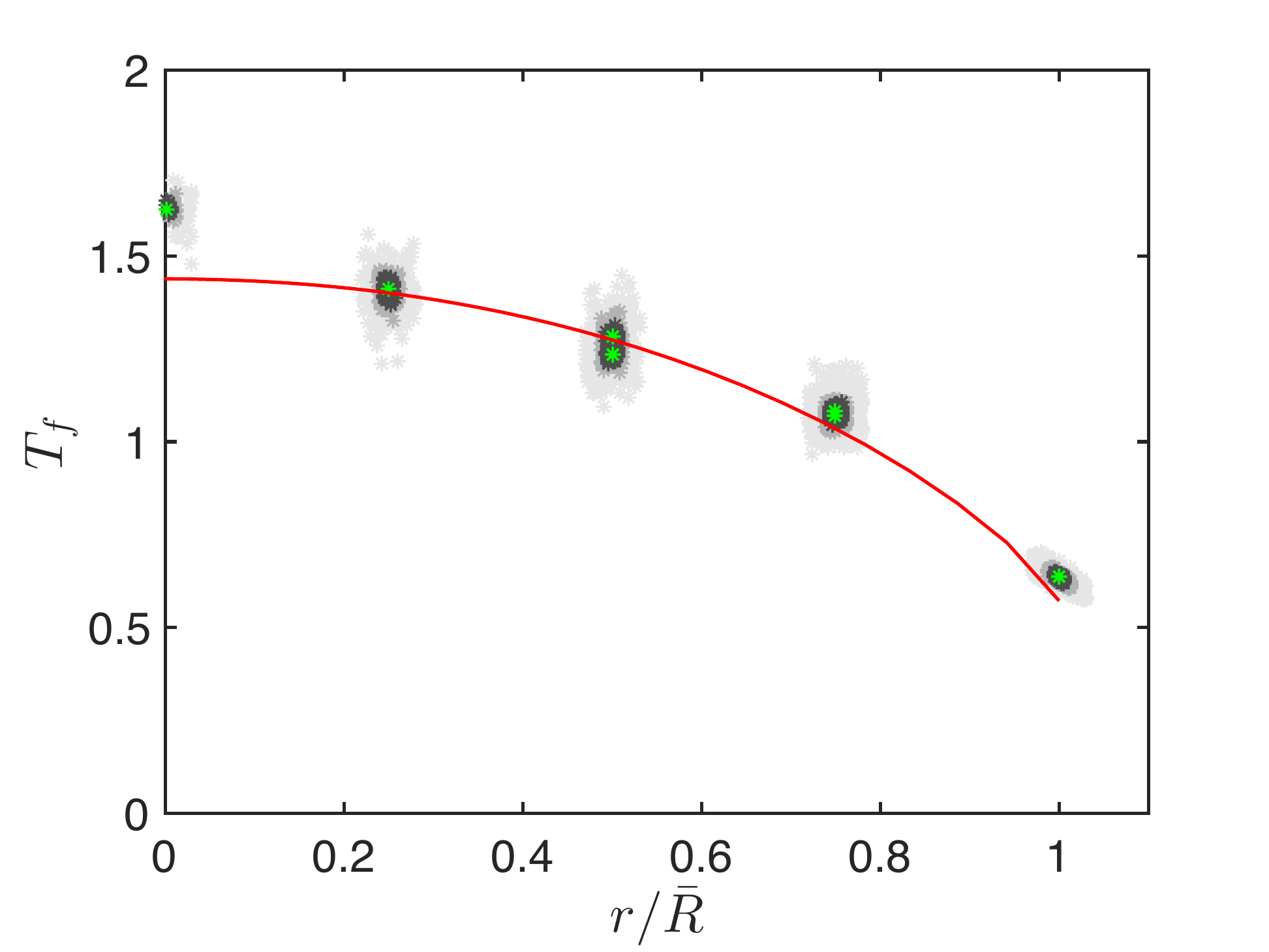} }&
\subfigure[~$N_l=10$, $d=4$ ($\tilde\sigma=2.40$)]{\includegraphics[height=6.3cm]{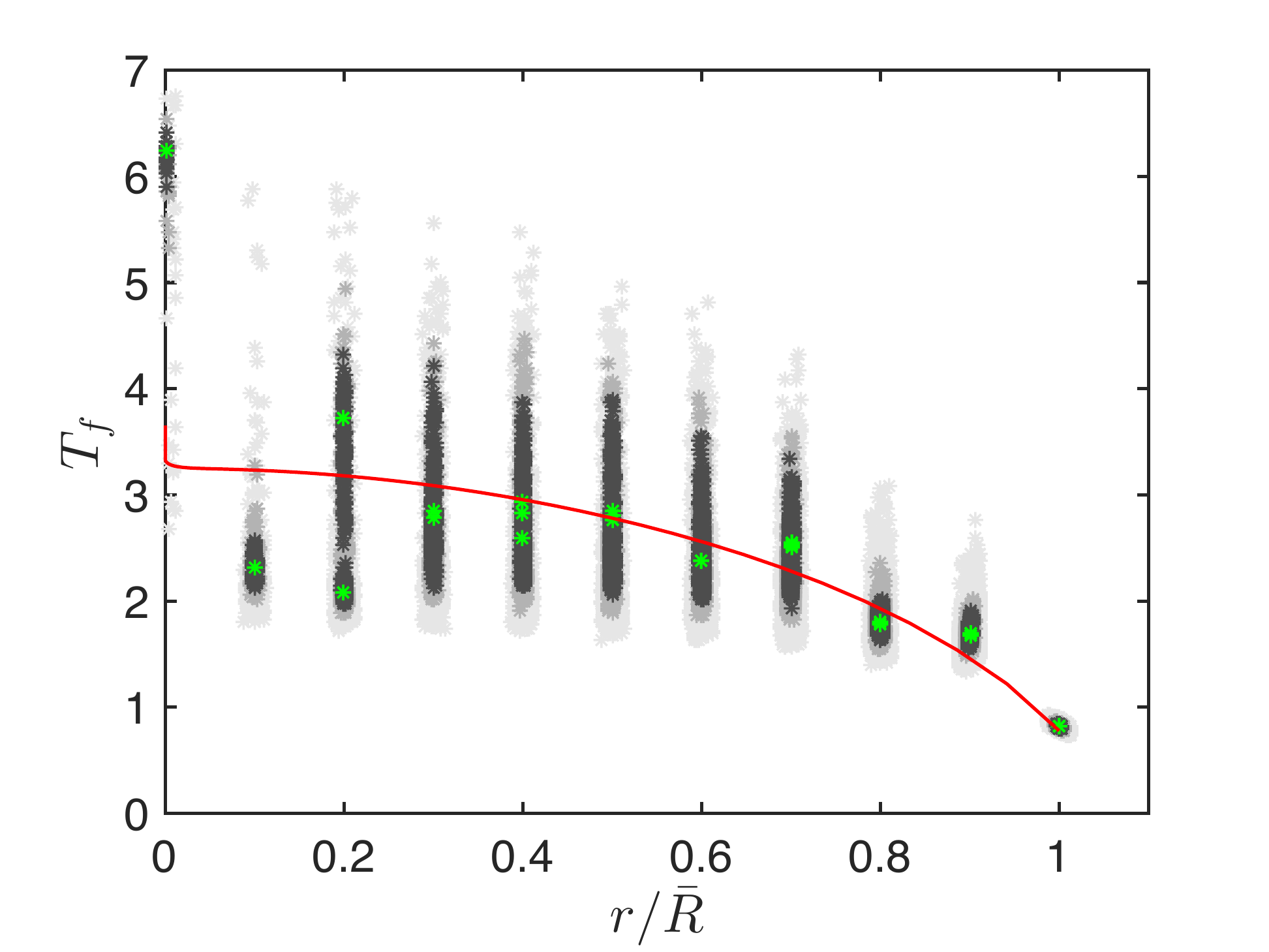} }
\end{tabular}
\caption{Influence of fluctuations in the bubble position on the dissolution time distribution $T_f(r)$ for circular lattices with $N_l=4$ (left) and $N_l=10$ layers (right). In each case, three different levels of noise ($10\%$, $20\%$ and $50\%$ bubble radius corresponding to decreasing shades of grey) are represented with 20 different configurations for each level. Each bubble's lifetime is represented as a function of the bubble's position within the lattice. The results for the regular circular lattice (green symbols) are shown as well as the predictions of the continuum model for the corresponding reduced density $\tilde\sigma$ (solid red line).}\label{fig:2d_noise}
\end{center}
\end{figure}

\begin{figure}[h!]
\begin{center}
\includegraphics[width=.8\textwidth]{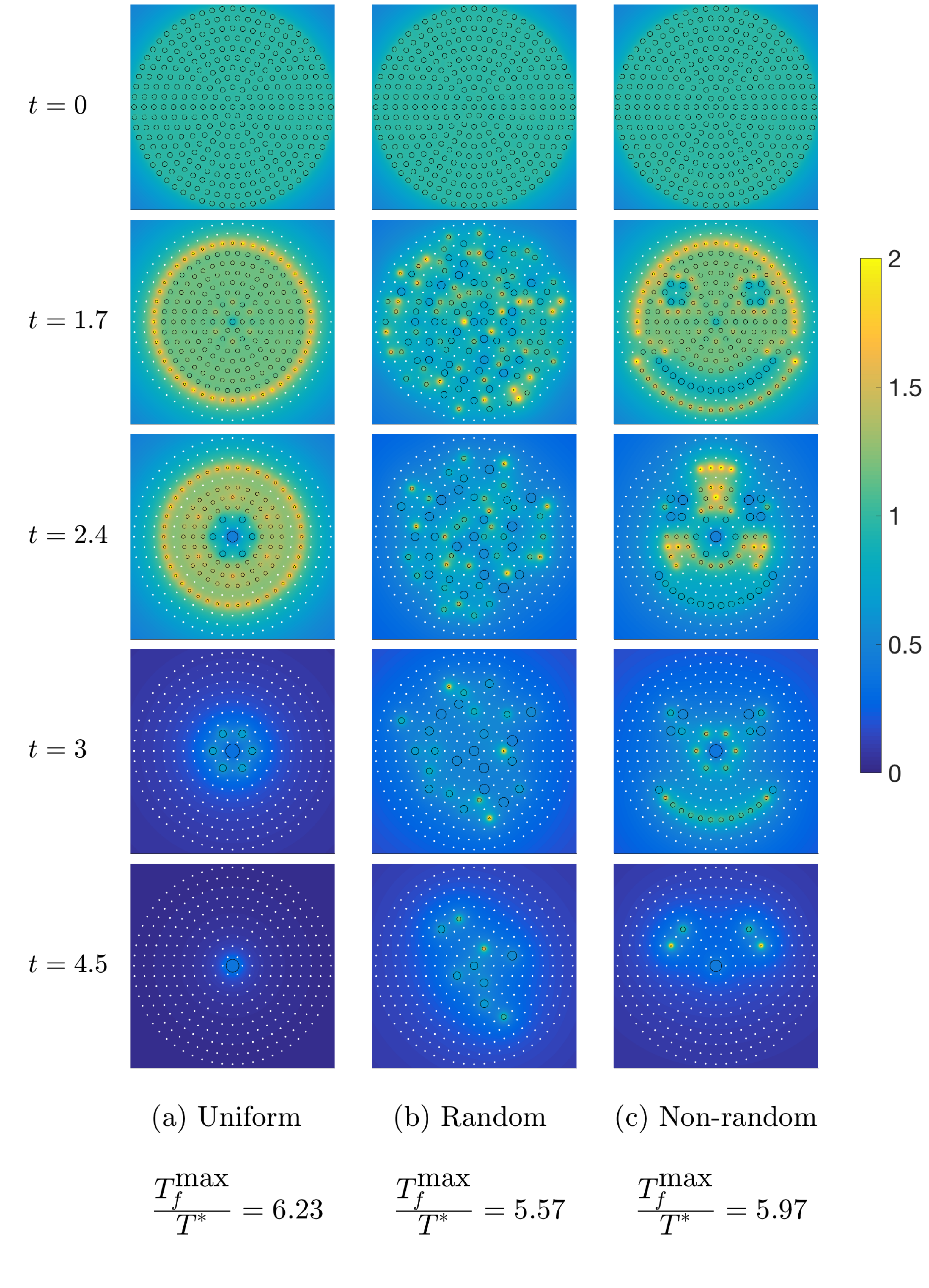} 
\caption{Sensitivity of the dissolution pattern to the initial sizes of the bubbles in a circular lattice ($N_l=10$, $d=4$): (a) Uniform ($a_i^0=1$), (b) random fluctuations and (c) non-random fluctuations in the shape of a smiley face, with $(\delta a_i^0)_\textrm{rms}=6\,10^{-3}$ in both cases. The final dissolution time $T_f^\textrm{max}$ is compared for each configuration to the reference dissolution time of a single bubble $T^*_f=1/4$, and  white dots indicate the   locations   of dissolved bubbles.  Corresponding videos of the dissolution process are available as supplementary material~\cite{SI}.}\label{fig:size_noise}
\end{center}
\end{figure}

These conclusions  do not hold for large reduced densities,  $\tilde\sigma\gtrsim 1.5$, for which the predictions of the continuum model are no longer accurate (Fig.~\ref{fig:2d_max_time}b). We further illustrate in Fig.~\ref{fig:timelapse_Nl10}   an instability in the inward-propagating dissolution front where bubble layers no longer   dissolve regularly anymore but alternatively. This leap-frogging process can be summarised as follows:
\begin{itemize}
\item[(a)]{the outer most layer of bubbles (labelled $P$ for clarity) experiences the highest diffusive flux and shrinks fastest;}
\item[(b)]{when the next layer ($P-1$) is located close enough (i.e.~large enough value of the reduced density $\tilde\sigma$), it is not only protected from excess diffusion by the outer layer but can also absorb some of the dissolved gas, in a fashion akin to the asymmetric dissolution of two different-sized bubbles (see \S~\ref{sec:bispherical});}
\item[(c)]{once layer $P$ has disappeared, the contrast in size between layer $P-2$ and $P-1$ introduced by the previous step leads to faster dissolution of layer $P-2$ that disappears before layer $P-1$;}
\item[(d)]{layer $P-1$ then dissolves, and the process is repeated until the innermost layers are reached.}
\end{itemize}

The dissolution dynamics becomes then very sensitive to the local arrangement  of the bubbles, as illustrated on Fig.~\ref{fig:2d_noise} for circular lattices to which a small amount of white noise is added in the original position of the bubbles. For small values of the reduced density $\tilde\sigma$, the overall  dynamics is not modified but at large $\tilde\sigma$, the addition of noise   leads to somewhat chaotic dissolution patterns, that completely depart from the predictions of the continuum model, which is not meant to reproduce dynamics where bubble charateristics vary significantly from one bubble to its immediate neighbour (see \S~\ref{sec:1dcontinuous}). This sensitivity to fluctuations in bubble position is not present for smaller or less dense lattices, and characterises large and relatively dense lattices.

The sensitivity to noise may also be observed by introducing   variability in the initial size of the bubbles. We illustrate  in Fig.~\ref{fig:size_noise} and the corresponding video~\cite{SI}  the dissolution patterns of three circular lattices of $N_l=10$ bubble layers with the exact same regular spatial arrangement (inter-bubble distance $d=4$), but slightly different initial bubble size distributions: for the first one, all bubbles have exactly the same (unit) initial size, while the others include random or non-random (in the shape of a  smiley face) fluctuations in size $a_j^0=1+\delta a_j$ with $(\delta a)_\textrm{rms}=0.6\%$. Although indistinguishable initially, the three lattices follow strikingly different dissolution patterns, showing transient amplification of initial perturbations before all bubbles finally dissolve. 

\subsection{Dissolution of three-dimensional bubble arrangements}
\label{sec:3d}
We finish  by turning to three-dimensional (3D) distributions of microscopic bubbles. For simplicity, we focus  on a regular spherical lattice generalising the circular lattice used in the previous section. 
Around a central bubble, $N_l$ spherical layers of bubbles are arranged, the $k$-th layer ($1\leq k\leq N_l$) having a radius $kd$ and including $12k^2$ bubbles. The position of these bubbles on a given layer are obtained so as to maximise their relative distances (using a repulsive particle algorithm) so that the mean distance $d_m$ of the bubbles within each layer is uniform, $d_m\approx 1.05\,d$. The total number of bubbles in the lattice is $N=4N_l^3+6N_l^2+2N_l+1$ so that $\bar{R}\sim N^{1/3}$.

\begin{figure}
\begin{center}
\begin{tabular}{cc}
\subfigure[~Lifetime distribution within the lattice]{\includegraphics[height=5.9cm]{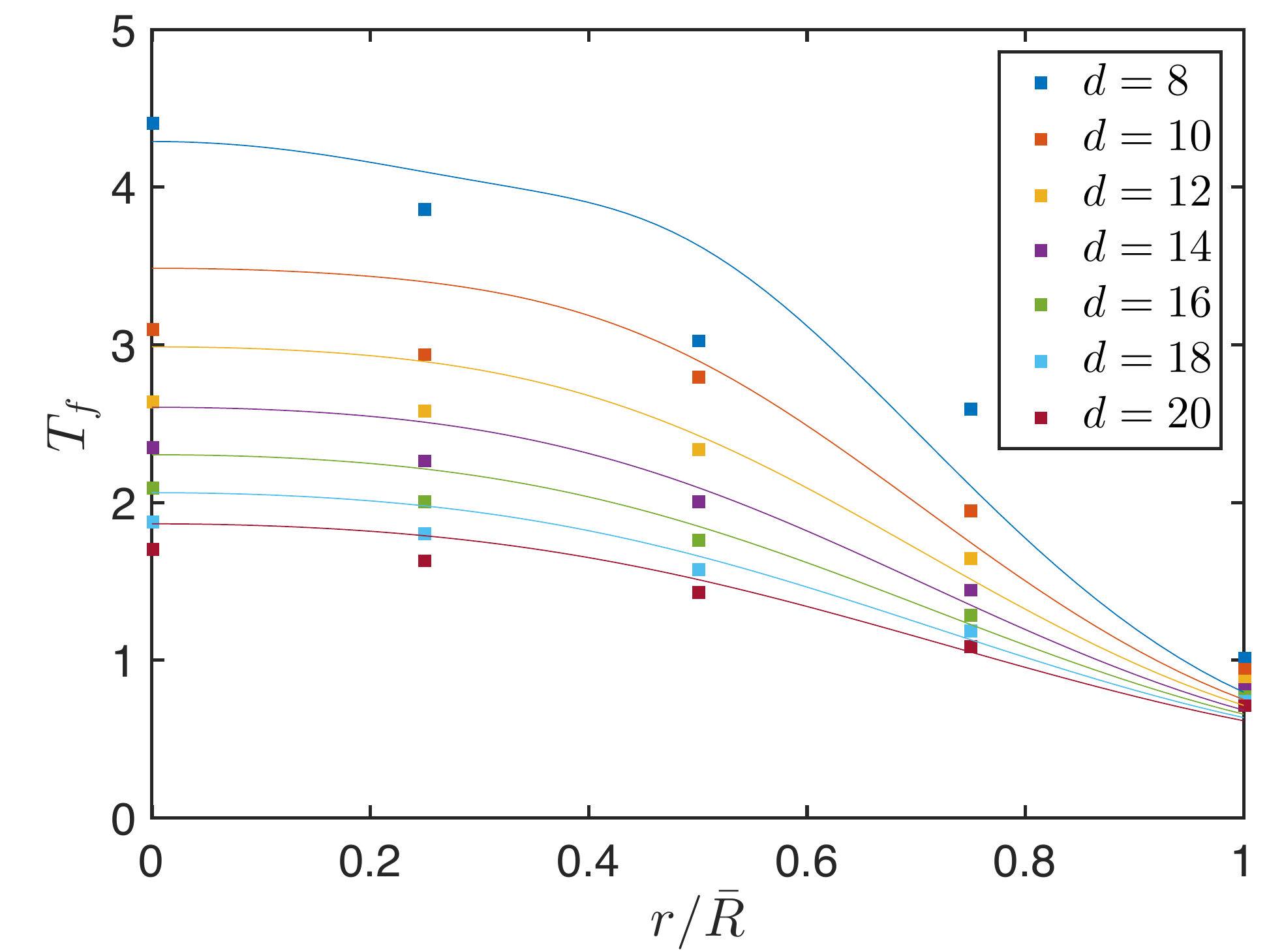}} & 
\subfigure[~Final dissolution time]{\includegraphics[height=5.9cm]{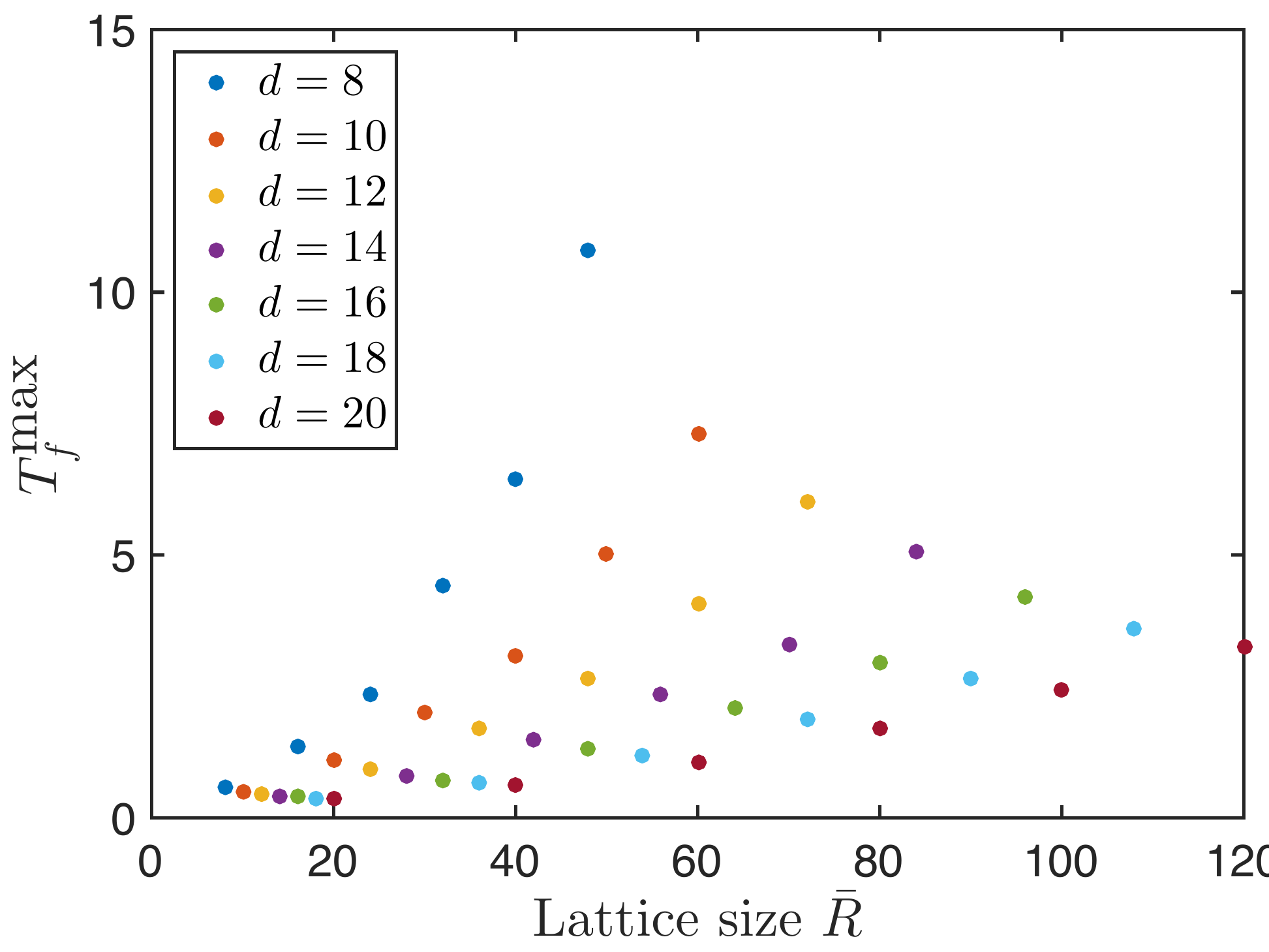}} \\
\multicolumn{2}{c}{\subfigure[~Final dissolution time (Model)]{\includegraphics[height=5.9cm]{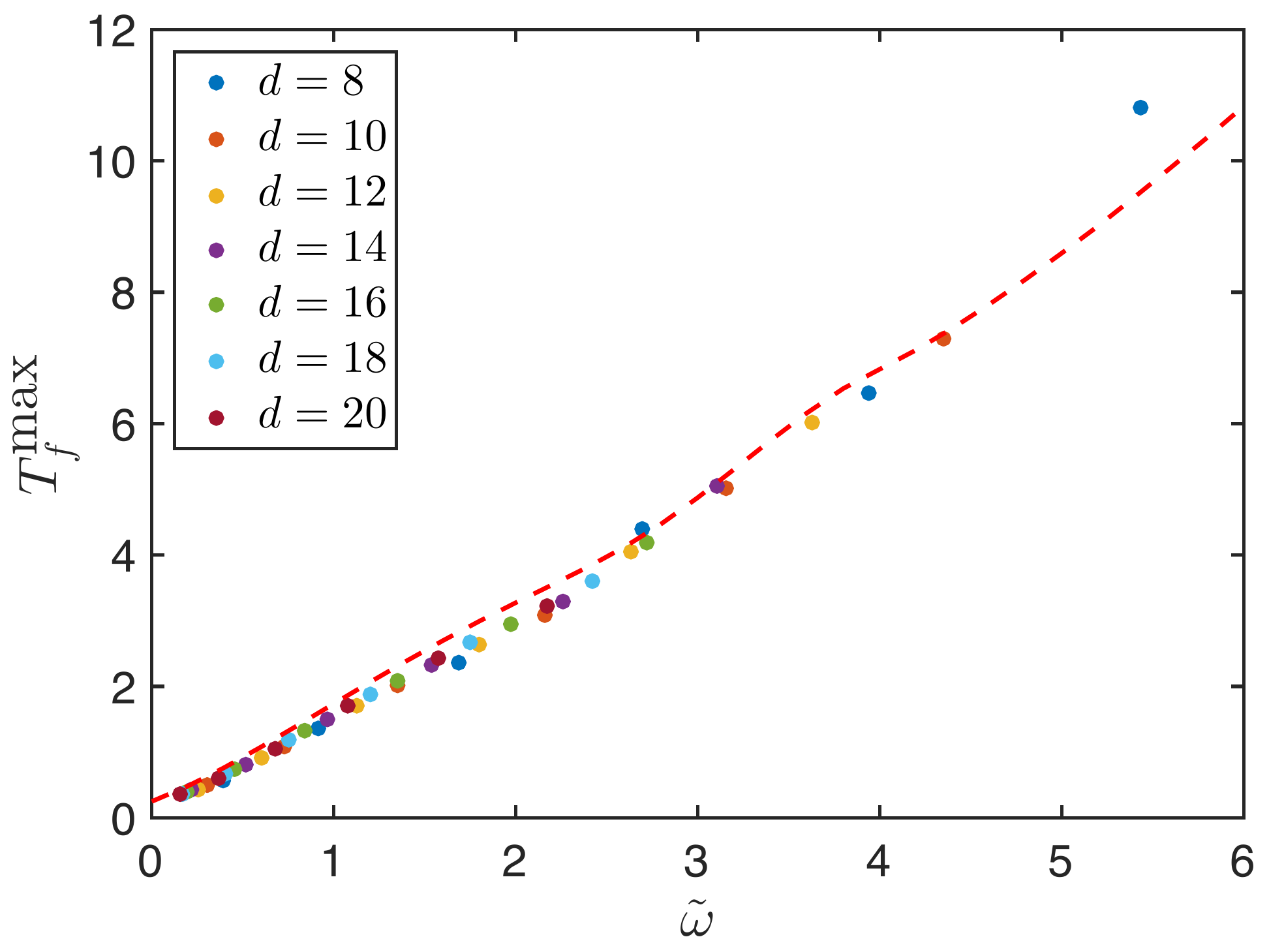}}}
\end{tabular}
\caption{(a) Distribution of bubble lifetime $T_f$ within the three-dimensional spherical lattice for $N_l=4$ layers, and (b) dependence  of the total dissolution time of the lattice on its radius $\bar{R}=N_l d$ (for varying inter-bubble distance $d$ and number of layers $1\leq N_l\leq 6$). In (a), the results are compared to the predictions of the isotropic continuum model (solid lines) with the reduced volume density of bubbles, $\tilde\omega=\omega\bar{R}^2$, where $\omega=3N/(4\pi\bar{R}^3)$ is the lattice volume density for each case (no fit). (c) The results of (b) are presented in terms of the reduced density, $\tilde{\omega}$, and compared to the predictions of the continuum isotropic model  (dashed red line). }\label{fig:3d_max_time}
\end{center}
\end{figure}

The dissolution process follows the same qualitative pattern as for one- and two-dimensional lattices, with the outer-most bubbles dissolving first, thereby shielding the central ones which experience a much extended lifetime (the corresponding video of the dissolution process is available as supplementary material~\cite{SI}). The results for the dissolution time,  $T_f$, as a function of a bubble's position within lattice are shown in Figs~\ref{fig:3d_max_time}a for a fixed number of bubble layers ($N_l=4$) and varying $d$, while the final dissolution time of the lattice, $T_f^\textrm{max}$, is shown for various number of layers $1\leq N_l\leq 6$ and inter-bubble distances $d$ in Fig.~\ref{fig:3d_max_time}b. 

Following the approach in the previous section, a continuum model can be proposed   where (i) the dissolution of each bubble is assumed to be  identical to that of an isolated bubble within a background concentration $c_\textrm{back}(\xb,t)$ and (ii) the background concentration $c_\textrm{back}(\xb,t)$ is obtained by considering the influence of a continuum distribution of bubbles acting as isolated sources of intensity $q(\xb,t)$, i.e.
\begin{equation}
q(\xb,t)=-2(1-a(\xb,t)c_\textrm{back}(\xb,t)),\quad c_\textrm{back}(\xb,t)=-\frac{\omega}{2}\int_\Omega\frac{q(\xib,t)\dd \Omega(\xib)}{|\xb-\xib|},\label{eq:3d_continuous}
\end{equation}
where $\omega$ is the volume density of bubble (considered uniform here for simplicity) and the integral is performed over the entire volume of the bubble lattice. A first approximation of the spherical lattice considered here is an isotropic model where $a(\xb,t)$ and $q(\xb,t)$ are functions of the distance to the lattice's centre, $r=|\xb|$, only. In that case, the integral equation simplifies as
\begin{equation}
q(r,t)+4\tilde\omega a(r,t)\int_0^1\frac{q(\rho\bar{R},t)\rho^2\dd\rho}{\textrm{max}(r/\bar{R},\rho)}=-2,\quad q(r,t)=a(r,t)\dot{a}(r,t),
\end{equation}
and the problem is now determined by a single parameter, namely the reduced volume density $\tilde\omega=\omega \bar{R}^2a_0\sim N^{2/3}(a_0/d)$. Note that the kernel involved in the previous integral equation is now regular. Solving this integro-differential equation numerically provides a prediction for the final dissolution time $T_f^\textrm{max}$ of the lattice and  the propagation of the dissolution front. Those are compared to the full model in Fig.~\ref{fig:3d_max_time}. While some discrepancies are observed on the detailed dissolution pattern (probably due to the non-uniform and non-isotropic bubble density in the actual lattice), an excellent agreement is observed for the final dissolution time which confirms that for such 3D lattices, the dissolution time now scales with the reduced volume density $\tilde\omega\sim N^{2/3}/d$, where $N$ is the total number of bubbles and $d$ their minimum relative distance.

\section{Conclusions}
\label{sec:conclusions}
The results presented here provide   quantitative insight on the fundamental physics of  diffusive shielding  in the collective dissolution of microbubbles. Each bubble, acting as a source of dissolved gas, reduces the effective under-saturation of the fluid around its neighbours  and slows down their dissolution. While all bubbles still dissolve in finite time, the final dissolution times of large bubble lattices may be orders of magnitude larger than the typical lifetime of an isolated bubble. This gain in dissolution time is inversely proportional to the typical inter-bubble distance $d$ and follows a different scaling with the number  $N$  of bubbles in the assembly depending on the  dimensionality of the lattice, namely $\log(N)$ for 1D lattices,  $N^{1/2}$ for 2D and $N^{2/3}$ for 3D bubble lattices.

Regular dissolution patterns are characterised by an inward-propagating dissolution front where outer bubbles dissolve first as they experience the weakest shielding and inner bubbles experience the longest lifetime. This regular cascade breaks down for large (or dense) bi-dimensional lattices, for which regular lattices exhibit leapfrogging patterns with successive bubble layers dissolving alternatively. This complex behaviour arises together with a critical sensitivity to fluctuations in the  spatial arrangement of the bubbles,  or their size distribution, leading to chaotic dissolution patterns of the lattice.

The dissolution dynamics is almost entirely controlled by diffusion of the dissolved gas within the liquid phase. While shrinking bubbles also create a flow that induce a relative motion of the bubble lattice,  this effect is completely negligible provided the contact distance between bubbles is comparable to, or greater than, their respective radii. Hydrodynamically, each bubble acts as a sink and the relative magnitude between their displacement and their surface motion scales therefore  as the square of radius over distance, $(a/d)^2$, which can be quickly neglected.

Accurate simulations of the   dissolution process pose   a fundamental challenge in the case of many bubbles. The present study provides in that regard two powerful analytical tools to treat problems involving a large number of bubbles. These are based on the classical method of reflections, whose validity is precisely characterised, and a self-consistent continuum framework, respectively. Using only two reflections provides   accurate estimates of the diffusive mass flux, provided the bubble contact distance $d_c$ is greater than a fraction of bubble radius, as carefully demonstrated for two-bubble systems for which a semi-analytical solution is available. These asymptotic  frameworks could open the door to many different applications, e.g.~as an attractive alternative to computationally-expensive numerical simulations to study suspension dynamics. 

The present analysis was formulated to tackle bubble dissolution in an infinite liquid, but it could easily be extended to confined geometries (e.g.~bubbles near a wall). A confinement-induced shielding effect is indeed expected for the same physical argument since, by restricting the diffusion of dissolved gas away from the bubble, a confining and insulating boundary reduces the diffusive mass flux and extends the lifetime of the bubbles. Furthermore, a similar physics (in reverse) would be expected to take place for groups of bubbles in a super-saturated situation  which would be subject to collective bubble growth.

The capillary-driven dissolution of a single bubble considered here is formally quite similar to liquid droplet dissolution with one important mathematical difference. In the case of droplets,  as for  bubbles  in the limit of negligible surface tension \cite{epstein1950},  the gas concentration at the surface takes a fixed equilibrium value, and so does the droplet density, whereas both quantities are inversely proportional to the bubble radius in the regime considered here. 
The equation of evolution for a single droplet is therefore identical but that governing collective dynamics (e.g.~a generalization of Eq.~\eqref{eq:mor_flux}) are not. Similar remarks would also apply to the heat diffusion   and   the evaporation/condensation dynamics of vapour bubbles. Collective effects for such applications therefore deserve further analysis in particular for large lattices which were identified here as highly sensitive to fluctuations. 

Experimentally, our work makes a  number of predictions which could be directly testable, including  the lifetimes of individual bubbles, the scaling of lifetimes of lattices with the total number of bubbles, and their spatial distributions. Clearly, however, the mathematical setup considered in a paper is idealised. An experimental investigation of our results is likely to involve surface-attached bubbles which instead of the spherical shapes considered in our work would take the shape of spherical caps. While this would render the solution to the diffusion and hydrodynamic problems more complex, we would expect our results to remain qualitatively correct.

\acknowledgements{This project has received funding from the European Research Council (ERC) under the European Union's Horizon 2020 research and innovation programme under grant agreements 714027 (SM) and 682754 (EL). The financial support of the French Embassy in the United Kingdom, Churchill College and Trinity College, Cambridge is also gratefully acknowledged.}


\appendix

\section{On the quasi-steady approximation of the bubble dissolution dynamics}\label{app:quasisteady}
The approach proposed in this work focuses on gases with very low solubility so that $\Lambda\sim c_s/\rho_g\ll 1$, which effectively decouples the diffusion and dissolution dynamics, and reduces the dissolved gas dynamics to a purely steady diffusion. This is demonstrated  here in more details by considering the unsteady diffusion of the dissolved gas around a single dissolving bubble, in a homogeneous environment with pressure $P_\infty$ and dissolved gas concentration $C_\infty$. In particular, the error introduced by the quasi-steady assumption on the dissolution time is quantified. 

In doing so, the following analysis still remains within the so-called quasi-stationary framework considered in the classical work of Epstein \& Plesset where advection by the flow is neglected and the bubble radius is considered constant in the boundary condition on the surface, Eq.~\eqref{eq:bcep} below. The reader is referred to Refs.~\cite{weinberg1980,penaslopez2016} for more in-depth discussion of the quasi-stationary assumption, which is relevant in the limit of low solubility (small $\Lambda$) considered here.

Using the reference scales introduced in Section~\ref{sec:single}, the non-dimensional diffusion and dissolution problems write
\begin{align}
\Lambda\pard{c}{t}&=\frac{1}{r^2}\pard{}{r}\left(r^2\pard{c}{r}\right),\\
c(r=a(t),t)&=r_0(1-\zeta)+\frac{1}{a},\label{eq:bcep}\\
c(r>a(t),t=0)&=c(r\rightarrow\infty,t)=0,\\
\left(1+\frac{3r_0a}{2}\right)\dot{a}&=  2a\left.\pard{c}{r}\right|_{r=a},
\end{align}
with $\Lambda=3\mathcal{R}T/4K_H$, $r_0=P_\infty R_0/2\gamma$ and $\zeta=K_HC_\infty/P_\infty$. This problem can be solved analytically for $c(r,t)$ as in Ref.~\cite{epstein1950}:
\begin{align}
c(r,t)&=r_0(1-\zeta)+\frac{1}{r}-\frac{1}{2r}\sqrt{\frac{\Lambda}{\pi t}}\int_{a(t)}^\infty\left(r_0(1-\zeta)s+1\right)\left[\ee^{-\displaystyle\frac{\Lambda(r-s)^2}{4t}}-\ee^{-\displaystyle\frac{\Lambda(r+s-2a)^2}{4t}}\right]\dd s,
\end{align}
and the dissolution dynamics follows as
\begin{equation}
\left(1+\frac{3r_0a}{2}\right)a\dot{a}=-2\left(1+r_0(1-\zeta)a\right)\left(1+a\sqrt{\frac{\Lambda}{\pi t}}\right),\label{eq:single_bubble_dyn_app}
\end{equation}
which is identical to Eq.~\eqref{eq:single_bubble_dyn} but for the same multiplicative factor as in the original derivation of Epstein \& Plesset. This correction is only significant initially and until $t\sim \Lambda a^2$, so that for $\Lambda\ll 1$, the quasi-steady result is recovered for $a(t)$ and for the final dissolution time $T_f$. 

More quantitatively, Eq.~\eqref{eq:single_bubble_dyn_app} can be recast as an equation for $t=T(a)$ with $T'(a)=1/\dot{a}$, and $T(1)=0$ and $T(0)=T_f$, the final dissolution time. For small $\Lambda$, this solution can be expended as a series in powers of $\Lambda^{1/2}$:
\begin{equation}
T(a)=T^{(0)}(a)+\Lambda^{1/2} T^{(1)}(a)+\Lambda T^{(2)}(a)+...
\end{equation}
with
\begin{align}
\totd{T^{(0)}}{a}&=-\frac{\left(1+\displaystyle\frac{3r_0a}{2}\right)a}{2(1+r_0(1-\zeta)a)},\qquad T^{(0)}(1)=0,\quad T^{(0)}(0)=T_f^{(0)},\label{eq:lambda0}\\
\totd{T^{(1)}}{a}&=\frac{\left(1+\displaystyle\frac{3r_0a}{2}\right)a^2}{2(1+r_0(1-\zeta)a)}\frac{1}{\sqrt{\pi T^{(0)}(a)}},\qquad T^{(1)}(1)=0,\quad T^{(1)}(0)=T_f^{(1)}.\label{eq:lambda1}
\end{align}
The first problem, Eq.~\eqref{eq:lambda0}, is exactly the quasi-steady limit, and leads to the dissolution time in Eq.~(8). Solving Eq.~\eqref{eq:lambda1} provides the leading-order correction, $T^{(1)}_f$, to the final dissolution time. In particular, in both the capillarity-dominated ($r_0\ll 1$, $T^{(0)}_f=1/4$) and negligible capillarity regimes ($r_0\gg 1$, $T^{(0)}_f=3/[8(1-\zeta)]$), the quasi-steady solution is simply $T^{(0)}(a)=T^{(0)}_f(1-a^2)$ and keeping only the dominant correction:
\begin{equation}
T_f=T^{(0)}_f\left(1-\sqrt{\frac{\pi\Lambda}{ T^{(0)}_f}}\right)+O(\Lambda).
\end{equation}
This shows that the relative error on the dissolution time introduced by neglecting unsteady effects in the gas diffusion is $O(\sqrt{\Lambda/T_f})$, and the quasi-steady framework is therefore especially relevant for gases with low solubility ($\Lambda$ small) or for long dissolution time (e.g. almost saturated conditions $\zeta\approx 1$ for larger bubbles -- or droplets), as expected qualitatively.

\section{Dissolved gas concentration and mass flux for two bubbles}
\label{ap:laplace}
The general solution of Laplace's equation which is regular outside both bubbles and  decays at infinity can be written as
\begin{equation}
c=\sqrt{\cosh\eta-\mu}\sum_{n=0}^\infty P_n(\mu)\left(\alpha_n\ee^{-(n+\frac{1}{2})\eta}+\beta_n\ee^{(n+\frac{1}{2})\eta}\right),
\end{equation}
with $\alpha_n$ and $\beta_n$ two sets of constants that are determined upon enforcing the boundary conditions at the surface of each bubble, Eq.~\eqref{eq:2b_bc1}. At the surface of bubble $i$, $\eta=\eta_i$ and $c=1/a_i$, or
\begin{equation}
\sum_{n=0}^\infty P_n(\mu)\left(\alpha_n\ee^{-(n+\frac{1}{2})\eta_i}+\beta_n\ee^{(n+\frac{1}{2})\eta_i}\right)=\frac{1}{a_i\sqrt{\cosh\eta_i-\mu}}=\frac{\sqrt{2}}{a_i}\sum_{n=0}^\infty \frac{P_n(\mu)\ee^{-(n+\frac{1}{2})|\eta_i|}}{2n+1},
\end{equation}
Projecting along $P_n(\mu)$ provides a linear system of two equations for $(\alpha_n,\beta_n)$ that can be solved for, as in Eq.~\eqref{eq:alphan_betan}. 

Remembering that $\nb=\eb_\eta$ (resp. $\nb=-\eb_\eta$) for $\eta=\eta_2<0$ (resp. $\eta=\eta_1>0$), and noting $h_\eta$, $h_\mu$ and $h_\phi$ the metric coefficients of the bispherical coordinate system, the diffusive flux on the boundary of bubble $i$ is now computed as
\begin{align}
q_i=&\mp\int_{-1}^1\frac{h_\mu h_\phi}{h_\eta}\left.\pard{c}{\eta}\right|_{\eta=\eta_i}\dd \mu\nonumber\\
=&\mp \sum_{n=0}^\infty\int_{-1}^1kP_n(\mu)\left[\frac{\sinh\eta_i }{2(\cosh\eta_i-\mu)^{3/2}}\left(\alpha_n\ee^{-(n+\frac{1}{2})\eta_i}+\beta_n\ee^{(n+\frac{1}{2})\eta_i}\right)\right.\nonumber\\
&+\left.\frac{(2n+1)}{2(\cosh\eta_i-\mu)^{1/2}}\left(-\alpha_n\ee^{-(n+\frac{1}{2})\eta_i}+\beta_n\ee^{(n+\frac{1}{2})\eta_i}\right)\right]\nonumber\\
=&\mp k\sqrt{2}\sum_{n=0}^\infty\ee^{-(n+\frac{1}{2})|\eta_i|}\left[\frac{\sinh\eta_i}{\sinh|\eta_i|}\left(\alpha_n\ee^{-(n+\frac{1}{2})\eta_i}+\beta_n\ee^{(n+\frac{1}{2})\eta_i}\right)-\alpha_n\ee^{-(n+\frac{1}{2})\eta_i}+\beta_n\ee^{(n+\frac{1}{2})\eta_i}\right],
\end{align}
with the $\mp$ sign taken respectively on the surface of bubble 1 and 2. Simplification of this result for $\eta=\eta_1>0$ and $\eta=\eta_2<0$  leads to the final result in Eq.~\eqref{eq:2bubble_fluxes}.

\section{Viscous flow outside two growing/shrinking bubbles}
\label{ap:hydro}
The classical solution derived by Stimson~\cite{stimson1926} for the axisymmetric flow in bispherical geometry is only valid for volume-preserving boundary conditions at the surface of the sphere and cannot account for the non-zero mass flux through the surface $\eta=\eta_{1,2}$ in the present problem. The unique solution to the Stokes flow problem presented in Eqs.~\eqref{eq:laplace_stokes}, \eqref{eq:2b_bc2} and \eqref{eq:2b_bc3} for given $\dot{a}_i$ and $\hat{W}_i$ is therefore sought as the superposition of a potential source flow and the general viscous solution,
\begin{equation}
\ub=\ub_\textrm{visc}+\ub_\textrm{pot},
\end{equation}
with
\begin{align}
\ub_\textrm{pot}&=\grad\varphi,\qquad \varphi=-\frac{(\cosh\eta-\mu)^{1/2}}{k\sqrt{2}}\left(Q_1\,\ee^{\eta/2}+Q_2\,\ee^{-\eta/2}\right),\label{eq:bispherical_pot}\\
\ub_\textrm{visc}&=-\frac{(\cosh\eta-\mu)^2}{k^2}\pard{\psi}{\mu}\eb_\eta+\frac{(\cosh\eta-\mu)^2}{k^2\sqrt{1-\mu^2}}\pard{\psi}{\eta}\eb_\mu,\\
\psi&=(\cosh\eta-\mu)^{-3/2}\chi(\eta,\mu),\quad \textrm{with   } \chi=\sum_{n=1}^\infty V_n(\mu)U_n(\eta),\label{eq:bispherical_visc}
\end{align}
where $Q_i=a_i^2\dot{a}_i$ and, following Ref.~\citep{stimson1926},
\begin{align}
V_n(\mu)&=P_{n-1}(\mu)-P_{n+1}(\mu)=\frac{2n+1}{n(n+1)}(1-\mu^2)P_n'(\mu),\\
U_n(\eta)&=A_n\cosh\left(n-\frac{1}{2}\right)\eta+B_n\sinh\left(n-\frac{1}{2}\right)\eta\notag\\
&\quad +C_n\cosh\left(n+\frac{3}{2}\right)\eta+D_n\sinh\left(n+\frac{3}{2}\right)\eta.
\end{align}
The kinematic boundary condition at the surface of sphere $i$, Eq.~\eqref{eq:2b_bc2}, imposes
\begin{align}
\pard{\psi}{\mu}=\pm\frac{Q_i\sinh^2\eta_i}{(\cosh\eta_i-\mu)^2}&-\frac{k^2\hat{W}_i(1-\mu\cosh\eta_i)}{(\cosh\eta_i-\mu)^3}+\frac{k}{(\cosh\eta_i-\mu)}\left.\pard{\varphi}{\eta}\right|_{\eta=\eta_i}\nonumber\\
=\pm\frac{Q_i\sinh^2\eta_i}{(\cosh\eta_i-\mu)^2}&-\frac{k^2\hat{W}_i(1-\mu\cosh\eta_i)}{(\cosh\eta_i-\mu)^3}-\frac{\sinh\eta_i}{2\sqrt{2}(\cosh\eta_i-\mu)^{3/2}}\left(Q_1\ee^{\eta_i/2}+Q_2\ee^{-\eta_i/2}\right)\nonumber\\
&-\frac{1}{2\sqrt{2}(\cosh\eta_i-\mu)^{1/2}}\left(Q_1\ee^{\eta_i/2}-Q_2\ee^{-\eta_i/2}\right).
\end{align}
In the equation above, the $\pm$ sign refers to $i=1$ and $2$, respectively, and stems from the definition of the outward normal vector on the bubbles' surface as $\nb=\mp \eb_\eta$. Integrating the previous equation with respect to $\mu$, with $\psi(\eta,\mu=-1)=0$ (the axis of symmetry is always a streamline), leads to
\begin{align}
\sum_{n=1}^\infty U_n(\eta_i)V_n(\mu)&=\pm Q_i\sinh^2\eta_i\left((\cosh\eta_i-\mu)^{1/2}-\frac{(\cosh\eta_i-\mu)^{3/2}}{\cosh\eta_i+1}\right)-\frac{k^2\hat{W}_i(1-\mu^2)}{2(\cosh\eta_i-\mu)^{1/2}}\nonumber\\
&-\frac{\sinh\eta_i\left(Q_1\ee^{\eta_i/2}+Q_2\ee^{-\eta_i/2}\right)}{\sqrt{2}}\left[(\cosh\eta_i-\mu)-\frac{(\cosh\eta_i-\mu)^{3/2}}{(\cosh\eta_i+1)^{1/2}}\right]\nonumber\\
&+\frac{\left(Q_1\ee^{\eta_i/2}-Q_2\ee^{-\eta_i/2}\right)}{\sqrt{2}}\left[(\cosh\eta_i-\mu)^{2}-(\cosh\eta_i+1)^{1/2}(\cosh\eta_i-\mu)^{3/2}\right].\label{eq:chi_surf}
\end{align}
Projecting the previous equation onto $P_n'(\mu)$ and applying classical properties of Legendre polynomials leads to Eq.~\eqref{eq:Un}.

The dynamic boundary condition on the surface of the bubbles imposes $\sigma_{\eta\mu}=0$ for $\eta=\eta_i$. Computing the velocity field gradient from Eqs.~\eqref{eq:bispherical_pot} and \eqref{eq:bispherical_visc}, we obtain
\begin{align}
\sigma^\textrm{visc}_{\eta\mu}&=\frac{(\cosh\eta-\mu)^{3/2}}{k^3\sqrt{1-\mu^2}}\sum_{n=1}^\infty V_n(\mu)\left[U_n''(\eta)+n(n+1)U_n(\eta)-\frac{3U_n(\eta)}{4}\left(1+\frac{2\sinh^2\eta}{(\cosh\eta-\mu)^2}\right)\right],\\
\sigma^\textrm{pot}_{\eta\mu}&=\frac{3\sqrt{1-\mu^2}}{2k^3\sqrt{2}}\left[\sinh\eta(\cosh\eta-\mu)^{1/2}(Q_1\ee^{\eta/2}+Q_2\ee^{-\eta/2})\notag\right.\\
&\qquad\qquad\qquad \left.+(\cosh\eta-\mu)^{3/2}(Q_1\ee^{\eta/2}-Q_2\ee^{-\eta/2})\right].
\end{align}
Projecting $\sigma^\textrm{pot}_{\eta\mu}+\sigma^\textrm{visc}_{\eta\mu}=0$ along $P_n'(\mu)$ leads to Eq.~\eqref{eq:Unddot}. 

Finally, the force-free condition must be applied on each bubble to determine $\hat{W}_i$ in terms of the change in radii. For axisymmetric Stokes flow, the total axial force can be computed as~\citep{stimson1926,happel}
\begin{equation}\label{eq:force_gen}
F_z=\pi k^2\int_{-1}^1\frac{\sqrt{1-\mu^2}\dd\mu}{(\cosh\eta-\mu)^3}\pard{}{\eta}\left[(\cosh\eta-\mu)\omega_\phi\right],
\end{equation}
with $\omega_\phi=\eb_\phi\cdot(\grad\times\ub)$ the azimuthal vorticity. This shows that the potential part $\ub_\textrm{pot}$ does not contribute to the force that is solely given in terms of the viscous contribution and can be computed directly using Stimson's result as in Eq.~\eqref{eq:2bubble_motion}.

\section{Method of reflections -- Laplace problem}
\label{ap:reflections}
The zeroth-order reflection for the Laplace problem outside bubble $l$ is 
\begin{equation}
c_l^0=-\frac{q_l}{2r_l}\cdot
\end{equation}
Rewriting this solution close to bubble $k\neq l$,
\begin{equation}
c_l^0(|\rb_k|=a_k)=-\frac{q_l}{2a_k}\sum_{p=0}^\infty\left(\frac{a_k}{d_{kl}}\right)^{p+1}P_p\left(\nb_k\cdot\eb_{kl}\right),
\end{equation}
and the solution to the first reflection problem is found as
\begin{equation}
c_k^1(\rb_k)=\sum_{l\neq k}\sum_{p=1}^\infty\frac{q_l}{2a_k}\left(\frac{a_k}{d_{kl}}\right)^{p+1}\left(\frac{a_k}{r_k}\right)^{p+1}P_p\left(\frac{\eb_{kl}\cdot\rb_k}{r_k}\right).
\end{equation}

The correction to surface concentration arising from the first two reflections is therefore
\begin{equation}
c_j^{s}=-\frac{q_j}{2a_j}-\sum_{k\neq j}\frac{q_k}{2d_{jk}}+\sum_{\substack{k\neq j\\l\neq k}}\sum_{p=1}^\infty \frac{q_l}{2a_k}\left(\frac{a_k}{d_{kl}}\right)^{p+1}\left(\frac{a_k}{d_{jk}}\right)^{p+1}P_p\left(\eb_{kl}\cdot\eb_{kj}\right).
\end{equation}
The third reflection will provide corrections $O(\varepsilon^7)$, and therefore truncating the previous result consistently provides the result in Eq.~\eqref{eq:mor_flux}.

\section{Method of reflections -- Validation}
\label{ap:validation}
The predictions of the method of reflections are quantitatively compared to the analytical solution for two bubbles in Table~\ref{tab:accuracy}.

\begin{table}
\begin{center}
\begin{tabular}{|c|p{2cm}|p{2cm}|p{2cm}|p{2cm}|p{2cm}|p{2cm}|}
\hline
\multirow{2}{*}{$d_c$} & \multicolumn{2}{c|}{Flux accuracy  $\left|\frac{q-q_\textrm{exact}}{q_\textrm{exact}-q^*}\right|$} & \multicolumn{2}{c|}{Velocity accuracy $\left|\frac{U-U_\textrm{exact}}{U_\textrm{exact}}\right|$} & \multicolumn{2}{c|}{Dissolution time $\left|\frac{T_M-T_M^\textrm{exact}}{T_M^\textrm{exact}-T_0}\right|$}\\
\cline{2-7}
& $S_4$ & $S_6$ & $S_3$ & $S_6$ & $S_3$ & $S_6$\\
\hline
$10^{-2}$ &$8.5\%$&$0.8\%$&$>20\%$&$>20\%$& $4.4\%$ & $1.7\%$ \\
$10^{-1}$ &$7.7\%$&$0.48\%$&$>20\%$&$>20\%$ & $4.7\%$  & $0.8\%$ \\
$1$ & $3.0\%$& $0.08\%$& $>20\%$ & $3.1\%$ & $1.9\%$ & $0.04\%$\\
$2$ & $1.3\%$& $0.04\%$& $20\%$ & $0.75\%$ & $0.8\%$ & $0.02\%$\\
$5$ & $0.25\%$ &$0.003\%$& $13.5\%$ & $0.06\%$ & $0.16\%$ & $0.01\%$\\
\hline
\end{tabular}
\vspace{.3cm}

(a) Identical bubbles, $a_2/a_1=1$

\vspace{1cm}
\begin{center}
\begin{tabular}{|c|p{2cm}|p{2cm}|p{2cm}|p{2cm}|p{2cm}|p{2cm}|}
\hline
\multirow{2}{*}{$d_c$} & \multicolumn{2}{c|}{Flux accuracy  $\left|\frac{q-q_\textrm{exact}}{q_\textrm{exact}-q^*}\right|$} & \multicolumn{2}{c|}{Velocity accuracy $\left|\frac{U-U_\textrm{exact}}{U_\textrm{exact}}\right|$} & \multicolumn{2}{c|}{Dissolution time $\left|\frac{T_M-T_M^\textrm{exact}}{T_M^\textrm{exact}-T_0}\right|$}\\\cline{2-7}
& $S_4$ & $S_6$ & $S_3$ & $S_6$ & $S_3$ & $S_6$\\
\hline
$10^{-2}$ &$>20\%$&$>20\%$&$>20\%$&$>20\%$ & $4\%$ & $3.8\%$\\
$10^{-1}$ &$>20\%$&$15.6\%$&$>20\%$&$>20\%$ &$1.4\%$ & $1.3\%$\\
$1$ & $1.2\%$& $0.3\%$& $4.8\%$ & $1.3\%$ & $0.01\%$ & $0.01\%$\\
$2$ & $0.2\%$& $0.03\%$& $5.2\%$ & $0.2\%$ & $0.16\%$ & $0.15\%$\\
$5$ & $0.01\%$ &$<0.001\%$& $3.4\%$ & $0.02\%$& $0.2\%$&$0.2\%$\\
\hline
\end{tabular}
\vspace{.3cm}

(b) Bubbles of different sizes, $a_2/a_1=1/4$

\end{center}
\caption{ Relative error on the diffusive flux,  velocity and relative dissolution time of bubble $1$ obtained using two different orders of approximation of the method of reflections for (a) two identical bubbles and (b) two  bubbles of different radii. Note that flux and velocities are instantaneous quantities (i.e.~for fixed radii $a_j$ at a given time) while the latter is a global quantity (i.e.~for given initial radii $a_j^0$). Furthermore, the errors are measured relative to the exact change introduced by the presence of the second bubble (i.e.~$|q_\textrm{exact}-q^*|$, $|U_\textrm{exact}|$ and $|T_M^\textrm{exact}-T_0|$).}\label{tab:accuracy}
\end{center}
\end{table}


\end{document}